\documentclass[11pt]{article}

\usepackage[letterpaper]{geometry}  
\usepackage[english]{babel}         
\usepackage{graphicx}               
\usepackage{float}                  
\usepackage[colorlinks=true, allcolors=blue]{hyperref}  
\usepackage{amsmath}
\usepackage{mathrsfs}
\usepackage{amsfonts}
\usepackage{csvsimple}
\usepackage{tabularx}
\usepackage{amsthm}
\usepackage{subcaption}
\usepackage{xcolor}
\usepackage{authblk}
\usepackage{float}
\usepackage{makecell}
\usepackage{hyperref}
\usepackage{array}
\numberwithin{equation}{section}
\graphicspath{ {images/} } 

\title{\bf\Large {New quasi-exactly solvable systems from SUSYQM and Bethe Ansatz}}

\author[1]{Siyu Li \footnote{Siyu.Li@latrobe.edu.au}}
\author[2]{Ian Marquette \footnote{i.marquette@latrobe.edu.au}}
\author[3]{Yao-Zhong Zhang \footnote{yzz@maths.uq.edu.au}}

\affil[1]{\em Department of Mathematical and Physical Sciences, La Trobe University, Bundoora, VIC 3086, Australia}
\affil[2]{\em Department of Mathematical and Physical Sciences, La Trobe University, Bendigo, VIC 3552, Australia}
\affil[3]{\em School of Mathematics and Physics, The University of Queensland, Brisbane, QLD 4072, Australia}

\begin{document}

\maketitle
\begin{abstract}
\noindent We give a systematic construction of new quasi-exactly solvable systems via Bethe ansatz and supersymmetric quantum mechanics (SUSYQM). Methods based on the intertwining of supercharges have been extensively used in the literature for exactly solvable systems. We generalize the state-deleting (Krein-Adler) supersymmetric transformations to quasi-exactly exactly solvable (QES) systems building on the Bethe ansatz method and related Bethe roots. This enables us to construct superpartners for a wide class of known QES systems classified previously through a hidden $sl(2)$ algebra. We present our constructions of factorizations and intertwining relations related to  1st-order SUSYQM and the $n=1$ state for 10 nonequivalent types, denoted I,...,X. In order to have a unified treatment we rely on their ODE standard form as this is also the appropriate setting to obtain the Bethe ansatz equations which constrain the polynomial solutions. This setting also allows one to deal with systems with $n$ states in a unified manner, using  analysis based on the Bethe ansatz equations to build the supersymmetric transformations in terms of the Bethe ansatz roots. We derive the Schr\"odinger potentials for the $n=1$  superpartners of the 10 QES cases and give closed-form solutions for the spectra and wavefunctions of the corresponding QES SUSYQM systems. Furthermore, we present numerical results for higher excited states up to the $n=10$ level. The results obtained may have wider applicability as our framework is built on ODEs with polynomial coefficients. 
\end{abstract}

\section{Introduction}
 Supersymmetric quantum mechanics (SUSYQM) comes in a variety of forms, e.g., first- and higher-order, confluent, or energy-dependent transformations, and provides a powerful tool for constructing and solving a range of quantum systems \cite{cooper1995supersymmetry, david2010supersymmetric}. The theory found its roots in the early work of Darboux and Moutard on differential equations and was later used in the context of toy models for quantum field theory. It is now one of the standard methods for the construction of deformations of quantum mechanical systems \cite{cooper1995supersymmetry, david2010supersymmetric}.  

Among the properties and features shared by the different transformations, an important one is that they allow one to obtain the wavefunctions of the superpartners.  This includes eigenfunctions that involve orthogonal polynomials beyond classical orthogonal polynomials, often referred to as exceptional orthogonal polynomials. The methods used in SUSYQM include Darboux-Crum (state adding), Krein-Adler (state deleting) transformations and combinations of different types of transformations. From a mathematical and physical point of view, there are various reasons why the deformations of Hamiltonians obtained via SUSY are interesting. One of them is that the models constructed via such transformations are isospectral (or more generally almost isospectral) \cite{cooper1995supersymmetry, david2010supersymmetric}. This framework has been extended to a more general setting \cite{nieto1984relationship} referred to as higher order supersymmetric quantum mechanics. In such constructions, the superpotentials and wavefunctions of the superpartners are determined using formulas that involve the Wronskian of seed solutions \cite{bagrov1995darboux, suzko2009supersymmetry, suzko2009darboux, marquette2014combined}. The seed solutions can be physical or nonphysical states of an initial Hamiltonians. At first order, the factorization of the Hamiltonian is the key property. However, for higher-order transformations, the intertwining relations usually provide simpler determining equations. Some particular cases occur, such as the cases in which the superpartner potentials preserve the form of the initial ones (i.e. shape invariance). This happens when the initial systems are exactly solvable and the additive case classification has been established \cite{dutt1988supersymmetry, bougie2010generation}. If the super partner does not have the shape invariance, then the procedure would produce an infinite set of potentials \cite{fellows2009factorization}. This set may contain both singular and regular potentials. The nature of the Hamiltonians in such chains usually depends on how the transformations are implemented. For certain choices of seed solutions, only the final Hamiltonian may be regular. Such a characterization can be achieved for different Darboux-Crum and Krein-Adler chains, as the wavefunctions of exactly solvable systems have regularity property and the structure of the zeroes of their related orthogonal polynomials is well-known.

Quasi-exactly solvable (QES) systems are models of which only a finite number of eigenvalues and eigenfunctions can be obtained exactly by algebraic means \cite{turbiner1988quasi, turbiner2016one, ushveridze2017quasi}. Unlike exactly-solvable models, they contain many model parameters. The presence of such model parameters significantly complicates the singularity analysis in extending the procedure mentioned above to find superpartner potentials. Other differences occurs as even for simpler QES systems the polynomial solutions of the corresponding ODE are not orthogonal polynomial in usual sense and were refereed as weak orthogonal and Bender-Dunne polynomials. In this paper, we present a systematic approach to obtain new QES systems by applying SUSY transformations to existing QES systems with $sl(2)$ algebraization of the classification in \cite{turbiner1988quasi, turbiner2016one, ushveridze2017quasi}. Some isolated results along this line of research have previously been reported in the literature \cite{marquette2014combined, marquette2013two, agboola2014new, ContrerasAstorga2024, ContrerasAstorga2025}. Our approach is based on the Bethe ansatz and SUSYQM, by implementing SUSY transformations to Bethe ansatz solutions of QES models. It allows for a systematic and unified construction and classification of the SUSY partners of systems from that classification and the corresponding list of potential. In this setting the same parameter constraints occur for the initial Hamiltonian and its superpartner and then simplify proving the quasi-exact solvability property.
\begin{table}[h]
    \centering
    \renewcommand{\arraystretch}{1.5}
    \setlength{\tabcolsep}{3pt}      
    \begin{tabular}{|c|c|}
    \hline
         &  Potential $V$\\
    \hline
       \hyperref[Morse-type potential I]{Morse-type potential I} &  $V=a^2 e^{-2\alpha x}-a[2b+\alpha (2n+1)]e^{-\alpha x}+c(2b-\alpha) e^{\alpha x}+c^2 e^{2\alpha x}$\\
    \hline
        \hyperref[Morse-type potential II]{Morse-type potential II} & \makecell[l]{$ V=d^2 e^{-4\alpha x}+2ade^{-3\alpha x}+[a^2-2d(b+\alpha+n \alpha)]e^{-2\alpha x}$\\ $\qquad-(2ab+\alpha a)e^{-\alpha x}+b^2$} \\
      \hline
        \hyperref[Morse-type potential III]{Morse-type potential III} & $V=d^2e^{4\alpha x}+2ade^{3\alpha x}+[a^2-2d(\alpha+b)]e^{2\alpha x}+b^2+\alpha n(\alpha n-2b)$\\
      \hline
    \hyperref[Poschl-Teller-type potential IV]{P\"oschl-Teller-type potential IV} & \makecell[l]{$V=c^2 \cosh{\alpha x}^4-c(c+2\alpha -2a)\cosh{\alpha x}^2-\{a^2+a\alpha)$\\ $\qquad+\alpha(2n+p)[\alpha(2n+p+1)+2a]\}\cosh{\alpha x}^{-2}
        +a^2+c\alpha-2ac$}\\
    \hline
     \hyperref[Poschl-Teller-type potential V]{P\"oschl-Teller-type potential V}& \makecell[l]{$V=-b^2\cosh{\alpha x}^{-6}-b(2a+3b+\alpha+4n\alpha+2p\alpha)\cosh{\alpha x}^{-4}$\\ $\qquad-\big[a^2+2ab+a\alpha(2p-2n-1)
        +\alpha(2b(n+p-1)+\alpha(n+2n^2$\\ $\qquad+2np+(p-1)p)\alpha)\big]\cosh{\alpha x}^{-2}+[a+2b+\alpha(p-1)]^2$} \\
    \hline
    \hyperref[Sextic potential VI]{Sextic potential VI} & $V=a^2x^6+2abx^4+[b^2-a(4n+3)]x^2$\\
    \hline
     \hyperref[Sextic potential VII]{Sextic potential VII} & \makecell[l]{$ V=a^2r^6+2abr^4+[b^2-a(4n+2l+d-2c+2)]r^2$\\ $\qquad+\frac{1}{r^2}\big[c(c-2l-d+2)+2l(l+d-2)\big]$ }\\
    \hline
     \hyperref[Coulomb-type potential VIII]{Coulomb-type potential VIII} & \makecell[l]{$V=a^2r^2+2abr-\frac{b(Dc-1)}{r}+\frac{c(c-2l-d+2)+2l(d+l-2)}{r^2}+b^2-a(Dc+2n)$}\\
    \hline
     \hyperref[Coulomb-type potential IX]{Coulomb-type potential IX} & \makecell[l]{$V=\frac{b^4}{r^4}+\frac{b(Dc-3)}{r^3}+\frac{c(c-2l-d+2)+2ab+2l(l+d-2)}{r^2}-\frac{a(2n+Dc-1)}{r}+a^2$}\\
    \hline
     \hyperref[Non-singular periodic potential X]{Non-singular periodic potential X} & $V=\alpha^2[a^2\sin{(\alpha x)}^2-(2n+1)a\cos{\alpha x}]$\\
    \hline
    \hyperref[A1 Lame potential XI]{$A_1$ Lam\'e potential XI} & $V=m(m+1)\wp(z)$ \\ 
    \hline
     \hyperref[BC1 Lame potential XII]{$BC_1$ Lam\'e potential XII} & \makecell[l]{$V=(\kappa_2\wp(2z)+\kappa_3\wp(z))|_{\tau=\wp(z)}=\frac{\kappa_2+4\kappa_3}{4}\tau+\frac{\kappa_2}{16}\frac{12g_2\tau^2+36g_3\tau+g_2^2}{4\tau^3-g_2\tau-g_3}$}\\
    \hline
    \end{tabular}
    
    \caption{QES potentials}
    \label{QES potential table}
\end{table}

In order to extend SUSYQM construction to those QES models in a unified way, we  consider the gauge-transformed Schr\"odinger-like equations and standard form for their ODE. Supersymmetry in the ODE setting was used in other contexts and in particular for exactly solvable systems and for the case of confluent transformations
\cite{bermudez2012wronskian, bermudez2016wronskian, schulze2018generalized} and their Jordan chain \cite{contreras2015integral, schulze2018generalized}. Our context concerns quasi-exactly solvable systems and state deleting transformations. 

We consider the general Hamiltonian of QES models:

\begin{equation}\label{general Hamiltonian}
    H_{S \rho}=\rho(x)H_{S}=\rho(x)\left[-\frac{d^2}{dx^2}-\frac{d-1}{x}\frac{d}{dx}-\frac{l(l+d-2)}{x^2}+V_S(x)\right],
\end{equation}
This setting includes different cases of the classification. When the parameter $d=1$, (\ref{general Hamiltonian}) reduces to the standard form of Hamiltonian $H=\rho(x)\left(-\frac{d^2}{dx^2}+V(x)\right)$, where $V(x)=V_S(x)-\frac{l(l+d-2)}{x^2}$. For the radial type of Hamiltonian, we change the independent variable in (\ref{general Hamiltonian}) with $r=x, r>0$, and it is interpreted as a radial variable and the related possibly obtained from a higher dimensional system and separation of variables. They can be transformed into a unified form of ODEs by the gauged transformation. The Bethe ansatz method has given a proof of obtaining closed-form polynomial solutions for the ODE if the parameters in the QES potentials satisfy some certain constraints \cite{zhang2012exact}. 

This work is organized as follows. In Section 2, we present formula for the transformations between the different forms of differential equations, the supercharges and the seed solutions. Sec.\ref{SUSY transformation in QES} introduces the connection between the ODE setting and Schr\"odinger-like setting for QES cases. A unified approach is presented for Schr\"odinger and radial Schr\"odinger-like equation. In Section 3, We relate the Bethe Ansatz method to SUSYQM and provide the general from of SUSYQM of QES cases in terms of Bethe Ansatz roots. Sec.\ref{Type 1 QES}-\ref{Radial type QES} apply the construction to QES models proposed by Turbiner \cite{turbiner1988quasi, turbiner2016one} with the unified setting. We provide figures comparing initial potentials and wavefunctions of the partner potentials and wavefunctions. Sec.\ref{SUSYQM for higher level n} presents explicit results in the case where the method is applied to higher values of $n$. We consider the case $n=10$ and use numerical solutions for the roots of the Bethe Ansatz equations to present explicit results. We also present a comparison between the initial potentials and their wavefunctions with their SUSY partners and related wavefunctions.

As mentioned, in our approach, we consider the problem using the second-order ODE with a polynomial solution $\varphi(z)$. Let us provide some further details and use the following standard form 
\begin{equation}\label{QES ODE general ODE}
    P_4(z)\frac{d^2}{dz^2}\varphi(z)+P_3(z)\frac{d}{dz}\varphi(z)+P_2(z)\varphi(z)=0,
\end{equation}
where $P_4(z), P_3(z), P_2(z)$ are polynomials with degrees 4, 3 and 2, respectively. The study of the polynomial solution $\varphi(z)$ in (\ref{QES ODE general ODE}) can be considered as a study of the polynomial eigenfunction in the Heun equation. The Heun equation and its confluent forms have also been related to the question of solvability and quasi-exact solvability of Schr\"odinger potentials \cite{ishkhanyan2018schrodinger,chen2012heun,figueiredo2024schrodinger,figueiredo2022schr}. The class of Heun equations admitting a finite-dimensional invariant subspace of polynomial eigenfunction $\varphi(z)$ corresponds to the quasi-exactly solvable (QES) system \cite{zhang2012exact,turbiner2016one}. (\ref{QES ODE general ODE}) provides a unified form transformed, via a gauge transformation, from different types Schr\"odinger equation with QES potentials, such as Schr\"odinger-like equation, Schr\"odinger equation and Schr\"odinger-like equation in spherical coordinate, etc \cite{turbiner2016one,turbiner1988quasi,miller2015quasi}. Some new classes of Shcr\"odinger potentials have been found for which their Schr\"odinger equation, with appropriate transformation, can be transformed into the general Heun equation, or into the confluent Heun equation \cite{batic2013potentials, hortaccsu2018heun}. 

 Thus, to obtain the closed-form solutions of SUSY transformed QES models, we will use a unified ODE setting (\ref{QES ODE general ODE}). This is in part due to QES models taking different forms of the Schr\"odinger equations and also to how the Bethe ansatz method must be implemented. In the ODE setting, for obtaining polynomial solutions $\varphi(z)$ in (\ref{QES ODE general ODE}), we rely on the formulas obtained in the following \cite{zhang2012exact}.  We first need to obtain the polynomial eigenfunctions of the initial QES model and then apply the SUSY transformation which will itself be written using Bethe roots. Then the gauge transformation is used to return to the initial form and potential.

\section{State deleting transformation for QES systems}\label{SUSY transformation in QES}

In this Section, we will present the construction of supercharges and their related seeds functions. The supercharges will be used to build the superpartners and then eigenfunctions for the of the corresponding Schr\"odinger-like equations. 

\subsection{Transformations between the Schr\"odinger and ODE form}

The Schr\"odinger-like equation we considered in QES system is $\rho(x)H\psi(x)=E\psi(x)$, where $\rho(x)$ is a weight function \cite{turbiner2016one}. It can be reduced to the ordinary form of the Schr\"odinger equation $H\psi(x)=E\psi(x)$ with $\rho(x)=1$. 

The Schr\"odinger-like equation with the Hamiltonian (\ref{general Hamiltonian}) is

\begin{equation}\label{type 2 QES Schrodinger equation r}
    \rho(x)\left(-\frac{d^2}{dx^2}-\frac{d-1}{x}\frac{d}{dx}+V(x)\right)\psi(x)=E\psi(x), \quad V(x)=V_S(x)-\frac{l(l+d-2)}{x^2}. 
\end{equation}
For radial type of QES models, we set $r=x$ with $r>0$ in (\ref{type 2 QES Schrodinger equation r}). The wavefunctions corresponding (\ref{type 2 QES Schrodinger equation r}) respectively are

\begin{equation}\label{type 2 QES wavefunction x}
     \psi(x)=\varphi(x)e^{-R(x)},\quad R(x)=x^{-\frac{d-1}{2}}e^{-G(x)},\quad G(x)=\int K(x) dx,
\end{equation}
where $\varphi(x)$ are polynomials and $e^{-R(x)}$ are gauge factors. Applying a change of variable and a gauge transformation, we then obtain the ODE as
\begin{equation}\label{type 2 QES general ODE z}
     P_4(z)\varphi''(z)+P_3(z)\varphi'(z)+P_2(z)\varphi(z)=0,\quad z=z(x) \quad \text{or} \quad z=z(r).
\end{equation}

We denote the change of the independent variable as $z=z(x)$. Substituting (\ref{type 2 QES wavefunction x}) into (\ref{type 2 QES Schrodinger equation r}), the transformed ODE can be written into 
\begin{equation}\label{type 2 QES ODE z}
\begin{split}
    &z'(x)^2\rho(z)\varphi''(z)+\rho(z)\left[z''(x)-2K(z)z'(x)^2\right]\varphi'(z)\\
    &+\{E+\rho(z)[(K^2(z)-K'(z))z'(x)^2-z''(x)K(z)-V(z)]\}\varphi(z)=0,\quad z=z(x). 
\end{split}
\end{equation}
Comparing (\ref{type 2 QES general ODE z}) to (\ref{type 2 QES ODE z}), we have
\begin{enumerate}
    \item Change of variables
    \begin{equation}\label{type 2 QES change variable}
    \begin{split}
                x=\int \sqrt{ \frac{P_4(z)}{\rho(z)}} dz. 
    \end{split}
    \end{equation}

    \item Gauge factor
    \begin{equation}\label{type 2 QES gauge factor}
        \begin{split}
            &K(z)=-\frac{P_3(z)}{2P_4(z)}+\frac{1}{4}\left[\frac{P_4'(z)}{P_4(z)}-\frac{\rho'(z)}{\rho(z)}\right],\quad G(z)=\int -\frac{P_3(z)}{2P_4(z)}+\frac{1}{4}\left[\frac{P_4'(z)}{P_4(z)}-\frac{\rho'(z)}{\rho(z)}\right] dz.
        \end{split}
    \end{equation}

    \item Schro\"odinger potential 
    \begin{equation}\label{type 2 QES Schrodinger potential}
        \begin{split}
            &V(z)=\frac{E-P_2(z)}{\rho(z)}+\left[K^2(z)-K'(z)\right]\frac{P_4(z)}{\rho(z)}-\frac{1}{2}K(z)\left[\frac{P_4'(z)}{\rho(z)}-\frac{P_4(z)\rho'(z)}{\rho^2(z)}\right].
        \end{split}
    \end{equation}
\end{enumerate}

We must accommodate the presence of the weight function $\rho$ in (\ref{type 2 QES Schrodinger equation r}). The presence of this weight function requires some modifications to use supersymmetric quantum mechanics.

\subsection{ ODE SUSY transformation}

In order to apply the SUSY formalism of (\ref{type 2 QES Schrodinger equation r}), we consider the SUSY transformation directly in the ODE setting. Suppose that the gauged transformed ODE is given by (\ref{type 2 QES general ODE z}), and we set the polynomial $P_2(z)=E-V_1(z)$. The ODE is 
\begin{equation}\label{Prove gauged ODE z}
      P_4(z)\varphi''(z)+P_3(z)\varphi'(z)+(E-V_1(z))\varphi(z)=0,
\end{equation}
where $V_1(z)$ is the potential and $E$ is the energy. In order to build the first-order SUSY transformation, we use solutions $\phi_i(z)$ and $\phi_j(z)$ of (\ref{Prove gauged ODE z}) as seed functions to build partner solutions. The $\phi(z)$ satisfies the ODE with energy $\Lambda$. 
\begin{equation}\label{Prove gauged ODE Lambda_i z}
    P_4(z)\phi_i''(z)+P_3(z)\phi_i'(z)+(\Lambda_i-V_1(z))\phi_i(z)=0.
\end{equation}
\begin{equation}\label{Prove gauged ODE Lambda_j z} 
    P_4(z)\phi_j''(z)+P_3(z)\phi_j'(z)+(\Lambda_j-V_1(z))\phi_j(z)=0.
\end{equation}
We consider first, the Schr\"odinger-like equation with $d=1$ in (\ref{type 2 QES Schrodinger equation r}). The wavefunctions corresponding to non-seed function $\varphi(z)$ and seed function $\phi(z)$ are derived by a gauge transformation. 
\begin{equation}\label{Prove wavefunctions x}
    \psi_{\Lambda_i}(x)=\phi_i(x)e^{-G(x)},\quad\psi_{\Lambda_j}(x)=\phi_j(x)e^{-G(x)},\quad \psi_E(x)=\varphi(x)e^{-G(x)},\quad x=x(z),
\end{equation}

where $e^{-G(x)}$ is the gauge factors. The wavefunctions in (\ref{Prove wavefunctions x}) are the eigenfunctions of its Shr\"odinger form given as
\begin{equation}\label{Prove Schrodinger form equaions}
   \rho(x) H\psi_{\Lambda_i}(x)=\Lambda_i \psi_{\Lambda_i}(x), \quad\rho(x) H\psi_{\Lambda_j}(x)=\Lambda_j \psi_{\Lambda_j}(x),\quad \rho(x)H\psi_E(x)=E\psi_E(x),
\end{equation}
where $H=-\frac{d^2}{dx^2}+V(x)$ and $\Lambda_i$ and $E$ are different eigenvalues. The constraints of supercharges $\tilde{A}$ and $\tilde{B}$ acting on one of the seed wave functions $\psi_{\Lambda_i}$ and non-seed wave functions $\psi_E$ are
\begin{equation}\label{Prove supercharges and wavefunctions relation} 
               \tilde{A}\psi_{\Lambda_i}=0,\quad \tilde{A}\psi_{E}=\sqrt{\Lambda_i-E} \psi^{(2)}_{E}, \quad \tilde{B}\psi^{(2)}_{E}=\sqrt{\Lambda_i-E}\psi_{E},
\end{equation}
where $\psi^{(2)}_{E}$ is a partner function of the non-seed wave function $\psi_E$. The factorized Hamiltonians $ \bar{H}_1$ and $ \bar{H}_2$ and the intertwining relations take the form
\begin{equation}\label{Prove Schrodinger intertwines}
    \tilde{A} \bar{H}_1= \bar{H}_2\tilde{A},\quad \tilde{B} \bar{H}_2= \bar{H}_1\tilde{B}.
\end{equation}
Combining (\ref{Prove supercharges and wavefunctions relation}) with (\ref{Prove Schrodinger intertwines}), the supercharges in a Schr\"odinger-like setting can be written in terms of the seed wavefunctions $\psi_{\Lambda_{i,j}}$, $i\neq j$
\begin{align}
    &\begin{aligned}\label{Prove Schrodinger supercharges tilde_A}
        \tilde{A}=\sqrt{\rho(\Lambda_i-E)}\left(\frac{d}{dx}-\frac{d}{dx}\ln{\psi_{\Lambda_i}}\right),
    \end{aligned}\\
    &\begin{aligned}\label{Prove Schrodinger supercharges tilde_Ad}
     \tilde{B}=\sqrt{\frac{\rho}{\Lambda_i-E}}\left(\frac{d}{dx}+(\Lambda_i-\Lambda_j)\frac{\psi_{\Lambda_i}\psi_{\Lambda_j}}{\rho{\cal W}(\psi_{\Lambda_i},\psi_{\Lambda_j})}-\frac{d}{dx}\ln{{\cal W}(\psi_{\Lambda_i},\psi_{\Lambda_j})}-\frac{1}{2}\frac{d}{dx}\ln{\rho}+\frac{d}{dx}\ln{\psi_{\Lambda_i}}\right),
    \end{aligned}
\end{align}
where $\rho=\rho(x)$ and ${\cal W}(\psi_{\Lambda_i},\psi_{\Lambda_j})$ is the Wronskian of $\psi_{\Lambda_{i,j}}$. $\Lambda_{i,j}$ and $E$ are energies of seed wavefunctions $\psi_{\Lambda_{i,j}}$ and non-seed wavefunction $\psi_E$, respectively. 

Based on (\ref{Prove supercharges and wavefunctions relation}), (\ref{Prove Schrodinger supercharges tilde_A}) and (\ref{Prove Schrodinger supercharges tilde_Ad}), the SUSY transformation of seed wavefunction $\psi_{\Lambda_j}$ can also be obtained. The coefficients in $\tilde{A}$ and $\tilde{B}$ change to $\sqrt{\rho(\Lambda_i-\Lambda_j)}$ and $\sqrt{\frac{\rho}{\Lambda_i-\Lambda_j}}$, respectively. The relation between supercharges and states in (\ref{Prove supercharges and wavefunctions relation}) will be $\tilde{A}\psi_{\Lambda_j}=\sqrt{\Lambda_i-\Lambda_j} \psi^{(2)}_{\Lambda_j}$ and $ \tilde{B}\psi^{(2)}_{\Lambda_j}=\sqrt{\Lambda_i-\Lambda_j}\psi_{\Lambda_j}$. 
If $\rho(x)=1$, the supercharges $\tilde{A}$ and $\tilde{B}$ will go back to standard form for the supercharges 
\begin{align}
    \begin{aligned}\label{Prove Cooper supercharges}
        \tilde{A}=\sqrt{\Lambda_i-E}\left(\frac{d}{dx}-\frac{d}{dx}\ln{\psi_{\Lambda_i}}\right),\quad \tilde{B}=\sqrt{\frac{1}{\Lambda_i-E}}\left(\frac{d}{dx}+\frac{d}{dx}\ln{\psi_{\Lambda_i}}\right),\quad \rho(x)=1.
    \end{aligned}
\end{align}
The factorization of Hamiltonian and partner Hamiltonian by $\tilde{A}$ and $\tilde{B}$ expressed via the formulas
\begin{equation}\label{Prove supercharges and Hamiltonian relations}
        \bar{H}_1=-\tilde{B} \tilde{A},\quad  \bar{H}_2=-\tilde{A} \tilde{B},\quad  \bar{H}_1=\rho(x)H-\Lambda_i \quad  \bar{H}_2=\rho(x)H_2-\Lambda_i,
\end{equation}
where $H_2=-\frac{d^2}{dz^2}+V^{(2)}(x)$ is the partner Hamiltonian of $H$. $V^{(2)}$ is the partner potential. The partner wavefunctions satisfy $\rho H_2 \psi^{(2)}_E=E \psi^{(2)}_E$, $\rho H_2 \psi^{(2)}_{\Lambda_j}=\Lambda_j \psi^{(2)}_{\Lambda_j}$. 

It should be noted that $A$ and $B$ are not adjoint of each other, i.e., $\tilde{B} \neq \tilde{A}^{\dagger}$. 
The patner wavefunction $\psi^{(2)}_E (\psi^{(2)}_{\Lambda_j})$  is written in terms of the wavefunction $\psi_E (\psi_{\Lambda_j})$ and one of the seed wavefunctions $\psi_{\Lambda_i}$
\begin{equation}\label{Prove SUSY wavefunction}
    \psi_E^{(2)}=\sqrt{\rho}\frac{{\cal W}(\psi_{\Lambda_i},\psi_{E})}{\psi_{\Lambda_i}},\quad \psi_{\Lambda_j}^{(2)}=\sqrt{\rho}\frac{{\cal W}(\psi_{\Lambda_i},\psi_{\Lambda_j})}{\psi_{\Lambda_i}}. 
\end{equation}
The partner potential $V^{(2)}(x)$ is 
\begin{equation}\label{Prove SUSY potential}
    \begin{split}
        V^{(2)}=-V+\frac{2\Lambda_i}{\rho}-\frac{1}{4}\left(\frac{d}{dx}\ln{\rho}\right)^2-\frac{d}{dx}\ln{\rho}\frac{d}{dx}\ln{\psi_{\Lambda_i}}+2\left(\frac{d}{dx}\ln{\psi_{\Lambda_i}}\right)^2+\frac{1}{2\rho}\frac{d^2}{dx^2}\rho. 
    \end{split}
\end{equation}
With the construction of SUSY transformation in simple form of Schr\"odinger-like equation, We extend it to general Schr\"odinger-like equation in (\ref{type 2 QES Schrodinger equation r}). The Hamiltonian in (\ref{general Hamiltonian}) can be expressed as
\begin{equation}\label{general Hamiltonian relation}
    H_{S \rho}=h\rho(x)Hh^{-1},\quad h=x^{-\frac{d-1}{2}},
\end{equation}
Eq (\ref{Prove supercharges and wavefunctions relation}) - (\ref{Prove supercharges and Hamiltonian relations}) corresponding to (\ref{type 2 QES Schrodinger equation r}) are
\begin{equation}
    \tilde{A}_S \psi_{SE}(x)=\sqrt{\Lambda_i-E}\psi_{SE}^{(2)}(x),\quad \tilde{B}_S\psi_{SE}^{(2)}(x)=\sqrt{\Lambda_i-E} \psi_{SE}(x). 
\end{equation}
\begin{equation}
     \bar{H}_{S\rho1}=-\tilde{B}_S\tilde{A}_S, \quad  \bar{H}_{S\rho2}=-\tilde{A}_S \tilde{B}_S,\quad  \bar{H}_{S\rho1}=H_{S \rho}-\Lambda_i,\quad  \bar{H}_{S\rho2}=H_{S\rho2}-\Lambda_i, 
\end{equation}
where
\begin{equation}
\begin{split}
        &\tilde{A}_S=h\tilde{A}h^{-1}, \quad \tilde{B}_S=h\tilde{B}h^{-1},\quad \psi_{SE}(x)=h\psi_E(x),\quad \psi_{SE}^{(2)}(x)=h\psi^{(2)}_E(x),\\
        &H_{S\rho2}=\rho(x)\left[-\frac{d^2}{dr^2}-\frac{d-1}{x}\frac{d}{dx}-\frac{l(l+d-2)}{x^2}+V_S^{(2)}\right],\quad V_S^{(2)}= V^{(2)}(x)-\frac{d^2-4 d (l+1)-4 (l-2) l+3}{4x^2}. 
\end{split}
\end{equation}
$H_{S\rho2}$ is a partner Hamiltonian and $V_S^{(2)}$ is a partner potential in spherical coordinate. 

The SUSY transformation in ODE setting is derived by gauge transformation. We go back to (\ref{Prove Schrodinger supercharges tilde_A})-(\ref{Prove SUSY potential}). We change the variable to $z$ by (\ref{type 2 QES change variable}) and $f=e^{-G(z)}$ is the gauge factor, the ODE supercharges $A$ and $B$ in ODE setting are
\begin{align}
    &\begin{aligned}\label{Prove ODE supercharge A}
        A=f^{-1}\tilde{A}f=\sqrt{P_4(\Lambda_i-E)}\left(\frac{d}{dz}-\frac{d}{dz}\ln{\phi_i}\right),
    \end{aligned}\\
    &\begin{aligned}\label{Prove ODE supercharge Ad}
        B=f^{-1}\tilde{B}f=\sqrt{\frac{P_4}{\Lambda_i-E}}\bigg[\frac{d}{dz}+(\Lambda_i-\Lambda_j)\frac{\phi_i\phi_j}{P_4{\cal W}(\phi_i,\phi_j)}-\frac{1}{2}\frac{d}{dz}\ln{P_4}-\frac{d}{dz}\ln{{\cal W}(\phi_i,\phi_j)}+\frac{d}{dz}\ln{\phi_i}\bigg],
    \end{aligned}
\end{align}
where ${\cal W}(\phi_i,\phi_j)$ is the Wronskian of seed functions $\phi_i$ and $\phi_j$. The equivalent above equations (\ref{Prove supercharges and wavefunctions relation})-(\ref{Prove Schrodinger intertwines}) for polynomial solutions are derived by gauged transformation.
\begin{align}
    &\begin{aligned}\label{Prove supercharges & wavefunction}
            &A\phi_i=0,\quad A\varphi=\sqrt{\Lambda_i-E}\varphi^{(2)},\quad B\varphi^{(2)}=\sqrt{\Lambda_i-E}\varphi,\\
    \end{aligned}\\
    &\begin{aligned}\label{Prove ODE factorized by supercharges}
         \bar{T}_1=BA,\quad  \bar{T}_2=AB,
    \end{aligned}\\
    &\begin{aligned}\label{Prove intertwines in ODE & supercharges}
        A \bar{T}_1= \bar{T}_2A,\quad B \bar{T}_2= \bar{T}_1B,
    \end{aligned}
\end{align}
 where
\begin{equation}\label{Prove gauge transformation in every operator}
    \begin{split}
         &\phi_i=f^{-1}\psi_{\Lambda_i},\quad \phi_j=f^{-1}\psi_{\Lambda_j},\quad \varphi=f^{-1}\psi_E,\quad \varphi^{(2)}=f^{-1}\psi_E^{(2)}, \\
         & \bar{T}_1=-f^{-1}\bar{H}_1f=T_1+\Lambda_i,\quad  \bar{T}_2=-f^{-1}\bar{H}_2f=T_2+\Lambda_i,
    \end{split}
\end{equation}
where $\phi_i,\phi_j$ are seed functions and $\varphi$ is non-seed function. $T_1$ is the gauged transform ODE operator in (\ref{Prove gauged ODE z}). $T_2$ is the SUSY transformed ODE operator with SUSY partner potential $V_2(z)$. The partner function $\varphi^{(2)}$ is the SUSY transofrmed ODE solution. 
\begin{equation}\label{Prove SUSY1 transformed ODE z}
    P_4(z)\frac{d^2}{dz^2}\varphi^{(2)}(z)+P_3(z)\frac{d}{dz}\varphi^{(2)}(z)+(E-V_2(z))\varphi^{(2)}(z)=0,
\end{equation}
The partner solution $\varphi^{(2)} (\phi^{(2)}_j)$ of non-seed function $\varphi (\phi_j)$ can be obtained from (\ref{Prove SUSY wavefunction}) and (\ref{Prove gauge transformation in every operator})
\begin{equation}\label{Prove SUSY1 partner polynomial solution}
    \varphi^{(2)}=\sqrt{P_4}\frac{{\cal W}(\phi_i,\varphi)}{\phi_i},\quad \phi^{(2)}_j=\sqrt{P_4}\frac{{\cal W}(\phi_i,\phi_j)}{\phi_i}.
\end{equation}
To get the $V_2$, we substitute (\ref{Prove ODE supercharge A}) into the intertwine $A\bar{T}_1=\bar{T}_2A$ in (\ref{Prove intertwines in ODE & supercharges}). The partner potential $V_2$ is 
\begin{equation}\label{Prove SUSY1 potential z}
\begin{split}
        V_2=V_1+\frac{1}{4} \bigg[2 \left(\frac{d^2}{dz^2}P_4-2 \frac{d}{dz}P_3\right)+\left(2 P_3-\frac{d}{dz}P_4\right)\frac{d}{dz}\ln{P_4}-4 \frac{d}{dz}P_4 \frac{d}{dz} \ln{\phi_i}
        +8 P_4 \frac{d^2}{dz^2}\ln{\phi_i} \bigg].
\end{split}
\end{equation}
(\ref{Prove SUSY1 partner polynomial solution}) and (\ref{Prove SUSY1 potential z}) are consistent with the result by confluent SUSY algorithm for 1st order SUSY in \cite{schulze2018generalized}.

\section{SUSY for Bethe Ansatz equations for $n$ polynomial states}\label{SUSY for BAE n}
The unified setting of SUSYQM construction is provided. We also need to explore the closed-form eigenfunctions $\varphi$ and $\phi$ of QES cases. Thus, we apply the Bethe Ansatz method to obtain the analytic polynomial solutions. 
General 2nd order ODE is,
\begin{equation}\label{general ODE}
       P_4(z)\frac{d^2}{dz^2}\varphi(z)+P_3(z)\frac{d}{dz}\varphi(z)+P_2(z)\varphi(z)=0.,
\end{equation}
where $X(z), Y(z)$ and $Z(z)$ are polynomials of degrees 4,3 and 2 respectively,
\begin{align}\label{X Y Z}
   P_4(z)=\sum_{k=0}^4 a_k z^k, \quad P_3(z)=\sum_{k=0}^3 b_k z^k,\quad P_2(z)=\sum_{k=0}^2 c_k z^k.
\end{align}
The following two results were respectively proved in \cite{zhang2012exact,zhang2016hidden}. 
The solution of ODE $S(z)$ has a product form
\begin{equation}\label{S(z)}
    \varphi(z)=\prod_{i=1}^n (z-z_i),\qquad \varphi(z)\equiv 1~{\rm for}~ n=0,
\end{equation}
where $z_1,z_2,\ldots, z_n$ are distinct Bethe roots and determined by the Bethe Ansatz equations
\begin{equation}\label{general Bethe ansatz}
    \sum_{j\neq i}^n \frac{2}{z_i-z_j}+\frac{b_3 z_i^3+b_2 z_i^2+b_1 z_i+b_0}{a_4 z_i^4+a_3 z_i^3+a_2 z_i^2+a_1 z_i+a_0}=0, \quad i=1,2,\dots, n.
\end{equation}
The coefficients in $P_4(z), P_3(z)$ with Bethe roots $z_i$ determine the coefficients in $P_2(z)$ by parameters constrains
\begin{align}
    &\begin{aligned}\label{c2}
        c_2=-n(n-1) a_4-n b_3,
    \end{aligned}\\
    &\begin{aligned}\label{c1}
        c_1=-[2(n-1) a_4+b_3]\sum_{i=1}^n z_i-n(n-1) a_3-n b_2,
    \end{aligned}\\
    &\begin{aligned}\label{c0}
        c_0=&-[2(n-1)a_4+b_3] \sum_{i=1}^n z_i^2-2a_4 \sum_{i<j}^n z_i z_j\\
        &-[2(n-1)a_3+b_2]\sum_{i=1}^n z_i-n(n-1) a_2-n b_1.
    \end{aligned}
\end{align}
The above equations (\ref{general Bethe ansatz})-(\ref{c2}) give all polynomials $P_2(z)$ such that the ODE (\ref{general ODE}) has polynomial solutions of the form (\ref{S(z)}). There are $n+1$ eigenfunctions in the form of a polynomial $\varphi(z)$ for (\ref{general ODE}) in QES cases. To construct SUSYQM, we choose two of them $\phi(z)$ and $\varphi(z)$ and set $\phi(z)$ as a seed function. The closed-form of $\phi(z)$ and $\varphi(z)$ are
\begin{equation}
    \begin{split}\label{SUSY BAE n polynomial solutions}
        \varphi=\prod_{i=1}^n(z-z_i),\quad \phi=\prod_{j=1}^n(z-z_j), \quad i,j=1,2,\cdots,n. 
    \end{split}
\end{equation}
Substitute (\ref{SUSY BAE n polynomial solutions}) into (\ref{Prove SUSY1 partner polynomial solution}), the partner polynomial solutions is 
\begin{equation}\label{SUSY BAE n partner polynomial solution}
    \varphi^{(2)}=\sqrt{P_4}\prod_{i=1}^n(z-z_i)\left(\sum_{i=1}^n\frac{1}{z-z_i}-\sum_{j=1}^n\frac{1}{z-z_j}\right). 
\end{equation}
The partner potential in (\ref{Prove SUSY1 potential z}) is given by
\begin{equation}\label{SUSY BAE n partner potential z}
    \begin{split}
        V_2=V_1+\frac{1}{4} \bigg[2 \left(\frac{d^2}{dz^2}P_4-2 \frac{d}{dz}P_3\right)+\left(2 P_3-\frac{d}{dz}P_4\right)\frac{d}{dz}\ln{P_4}-4 \frac{d}{dz}P_4 \sum_{j=1}^n\frac{1}{z-z_j}
        \\-8 P_4 \sum_{j=1}^n\sum_{k=1}^n\frac{1}{(z-z_j)(z-z_k)} \bigg]. 
    \end{split}
\end{equation}
The supercharges operator for the ODE formnoted $A$ and $B$ in (\ref{Prove ODE supercharge A}) and (\ref{Prove ODE supercharge Ad}) are taking the form
\begin{align}
    &\begin{aligned}\label{SUSY BAE n ODE supercharge A}
        A=\sqrt{P_4(\Lambda-E)}\left(\frac{d}{dz}-\sum_{j=1}^n\frac{1}{z-z_j}\right),
    \end{aligned}\\
    &\begin{aligned}\label{SUSY BAE n ODE supercharge Ad}
        B=\sqrt{\frac{P_4}{\Lambda-E}}\Bigg\{\frac{d}{dz}+(\Lambda-E)\left[P_4\left(\sum_{i=1}^n\frac{1}{z-z_i}-\sum_{j=1}^n\frac{1}{z-z_j}\right)\right]^{-1}-\frac{1}{2}\frac{d}{dz}\ln{P_4}+\sum_{j=1}^n\frac{1}{z-z_j}\\-\left(\sum_{i=1}^n\frac{1}{z-z_i}\sum_{m \neq i}^n\frac{2}{z_i-z_m}-\sum_{j=1}^n\frac{1}{z-z_j}\sum_{k \neq j}^n\frac{2}{z_j-z_k}\right)\left(\sum_{i=1}^n\frac{1}{z-z_i}-\sum_{j=1}^n\frac{1}{z-z_j}\right)^{-1}\Bigg\},
    \end{aligned}
\end{align}
Transform (\ref{SUSY BAE n polynomial solutions})-(\ref{SUSY BAE n ODE supercharge Ad}) to Schr\"odinger form and change variable to $x$ by (\ref{type 2 QES change variable}), we have 
\begin{equation}
    \begin{split}\label{SUSY BAE n wavefunctions}
        \psi_{E}(x)=\varphi e^{-G(x)},\quad \psi_{\Lambda}(x)=\phi e^{-G(x)}, \quad x=x(z),
    \end{split}
\end{equation}
where $G(x)$ can be obtained by (\ref{type 2 QES gauge factor}) with $x=x(z)$. The partner wavefunction $\psi_E^{(2)}$ is 
\begin{equation}\label{SUSY BAE n SUSY wavefunction}
    \psi_E^{(2)}=\sqrt{\rho}\prod_{i=1}^n(x-x_i)\left(\sum_{i=1}^n\frac{1}{x-x_i}-\sum_{j=1}^n\frac{1}{x-x_j}\right)e^{-G(x)}. 
\end{equation}
The partner Schr\"odinger potential is 
\begin{equation}\label{SUSY BAE n SUSY potential}
    \begin{split}
        V^{(2)}=-V+\frac{2\Lambda}{\rho}-\frac{1}{4}\left(\frac{d}{dx}\ln{\rho}\right)^2+\frac{1}{2\rho}\frac{d^2}{dx^2}\rho-\frac{d}{dx}\ln{\rho}\left(\sum_{j=1}^n\frac{1}{x-x_j}-K(x)\right)\\+2\left(\sum_{j=1}^n\frac{1}{x-x_j}-K(x)\right)^2, 
    \end{split}
\end{equation}
where $K(x)=G'(x)$ in (\ref{type 2 QES gauge factor}). The supercharges $\tilde{A}$ and $\tilde{B}$ in (\ref{Prove Schrodinger supercharges tilde_A}) and (\ref{Prove Schrodinger supercharges tilde_Ad}) are 
\begin{align}
    &\begin{aligned}\label{SUSY BAE n Schrodinger supercharges tilde_A}
        \tilde{A}=\sqrt{\rho(\Lambda-E)}\left(\frac{d}{dx}-\sum_{j=1}^n\frac{1}{x-x_j}+K(x)\right),
    \end{aligned}\\
    &\begin{aligned}\label{SUSY BAE n Schrodinger supercharges tilde_Ad}
        \tilde{B}=\sqrt{\frac{\rho}{\Lambda-E}}\Bigg\{\frac{d}{dx}-\frac{1}{2}\frac{d}{dx}\ln{\rho}+K(x)+\sum_{j=1}^n\frac{1}{x-x_j}+(\Lambda-E)\left[\rho\left(\sum_{i=1}^n\frac{1}{x-x_i}-\sum_{j=1}^n\frac{1}{x-x_j}\right)\right]^{-1}\\-\left(\sum_{i=1}^n\frac{1}{x-x_i}\sum_{m \neq i}^n\frac{2}{x_i-x_m}-\sum_{j=1}^n\frac{1}{x-x_j}\sum_{k \neq j}^n\frac{2}{x_j-x_k}\right)\left(\sum_{i=1}^n\frac{1}{x-x_i}-\sum_{j=1}^n\frac{1}{x-x_j}\right)^{-1}
        \Bigg\},
    \end{aligned}
\end{align}

\section{QES with a Schr\"odinger form}\label{Type 1 QES}
We apply the construction of the SUSY transformation for QES potentials in the classification \cite{turbiner1988quasi}. We separate those potentials into 3 groups based on their Schr\"odinger form equation. Type 1 QES is the group of cases with ordinary Schr\"odinger equation.i.e. $H\psi(x)=E\psi(x)$.
\subsection{Morse-type potentials I}\label{Morse-type potentials I}
The Schr\"odinger potential is 
\begin{equation}\label{Morse 1 orginal potential x}
    V=a^2 e^{-2\alpha x}-a[2b+\alpha (2n+1)]e^{-\alpha x}+c(2b-\alpha) e^{\alpha x}+c^2 e^{2\alpha x}. 
\end{equation}
The wavefunction is 
\begin{equation}
    \psi(x)=\varphi(x)\exp\left[-\frac{a}{\alpha}e^{-\alpha x}-bx-\frac{c}{\alpha}e^{\alpha x}\right],
\end{equation}
where $\varphi(z)$ is a polynomial and satisfies ODE, which is 
\begin{equation}\label{Morse 1 orginal ODE z}
\begin{split}
        \alpha^2 z^2 \varphi''(z)+[2c\alpha -2a\alpha z^2+\alpha z (2b+\alpha)]\varphi'(z)+[E-V_1(z)]\varphi(z)=0,\quad z=e^{-\alpha x}, 
\end{split}
\end{equation}
\begin{equation}\label{Morse 1 orginal ODE potential z}
    V_1(z)=-(b^2-2ac+2na\alpha z), 
\end{equation}
where $V_1$ is ODE potential. The (\ref{Morse 1 orginal ODE z}) can be written into an operator acting on $\varphi(z)$ i.e. $T_1 \varphi(z)=0$, where $T_1$ is 
\begin{equation}\label{Morse 1 ODE operator T1}
    T_1=\alpha^2 z^2 \frac{d^2}{dz^2}+[2c\alpha -2a\alpha z^2+\alpha z (2b+\alpha)]\frac{d}{dz}-V_1(z). 
\end{equation}
Applying BAE (\ref{general Bethe ansatz})-(\ref{c0}), we get analytic solutions and energies for $n=1$
\begin{equation}\label{Morse 1 initial ODE solution z}
    \varphi_1(z)=z-\frac{2b+\alpha - \sqrt{16ac+(2b+\alpha)^2}}{4a},\quad \varphi_2(z)=z-\frac{2b+\alpha + \sqrt{16ac+(2b+\alpha)^2}}{4a}. 
\end{equation}
\begin{equation}\label{Morse 1 energy}
    E_1=-b^2+2ac-\alpha b - \frac{1}{2}\alpha (\alpha - \sqrt{16ac+(2b+\alpha)^2}),\quad E_2=-b^2+2ac-\alpha b - \frac{1}{2}\alpha (\alpha + \sqrt{16ac+(2b+\alpha)^2}) . 
\end{equation}
There are two independent polynomial solutions for $n=1$. We select $\varphi_1(z)$ as a seed solution to contribute SUSY partner. The SUSY polynomial solution derived by (\ref{Prove SUSY1 partner polynomial solution}) is
\begin{equation}\label{Morse 1 SUSYg z}
    \varphi^{(2)}(z)=\frac{2\alpha z\sqrt{16ac+(2b+\alpha)^2}}{4az-2b-\alpha+\sqrt{16ac+(2b+\alpha)^2}}. 
\end{equation}
Partner ODE potential derived by (\ref{Prove SUSY1 potential z}) is
\begin{equation}\label{Morse 1 SUSY ODE potential z}
\begin{split}
        V_2(z)=&-b^2+2ac+\frac{2\alpha c}{z}+\frac{32a^2\alpha^2 z^2}{(2b-4az+\alpha-\sqrt{4b^2+16ac+4b\alpha+\alpha^2})^2}\\
        &+\frac{8a\alpha^2 z}{2b-4az+\alpha-\sqrt{4b^2+16ac+4b\alpha+\alpha^2}}. 
\end{split}
\end{equation}
We can obtain SUSY ODE with partner potential (\ref{Morse 1 SUSY ODE potential z})
\begin{equation}\label{Morse 1 SUSY ODE z}
    \begin{split}
            \alpha^2 z^2 \frac{d^2}{dz^2}\varphi^{(2)}(z)+[2c\alpha -2a\alpha z^2+\alpha z (2b+\alpha)]\frac{d}{dz}\varphi^{(2)}(z)+[E-V_2(z)]\varphi^{(2)}(z)=0. 
    \end{split}
\end{equation}
We set $T_2\varphi^{(2)}(z)=0$ and $T_2$ is 
\begin{equation}\label{Morse 1 SUSY ODE operator T2}
    T_2=\alpha^2 z^2 \frac{d^2}{dz^2}+[2c\alpha -2a\alpha z^2+\alpha z (2b+\alpha)]\frac{d}{dz}-V_2(z).
\end{equation}
Compared (\ref{Morse 1 orginal ODE z}) and (\ref{Morse 1 SUSY ODE z}), forms of ODEs are exact same except ODE potentials.The ODE supercharges are derived by intertwines (\ref{Prove intertwines in ODE & supercharges}).

\begin{figure}[H]
  \centering
  \begin{subfigure}{0.48\linewidth}
    \centering
    \includegraphics[width=\linewidth]{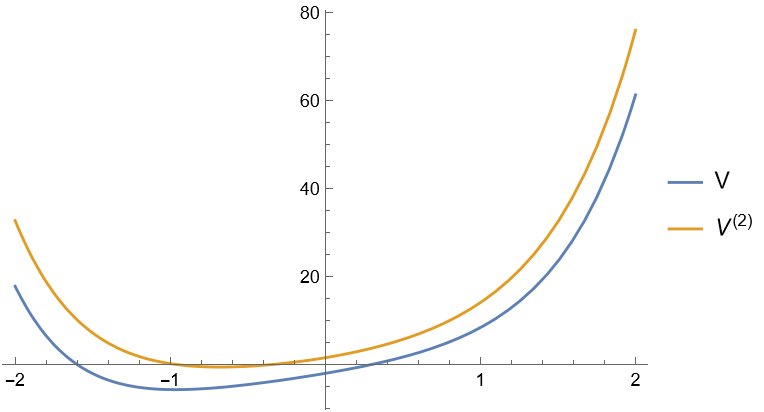}
    \caption{Schr\"odinger potential $V$ and partner Schr\"odinger potential $V^{(2)}$.}
    \label{f case1 V}
  \end{subfigure}
  \hfill
  \begin{subfigure}{0.48\linewidth}
    \centering
    \includegraphics[width=\linewidth]{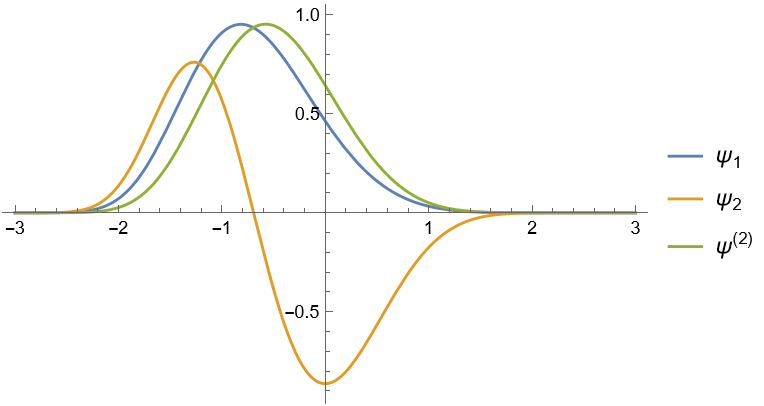}
    \caption{Wavefuntions $\psi_1, \psi_2$ and partner wavefunction $\psi^{(2)}$.}
    \label{f case1 wf}
  \end{subfigure}
  \caption{SUSY transformation in wavefunctions and Schr\"odinger potential.}
  \label{fig case1}
\end{figure}

The $A(z)$ is 
\begin{equation}\label{Morse 1 A(z)}
\begin{split}
       A=\frac{1}{\sqrt{\alpha } \left[16 a c+(\alpha +2 b)^2\right]^{1/4}}\bigg[\alpha  z \frac{d}{dz}-\frac{\alpha  z \left(-\sqrt{16 a c+(\alpha +2 b)^2}+\alpha +2 b\right)+4 \alpha c}{2 z (-2 a z+\alpha +2 b)+4 c}\\-2 a z+\alpha +2 b+\frac{2 c}{z}\bigg]. 
\end{split}
\end{equation}
$B(z)$ is 
\begin{equation}\label{Morse 1 Ad(z)}
    \begin{split}
    B=\alpha ^{3/2} \left[16 a c+(\alpha +2 b)^2\right]^{1/4} \left[z \frac{d}{dz}+\frac{4 c}{z \left(\sqrt{16 a c+(\alpha +2 b)^2}+\alpha +2 b\right)+4 c}-1\right].
    \end{split}
\end{equation}
It can be verified that the following relations hold
\begin{equation}
  \bar{T}_1=BA, \quad  \bar{T}_2=A(z)B(z), \quad  \bar{T}_1=T_1+E_1,\quad  \bar{T}_2=T_2+E_1. 
\end{equation}
We can also verify that 
\begin{equation}
    A\varphi_2(z)=\sqrt{E_2-E_1}\varphi^{(2)}(z)=\sqrt{\alpha} [16ac+(2b+\alpha)^2]^{1/4} \varphi^{(2)}(z),
\end{equation}
\begin{equation}
    B\varphi^{(2)}(z)=-\sqrt{E_2-E_1}\varphi_2(z)=-\sqrt{\alpha} [16ac+(2b+\alpha)^2]^{1/4}\varphi_2(z).
\end{equation}
Let's transform ODE to Schr\"odinger equation and change variable $z$ back to $x$. Wavefunctions of Schr\"odinger equation is 
\begin{equation}\label{Morse 1 wavefunction x}
\begin{split}
        &\psi_1(x)=\left(e^{-\alpha x}-\frac{2b+\alpha - \sqrt{16ac+(2b+\alpha)^2}}{4a}\right)\exp\left[-\frac{a}{\alpha}e^{-\alpha x}-bx-\frac{c}{\alpha}e^{\alpha x}\right],\\
        &\psi_2(x)=\left(e^{-\alpha x}-\frac{2b+\alpha + \sqrt{16ac+(2b+\alpha)^2}}{4a}\right)\exp\left[-\frac{a}{\alpha}e^{-\alpha x}-bx-\frac{c}{\alpha}e^{\alpha x}\right],
\end{split}
\end{equation}
SUSY wavefunction is 
\begin{equation}\label{Morse 1 SUSY wavefunction x}
    \psi^{(2)}(x)=\frac{\alpha e^{-\alpha x}\sqrt{16ac+(2b+\alpha)^2}}{2\left[ae^{-\alpha x}-\frac{1}{4}\left(\alpha+2b-\sqrt{16ac+(2b+\alpha)^2}\right)\right]}\exp\left[-\frac{a}{\alpha}e^{-\alpha x}-bx-\frac{c}{\alpha}e^{\alpha x}\right],
\end{equation}
Schr\"odinger SUSY partner potential is obtained by (\ref{type 2 QES Schrodinger potential})
\begin{equation}\label{Morse 1 SUSY potential x}
\begin{split}
        V^{(2)}(x)&=a^2e^{-2\alpha x}-a(2b+\alpha)e^{-\alpha x}+c(2b+\alpha)e^{\alpha x}+c^2e^{2\alpha x}\\
        &+\frac{32\alpha^2a^2e^{-2\alpha x}}{(2b-4ae^{-\alpha x}+\alpha-\sqrt{16ac+(2b+\alpha)^2})}+\frac{8\alpha^2 ae^{-\alpha x}}{2b-4ae^{-\alpha x}+\alpha-\sqrt{16ac+(2b+\alpha)^2}}. 
\end{split}
\end{equation}
We set parameter $a=b=c=\alpha=1$ and plot the figures. The Fig.\ref{fig case1} provides the potential and the wavefunctions of Schr\"odinger equation and their SUSY partners. The partner potential $V^{(2)}$ is similar to the initial potential $V$ and no singularity in Fig.\ref{f case1 V}. The partner wavefunction $\psi^{(2)}$ is similar to the seed function $\psi_1$ and $\psi_2$ has a zero in the Fig.\ref{f case1 wf}. 
We can write transformed Hamiltonian is 
\begin{equation}\label{Morse 1 SUSY Hamiltonian x}
    H_2(x)=-\frac{d^2}{dx^2}+V^{(2)}(x). 
\end{equation}
It is easy to check 
\begin{equation}
    H_2 \psi^{(2)}(x)=E_2 \psi^{(2)}(x). 
\end{equation}
The ODE supercharges are then transformed to the Schr\"odinger supercharges by applying (\ref{type 2 QES change variable}). We then have the formulas 
\begin{equation}
    \tilde{A}=fAf^{-1}, \quad \tilde{B}=fBf^{-1},\quad f=\exp\left[-\frac{a}{\alpha}e^{-\alpha x}-bx-\frac{c}{\alpha}e^{\alpha x}\right]. 
\end{equation}
It is easy to check
\begin{equation}
    \tilde{A} \psi_2(x)=\sqrt{E_2-E_1}\psi^{(2)}(x),\quad \tilde{B}\psi^{(2)}(x)=-\sqrt{E_2-E_1}\psi_2(x). 
\end{equation}
Then, we obtain the following relations between Hamiltonian and SUSY Hamiltonian with Schr\"odinger supercharges.
\begin{equation}
     \bar{H}_1=-\tilde{B}\tilde{A},\quad  \bar{H}_2=-\tilde{A}\tilde{B},\quad  \bar{H}_1=H-E_1,\quad  \bar{H}_2=H_2-E_1. 
\end{equation}

\subsection{P\"oschl-Teller-type potentials IV}\label{Poschl-Teller-type potentials IV}
\begin{figure}[H]
  \centering
  \begin{subfigure}{0.48\linewidth}
    \centering
    \includegraphics[width=\linewidth]{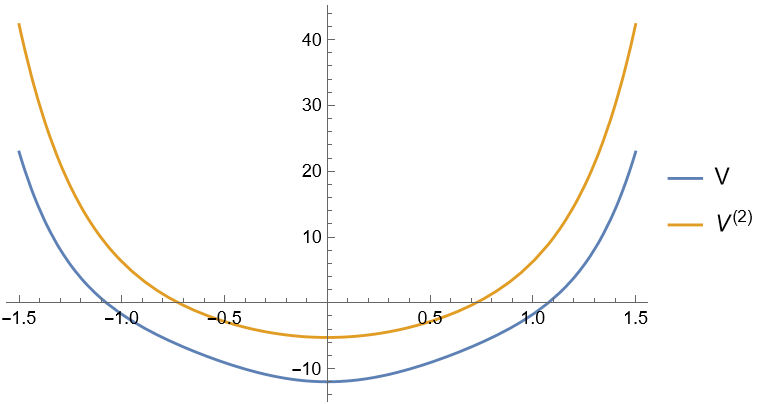}
    \caption{Schr\"odinger potential $V$ and partner Schr\"odinger potential $V^{(2)}$.}
    \label{f case4 V}
  \end{subfigure}
  \hfill
  \begin{subfigure}{0.48\linewidth}
    \centering
    \includegraphics[width=\linewidth]{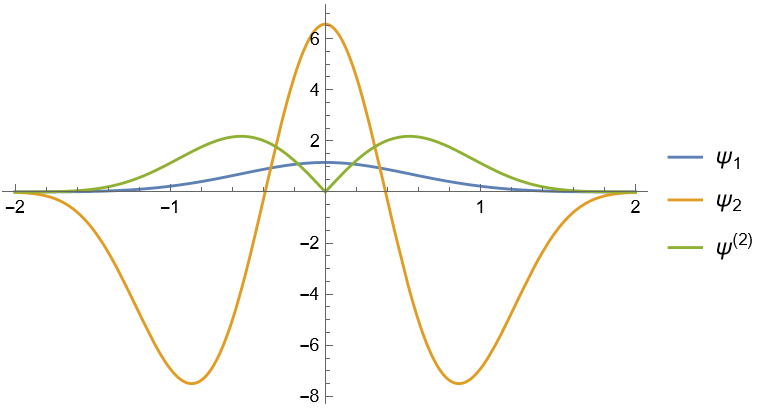}
    \caption{Wavefuntions $\psi_1, \psi_2$ and partner wavefunction $\psi^{(2)}$.}
    \label{f case4 wf}
  \end{subfigure}
  \caption{SUSY transformation in wavefunctions and Schr\"odinger potential.}
  \label{fig case4}
\end{figure}

The initial potential is 
\begin{equation}
\begin{split}
        V=&c^2 \cosh{\alpha x}^4-c(c+2\alpha -2a)\cosh{\alpha x}^2-\{a^2+a\alpha)+\alpha(2n+p)[\alpha(2n+p+1)+2a]\}\cosh{\alpha x}^{-2}\\
        &+a^2+c\alpha-2ac. 
\end{split}
\end{equation}
The Schr\"odinger equation with a weight function has the form $\rho(x)H\psi(x)=E\psi(x)$ and the weight function is 
\begin{equation}
    \rho(x)=\frac{1}{\alpha}. 
\end{equation}
As $\rho(x)$ is only a parameter in this case, the Schr\"odinger equation has a standard form. The wavefunction is 
\begin{equation}
    \psi(x)=\varphi(x)\exp\left[-\frac{c}{2\alpha}\cosh{\alpha x}^2+\frac{a}{2\alpha}\ln{\left(\cosh{\alpha x}^{-2}\right)}+\frac{p}{2}\ln{\left(2\tanh{\alpha x}^2\right)}\right],
\end{equation}
$\varphi(z)$ is a polynomial and satisfies ODE, which is 
\begin{equation}
\begin{split}
        &(4\alpha z^2-4\alpha z^3)\varphi''(z)+[4c+z(4a-4c+4\alpha)-z^2(4a+2\alpha(3+2p))]\varphi'(z)\\
        &+[E+2nz(2a+\alpha+2n\alpha+2p\alpha)-2cp]\varphi(z)=0, \quad z=\cosh{(\alpha x)^{-2}}. 
\end{split}
\end{equation}
The ODE operator is taking the form
\begin{equation}
\begin{split}
        T_1=(4\alpha z^2-4\alpha z^3)\frac{d^2}{dz^2}+[4c+z(4a-4c+4\alpha)-z^2(4a+2\alpha(3+2p))]\frac{d}{dz}\\+2nz(2a+\alpha+2n\alpha+2p\alpha)-2cp
\end{split}
\end{equation}
Applying BAE, we then get the polynomial solutions and energies for $n=1$ as follow
\begin{equation}\label{Poschl-teller IV polynomial solutions z}
\begin{split}
        &\varphi_1(z)=z-\frac{a-c+\alpha-\sqrt{(a+c)^2+2[a+2c(1+p)]\alpha+\alpha^2}}{2a+(3+2p)\alpha},\\
        &\varphi_2(z)=z-\frac{a-c+\alpha+\sqrt{(a+c)^2+2[a+2c(1+p)]\alpha+\alpha^2}}{2a+(3+2p)\alpha}.
\end{split}
\end{equation}
\begin{equation}\label{Poschl-teller IV energy}
\begin{split}
        &E_1=-2\left[a-c(1+p)+\alpha+\sqrt{(a+c)^2+2[a+2c(1+p)]\alpha+\alpha^2}\right],\\
        &E_2=-2\left[a-c(1+p)+\alpha-\sqrt{(a+c)^2+2[a+2c(1+p)]\alpha+\alpha^2}\right]. 
\end{split}
\end{equation}
The SUSY polynomial solution of the ODE is written as 
\begin{equation}\label{Poschl-teller IV SUSY polynomial z}
    \varphi^{(2)}(z)=\frac{4\sqrt{[(1-z)z^2 \alpha][(a+c)^2+2(a+2c(1+p))\alpha+\alpha^2]}}{[2a+(3+2p)\alpha]z-a+c-\alpha+\sqrt{(a+c)^2+2(a+2c(1+p))\alpha+\alpha^2}}. 
\end{equation}
The SUSY potential is expressed 
\begin{equation}\label{Poschl-teller SUSY potential z}
    \begin{split}
        V_2(z)&=2cp-4az-2(3+2p)z\alpha-4\left[\frac{2c}{z^3}+\frac{a}{z^2}+\frac{\alpha p}{(z-1)^2}\right](z-1)z^2+2(3z-a)\left(\frac{c}{z}+a+\frac{\alpha p z}{z-1}\right)\\
        &+\frac{4\alpha z(3z-2)[2a+\alpha(3+2p)]}{c-a+2az-\alpha+3\alpha z+2p\alpha z+\sqrt{a^2+c^2+4c\alpha(p+1)+\alpha^2+2a(\alpha+c)}}\\
        &-\frac{8\alpha z^2(z-1)}{\left[z-\frac{a-c+\alpha-\sqrt{(a+c)^2+2\alpha[a+2c(1+p)]+\alpha^2}}{2a+\alpha(3+2p)}\right]^2}. 
    \end{split}
\end{equation}
We have a partner ODE operator $T_2$ is
\begin{equation}
     T_2=(4\alpha z^2-4\alpha z^3)\frac{d^2}{dz^2}+[4c+z(4a-4c+4\alpha)-z^2(4a+2\alpha(3+2p))]\frac{d}{dz}-V_2(z)
\end{equation}
The polynomial supercharges are
\begin{align}
    &\begin{aligned}
    A=4 \lambda\left(\frac{d}{dz}+\frac{\sqrt{\alpha ^2+2 \alpha  (a+2 c (p+1))+(a+c)^2}-2 a z+a+\alpha -c-\alpha  (2 p+3) z}{2 (z-1) (a z+c)+\alpha  z ((2 p+3) z-2)}\right),\\
    \end{aligned}\\
    &\begin{aligned}
        B=-\frac{1}{\lambda}\bigg\{\alpha z^2(z-1)\frac{d}{dz}+\frac{2 c z (z-1) [2 a (z-1)+\alpha  (2 (p+1) z-1)]+2 c^2 (z-1)^2}{2 (z-1) (a z+c)+\alpha  z ((2 p+3) z-2)}\\+z^2\bigg[\frac{\alpha  \left[\alpha +\alpha z (p+1)  (2 p z+3 z-4)-(z-1) \sqrt{\alpha ^2+2 \alpha  (a+2 c (p+1))+(a+c)^2}\right]}{2 (z-1) (a z+c)+\alpha  z ((2 p+3) z-2)}\\+\frac{2 a^2 (z-1)^2+a \alpha  (z-1) ((4 p+5) z-3)}{2 (z-1) (a z+c)+\alpha  z [(2 p+3) z-2]}\bigg]\bigg\}. 
    \end{aligned}
\end{align}
where $\lambda=z\sqrt{\alpha (1-z)} \left[\alpha ^2+2 \alpha  (a+2 c (p+1))+(a+c)^2\right]^{1/4}$. 
The operator related to the ODE form can be rewritten in terms of the supercharges as
\begin{equation}
     \bar{T}_1=BA,\quad  \bar{T}_2=AB, \quad  \bar{T}_1=T_1+E_1,\quad  \bar{T}_2=T_2+E_1. 
\end{equation}
The action of the supercharges on the polynomial states and SUSY polynomial states are given by 
\begin{equation}
    \begin{split}
        &A\varphi_2(z)=\sqrt{E_2-E_1}\varphi^{(2)}(z)=2\left[\alpha ^2+2 \alpha  (a+2 c (p+1))+(a+c)^2\right]^{1/4}\varphi^{(2)}(z),\\
        &B\varphi^{(2)}(z)=-\sqrt{E_2-E_1}\varphi_2(z)=-2\left[\alpha ^2+2 \alpha  (a+2 c (p+1))+(a+c)^2\right]^{1/4}\varphi_2(z). 
    \end{split}
\end{equation}
Using the transformation back from the variable $z$ to $x$, we get the results in the Schr\"odinger setting. The wavefunctions are provided by
\begin{equation}\label{Poschl-teller IV wavefunction x}
\begin{split}
    \psi_1(x)&=\left(\cosh{\alpha x}^{-2}-\frac{a-c+\alpha-\sqrt{(a+c)^2+2[a+2c(1+p)]\alpha+\alpha^2}}{2a+(3+2p)\alpha}\right)\\
    &\cdot\exp\left[-\frac{c}{2\alpha}\cosh{\alpha x}^2+\frac{a}{2\alpha}\ln{\left(\cosh{\alpha x}^{-2}\right)}+\frac{p}{2}\ln{\left(2\tanh{\alpha x}^2\right)}\right],\\
    \psi_2(x)&=\left(\cosh{\alpha x}^{-2}-\frac{a-c+\alpha+\sqrt{(a+c)^2+2[a+2c(1+p)]\alpha+\alpha^2}}{2a+(3+2p)\alpha}\right)\\
    &\cdot\exp\left[-\frac{c}{2\alpha}\cosh{\alpha x}^2+\frac{a}{2\alpha}\ln{\left(\cosh{\alpha x}^{-2}\right)}+\frac{p}{2}\ln{\left(2\tanh{\alpha x}^2\right)}\right]. 
\end{split}
\end{equation}
The SUSY wavefunction is 
\begin{equation}\label{Poschl-teller IV SUSY wavefunction x}
\begin{split}
       \psi^{(2)}(x)&= \frac{4\sqrt{\alpha[(a+c)^2+2\alpha(a+2c(p+1))+\alpha^2]}\tanh{\alpha x}}{2a+(3+2p)\alpha+\left(c-a-\alpha+\sqrt{(a+c)^2+2\alpha(a+2c(p+1))+\alpha^2}\right)\cosh{\alpha x}^2}\\
       &\cdot\exp\left[-\frac{c}{2\alpha}\cosh{\alpha x}^2+\frac{a}{2\alpha}\ln{\left(\cosh{\alpha x}^{-2}\right)}+\frac{p}{2}\ln{\left(2\tanh{\alpha x}^2\right)}\right]. 
\end{split}
\end{equation}
The SUSY Schr\"odinger potential is 
\begin{equation}\label{Posch-teller IV SUSY potential x}
    \begin{split}
        V^{(2)}(x)=c(a+\alpha)\cosh{2\alpha x}+\frac{1}{8}c^2\cosh{4\alpha x}+p(1+p)\alpha^2 \sinh{\alpha x}^{-2}\\-\big[a^2+\alpha a(3+2p)
        +(6+p(p+3))\alpha^2\big]\cosh{\alpha x}^{-2}
        +a^2-ac-\frac{c^2}{8}\\
        +\frac{8\alpha^2[2a+(3+2p)\alpha]^2\cosh{\alpha x}^{-4}\tanh{\alpha x}^2}{\left[c-a-\alpha+\sqrt{(a+c)^2+2\alpha(a+2c(p+1))+\alpha^2}+(2a+\alpha(3+2p))\cosh{\alpha x}^{-2}\right]^2}\\
        -\frac{4\alpha^2[2a+\alpha(3+2p)](\cosh{2\alpha x}-2)\cosh{\alpha x}^{-4}}{c-a-\alpha+\sqrt{(a+c)^2+2\alpha(a+2c(p+1))+\alpha^2}+(2a+\alpha(3+2p))\cosh{\alpha x}^{-2}}. 
    \end{split}
\end{equation}
The Fig.\ref{fig case4} provides the potential and the wavefunctions of Schr\"odinger equation and their SUSY partners. In this case, we set $a=c=\alpha=1$ and $p=0$. The partner potential $V^{(2)}$ is similar to the initial potential $V$ and no singularity in Fig.\ref{f case4 V}. The partner wavefunction $\psi^{(2)}$ and wavefunctions $\psi_1$ and $\psi_2$ are symmetry at $y$-axis. $\psi^{(2)}$ has a zero at origin and $\psi_2$ has two zeros in Fig.\ref{f case4 wf}. 
From the partner potential, we can construct the partner Hamiltonian
\begin{equation}\label{Posch-teller IV Schrodinger SUSY Hamiltonian x}
    H_2=-\frac{d^2}{dx^2}+V^{(2)}(x), \quad \rho(x)H_2 \psi^{(2)}(x)=E_2 
    \psi^{(2)}(x). 
\end{equation}
The supercharges are expressed in Schr\"odinger form by 
\begin{equation}
\begin{split}
        &\tilde{A}=fAf^{-1}, \quad \tilde{B}=fBf^{-1},\\ 
        &f=\exp\left[-\frac{c}{2\alpha}\cosh{(\alpha x)}^2+\frac{a}{2\alpha}\ln{\left(\cosh{(\alpha x)}^{-2}\right)}+\frac{p}{2}\ln{\left(2\tanh{(\alpha x)}^2\right)}\right].
\end{split}
\end{equation}
We can verify that the following relations hold
\begin{equation}
    \tilde{A} \psi_2(x)=\sqrt{E_2-E_1}\psi^{(2)}(x),\quad \tilde{B}\psi^{(2)}(x)=-\sqrt{E_2-E_1}\psi_2(x). 
\end{equation}
\begin{equation}
     \bar{H}_1=-\tilde{B}\tilde{A}, \quad  \bar{H}_2=-\tilde{A}\tilde{B},\quad  \bar{H}_1=\rho(x)H-E_1,\quad  \bar{H}_2=\rho(x)H_2-E_1. 
\end{equation}

\subsection{Harmonic oscillator-type potentials VI}\label{Harmonic oscillator-type potentials VI}

\begin{figure}[H]
  \centering
  \begin{subfigure}{0.48\linewidth}
    \centering
    \includegraphics[width=\linewidth]{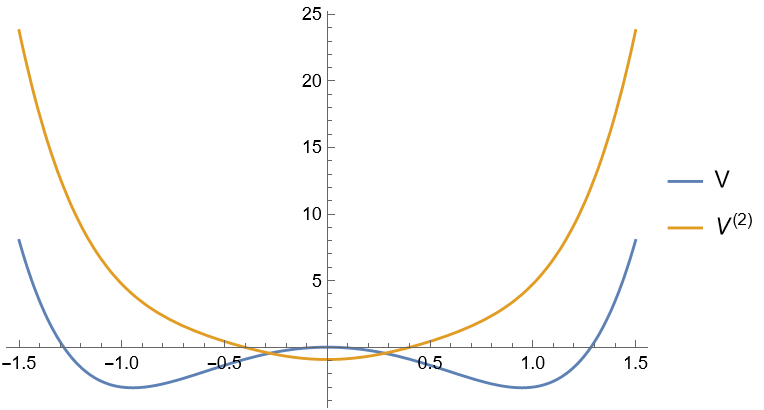}
    \caption{Schr\"odinger potential $V$ and partner Schr\"odinger potential $V^{(2)}$.}
    \label{f case6 V}
  \end{subfigure}
  \hfill
  \begin{subfigure}{0.48\linewidth}
    \centering
    \includegraphics[width=\linewidth]{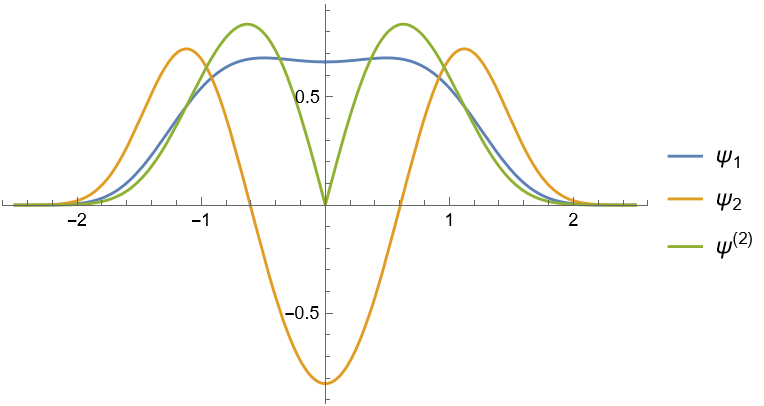}
    \caption{Wavefuntions $\psi_1, \psi_2$ and partner wavefunction $\psi^{(2)}$.}
    \label{f case6 wf}
  \end{subfigure}
  \caption{SUSY transformation in wavefunctions and Schr\"odinger potential.}
  \label{fig case6}
\end{figure}

The initial potential is given by the formula
\begin{equation}
    V=a^2x^6+2abx^4+[b^2-a(4n+3)]x^2. 
\end{equation}
and the wavefunctions are expressible as 
\begin{equation}
   \psi(x)=\varphi(x)\exp\left[-\frac{a}{4}x^4-\frac{b}{2}x^2\right],
\end{equation}
where $\varphi(z)$ is a polynomial that satisfies the ODE 
\begin{equation}
    4z\varphi''(z)+[2-4z(b+az)]\varphi'(z)-(b-E-4anz)\varphi(z)=0, \quad z=x^2. 
\end{equation}
The ODE operator $T_1$ is 
\begin{equation}
    T_1=4z\frac{d^2}{dz^2}+[2-4z(b+az)]\frac{d}{dz}+4anz-b,
\end{equation}
Applying BAE, we have polynomial solutions and energies for $n=1$
\begin{equation}\label{Sextic polynomial solutions z}
    \varphi_1(z)=z-\frac{1}{b-\sqrt{2a+b^2}},\quad \varphi_2(z)=z-\frac{1}{b+\sqrt{2a+b^2}}.
\end{equation}
\begin{equation}\label{Sextic energies}
    E_1=3b-2\sqrt{2a+b^2},\quad E_2=3b+2\sqrt{2a+b^2}. 
\end{equation}
The SUSY polynomial solution is given by 
\begin{equation}\label{Sextic SUSY polynomial z}
    \varphi^{(2)}(z)=\frac{2\sqrt{z(2a+b^2)}\left(b-\sqrt{2a+b^2}\right)}{a\left[\left(b-\sqrt{2a+b^2}\right)z-1\right]}
\end{equation}
while the partner potential is expressed as
\begin{equation}\label{Sextic SUSY potential z}
    V_2(z)=3b+2az+\frac{4\left(z+\frac{1}{b-\sqrt{2a+b^2}}\right)}{\left(z-\frac{1}{b-\sqrt{2a+b^2}}\right)^2}.
\end{equation}
The partner ODE operator $T_2$ takes the form
\begin{equation}
    T_2=4z\frac{d^2}{dz^2}+[2-4z(b+az)]\frac{d}{dz}-V_2(z),
\end{equation}
with the supercharges of ODE SUSY given by the formulas
\begin{align}\label{Sextic ODE supercharges z}
    \begin{aligned}
        &A=4(2a+b^2)^{1/4}\sqrt{z}\left(\frac{d}{dz}-\frac{2az+b-\sqrt{2a+b^2}}{2z(b+az)-1}\right),\\
        &B=\frac{\sqrt{z}}{(2a+b^2)^{1/4}}\left[\frac{d}{dz}-(b+az)-\frac{\sqrt{2a+b^2}-b - 2 a z}{2z(b+az)-1}\right].
    \end{aligned}
\end{align}
The ODE operator can be rewritten in terms of supercharges as follows.
\begin{equation}
     \bar{T}_1=BA,\quad  \bar{T}_2=AB, \quad  \bar{T}_1=T_1+E_1,\quad  \bar{T}_2=T_2+E_1. 
\end{equation}
The supercharges actions with the polynomial states and SUSY polynomial states are given by
\begin{equation}
    \begin{split}
        &A\varphi_2(z)=\sqrt{E_2-E_1}\varphi^{(2)}(z)=2(2a+b^2)^{1/4}\varphi^{(2)}(z),\\
        &B\varphi^{(2)}(z)=-\sqrt{E_2-E_1}\varphi_2(z)=-2(2a+b^2)^{1/4}\varphi_2(z). 
    \end{split}
\end{equation}
 The wavefunctions in the x variable can be expressed as
\begin{equation}\label{Sextic wavefunctions x}
    \psi_1(x)=\left(x^2-\frac{1}{b-\sqrt{2a+b^2}}\right)\exp\left[-\frac{a}{4}x^4-\frac{b}{2}x^2\right],\quad  \psi_2(x)=\left(x^2-\frac{1}{b+\sqrt{2a+b^2}}\right)\exp\left[-\frac{a}{4}x^4-\frac{b}{2}x^2\right]. 
\end{equation}
The SUSY wavefunction is taking the form
\begin{equation}\label{Sextic SUSY wavefunction x}
    \psi^{(2)}(x)=\frac{4x\sqrt{2a+b^2}}{2ax^2+b+\sqrt{2a+b^2}}\exp\left[-\frac{1}{4}x^2(ax^2+2b)\right],
\end{equation}
while the partner potential for the Schr\"odinger form is given by
\begin{equation}\label{Sextic SUSY Schrodinger potential x}
    V^{(2)}(x)=a^2x^6+2abx^4+(b^2-a)x^2+2b+\frac{8\left[a+b\left(b-\sqrt{2a+b^2}\right)\right]x^2-4\sqrt{2a+b^2}+4b}{\left[1+(\sqrt{2a+b^2}-1)x^2\right]^2}. 
\end{equation}
The Fig.\ref{fig case6} provides the potential and the wavefunctions of Schr\"odinger equation and their SUSY partners. In this case, we set $a=b=1$. The partner potential $V^{(2)}$ is similar to the initial potential $V$ and no singularity in Fig.\ref{f case6 V}. The partner wavefunction $\psi^{(2)}$ and wavefunctions $\psi_1$ and $\psi_2$ are symmetry at $y$-axis. $\psi^{(2)}$ has a zero at origin and $\psi_2$ has two zeros in Fig.\ref{f case6 wf}. 
Thus, the partner Hamiltonian is taking the form
\begin{equation}\label{Sextic partner Hamiltonian x}
    H_2=-\frac{d^2}{dx^2}+V^{(2)}(x),\quad H_2\psi^{(2)}(x)=E_2 \psi^{(2)}(x). 
\end{equation}
The supercharges are expressed in the Schr\"odinger form by
\begin{equation}
    \tilde{A}=fAf^{-1}, \quad \tilde{B}=fBf^{-1},\quad f=\exp\left[-\frac{a}{4}x^4-\frac{b}{2}x^2\right].
\end{equation}
We can verify that the following relations hold
\begin{equation}
    \tilde{A} \psi_2(x)=\sqrt{E_2-E_1}\psi^{(2)}(x),\quad \tilde{B}\psi^{(2)}(x)=-\sqrt{E_2-E_1}\psi_2(x). 
\end{equation}
\begin{equation}
     \bar{H}_1=-\tilde{B}\tilde{A}, \quad  \bar{H}_2=-\tilde{A}\tilde{B},\quad  \bar{H}_1=H-E_1,\quad  \bar{H}_2=H_2-E_1. 
\end{equation}

\subsection{Non-singular Periodic potential X}\label{Non-singular Periodic potential X}

\begin{figure}[H]
  \centering
  \begin{subfigure}{0.48\linewidth}
    \centering
    \includegraphics[width=\linewidth]{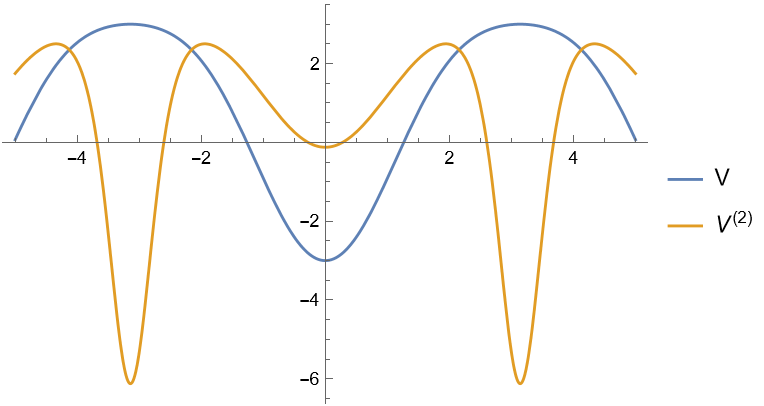}
    \caption{Schr\"odinger potential $V$ and partner Schr\"odinger potential $V^{(2)}$.}
    \label{f case10 V}
  \end{subfigure}
  \hfill
  \begin{subfigure}{0.48\linewidth}
    \centering
    \includegraphics[width=\linewidth]{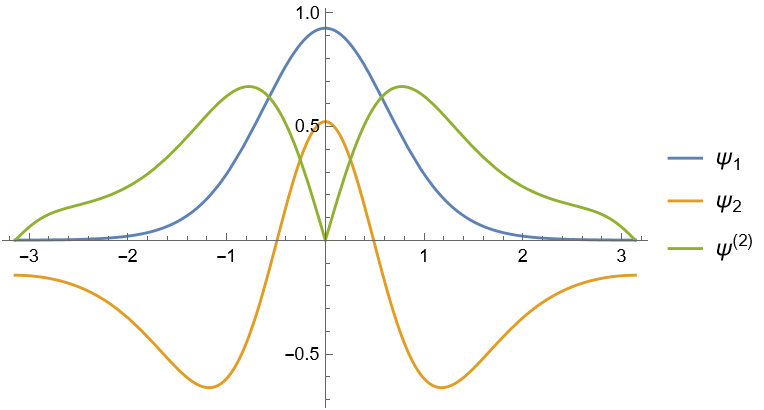}
    \caption{Wavefuntions $\psi_1, \psi_2$ and partner wavefunction $\psi^{(2)}$.}
    \label{f case10 wf}
  \end{subfigure}
  \caption{SUSY transformation in wavefunctions and Schr\"odinger potential.}
  \label{fig case10}
\end{figure}

In this case rhe initial potential can be written as
\begin{equation}
    V=\alpha^2[a^2\sin{(\alpha x)}^2-(2n+1)a\cos{\alpha x}]. 
\end{equation}
The wavefunctions are expressible as 
\begin{equation}
   \psi(x)=\varphi(x)e^{a\cos{\alpha x}},
\end{equation}
where $\varphi(z)$ is a polynomial and satisfies ODE, which is written aa
\begin{equation}
    \alpha(1-z^2)\varphi''(z)-\alpha^2[z+2a(z^2-1)]\varphi'(z)+(E+2a\alpha^2 nz)\varphi(z)=0, \quad z=\cos{\alpha x}. 
\end{equation}
The ODE operator $T_1$ is
\begin{equation}
    T_1=\alpha(1-z^2)\frac{d^2}{dz^2}-\alpha^2[z+2a(z^2-1)]\frac{d}{dz}+2a\alpha^2 nz.
\end{equation}
Applying BAE, we get two linear independent polynomial solutions and energies for $n=1$ as follow
\begin{equation}\label{Non-singular Periodic X ODE polynomial solutions z}
    \begin{split}
        \varphi_1(z)=z+\frac{1+\sqrt{1+16a^2}}{4a},\quad \varphi_2(z)=z+\frac{1-\sqrt{1+16a^2}}{4a}.
    \end{split}
\end{equation}
\begin{equation}\label{Non-singular Periodic X energies}
    E_1=\frac{1}{2}\left(1-\sqrt{1+16a^2}\right)\alpha^2,\quad E_2=\frac{1}{2}\left(1+\sqrt{1+16a^2}\right)\alpha^2.
\end{equation}
The polynomial part of the solution of the ODE is taking the form
\begin{equation}\label{Non-singular Periodic X ODE SUSY polynomial solutions z}
    \varphi^{(2)}(z)=\frac{2\alpha\sqrt{(1-z^2)(1+16a^2)}}{4az+1+\sqrt{1+16a^2}}.
\end{equation}
The SUSY potential of ODE form is then
\begin{equation}\label{Non-singular Periodic X ODE SUSY potential z}
    V_2(z)=\frac{8a\alpha^2\left(z+\sqrt{1+16a^2}z+4a\right)}{\left(1+\sqrt{1+16a^2}+4az\right)^2}.
\end{equation}
The SUSY partner ODE operator $T_2$ is
\begin{equation}
    T_2=\alpha(1-z^2)\frac{d^2}{dz^2}-\alpha^2[z+2a(z^2-1)]\frac{d}{dz}-V_2(z).
\end{equation}
The supercharges of ODE setting are then given by the formulas
\begin{align}\label{Non-singular Periodic X ODE supercharges z}
    &\begin{aligned}
        A=\alpha^2(1+16a^2)^{1/4}\sqrt{z^2-1}\left[\frac{d}{dz}+\frac{\sqrt{16 a^2+1}-4 a z-1}{2 \left(2 a z^2-2 a+z\right)}\right],
    \end{aligned}\\
    &\begin{aligned}
        B=\frac{\sqrt{z^2-1}}{(1+16a^2)^{1/4}}\left[-\frac{d}{dz}+\frac{\sqrt{16 a^2+1}-4 a z-1}{2 \left(2 a z^2-2 a+z\right)}-2 a\right].
    \end{aligned}
\end{align}
We obtain the following relations
\begin{equation}
     \bar{T}_1=BA,\quad  \bar{T}_2=AB, \quad  \bar{T}_1=T_1+E_1,\quad  \bar{T}_2=T_2+E_1. 
\end{equation}
The action of the supercharges with the polynomial states and SUSY polynomial states are given by 
\begin{equation}
    \begin{split}
        &A\varphi_2(z)=\sqrt{E_2-E_1}\varphi^{(2)}(z)=\alpha(1+16a^2)^{1/4}\varphi^{(2)}(z),\\
        &B\varphi^{(2)}(z)=\sqrt{E_2-E_1}\varphi_2(z)=-\alpha(1+16a^2)^{1/4}\varphi_2(z). 
    \end{split}
\end{equation}
The wavefunctions after a chang eof variable in the x variable given by the formulas
\begin{equation}\label{Non-singular Periodic X Schrodinger wavefunctions x}
    \begin{split}
        \psi_1(x)=\left(\cos{\alpha x}+\frac{1+\sqrt{1+16a^2}}{4a}\right)e^{a\cos{\alpha x}},\quad \psi_2(x)=\left(\cos{\alpha x}+\frac{1-\sqrt{1+16a^2}}{4a}\right)e^{a\cos{\alpha x}}.
    \end{split}
\end{equation}
The SUSY wavefunction is then
\begin{equation}\label{Non-singular Periodic X SUSY wavefuncion x}
\psi^{(2)}(x)=\frac{2\alpha\sqrt{1+16a^2} \sin{\alpha x}}{4a\cos{\alpha x}+1+\sqrt{1+16a^2}}.
\end{equation}
The SUSY potential in the Schr\"odinger from is 
\begin{equation}\label{Non-singular Periodic X Schrodinger SUSY potential x}
    \begin{split}
        V^{(2)}(x)&=\frac{2a\alpha^2\bigg[8a^3\cos{\alpha x}^4+4a^2\left(3+\sqrt{1+16a^2}\right)\cos{\alpha x}^3+5a\left(1+\sqrt{1+16a^2}\right)\cos{\alpha x}^2}{\left(4a\cos{\alpha x}+1+\sqrt{1+16a^2}\right)^2}\\
        &\frac{-\left((4a^2+3)\sqrt{1+16a^2}-4a^2+3\right)\cos{\alpha x}+a\left(8a^2+17+\sqrt{1+16a^2}\right)\bigg]}{\left(4a\cos{\alpha x}+1+\sqrt{1+16a^2}\right)^2}. 
    \end{split}
\end{equation}
The Fig.\ref{fig case10} provides the potential and the wavefunctions for the Schr\"odinger equation and their SUSY partners. In this case, we set $a=\alpha=1$. The partner potential $V^{(2)}$ is similar to the initial potential $V$ from the point of view of singularity, as no further singularities in Fig.\ref{f case10 V} are created. The partner wavefunction $\psi^{(2)}$ and the wavefunction $\psi_1$ and $\psi_2$ are periodic functions in Fig.\ref{f case10 wf}. 
We build the partner Hamiltonian which takes the form
\begin{equation}
    H_2=-\frac{d^2}{dx^2}+V^{(2)}(x),\quad H_2 \psi^{(2)}(x)=E_2\psi^{(2)}(x). 
\end{equation}
the Supercharges are expressed in Schr\"odinger form by
\begin{equation}
    \tilde{A}=fAf^{-1}, \quad \tilde{B}=fBf^{-1},\quad f=e^{a\cos{\alpha x}}.
\end{equation}
We can verify that 
\begin{equation}
    \tilde{A} \psi_2(x)=\sqrt{E_2-E_1}\psi^{(2)}(x),\quad \tilde{B}\psi^{(2)}(x)=-\sqrt{E_2-E_1}\psi_2(x). 
\end{equation}
\begin{equation}
     \bar{H}_1=-\tilde{B}\tilde{A}, \quad  \bar{H}_2=-\tilde{A}\tilde{B},\quad  \bar{H}_1=H-E_1,\quad  \bar{H}_2=H_2-E_1. 
\end{equation}

\subsection{$A_1$ Lam\'e equation XI}\label{A1 Lame equation XI}

The Scr\"odinger equation of the case $A_1$ quantum elliptic Calogero-Sutherland model, which also satisfies Lam\'e equation, is expressed as
\begin{equation}\label{A1 Lame Schrodinger equation}
   \left( -\frac{d^2}{dz^2}+m(m+1)\wp(z)\right)\psi(z)=E\psi(z),
\end{equation}
where $\wp(z) \equiv \wp(z|g_2,g_3)$ is the Weierstrass function. It is defined as
\begin{equation}\label{Weierstrass function definitaion}
    (\wp(z))'=4\wp(z)^3-g_2 \wp(z)-g_3=4\left(\wp(z)-e_1\right)\left(\wp(z)-e_2\right)\left(\wp(z)-e_3\right). 
\end{equation}
We consider the change of variables $\xi=\wp(z)+\frac{1}{3}\sum a_i$, which allows to write the Schr\"odinger equation as 
\begin{equation}\label{A1 Lame ODE 1}
    \psi''(\xi)+\frac{1}{2}\left(\frac{1}{\xi-a_1}+\frac{1}{\xi-a_2}+\frac{1}{\xi-a_3}\right)\psi'(\xi)-\frac{m(m+1)\xi-\varepsilon}{4(\xi-a_1)(\xi-a_2)(\xi-a_3)}\psi=0,
\end{equation}
where $\varepsilon=E-\frac{1}{3}m(m+1)\sum a_i$ is a coupling constant corresponding to the energy. There are 4 different types of solution for (\ref{A1 Lame ODE 1}). They belong to the even and odd sectors that correspond to the even value of $m$ and the odd value of $m$, respectively. We merge the 4 types of solution to a general form 
\begin{equation}\label{Lame wavefunction}
    \psi=(a_1-\xi)^{k_1/2}(a_2-\xi)^{k_2/2}(a_3-\xi)^{k_3/2}\varphi(\xi), \quad k_1, k_2, k_3=0,1, 
\end{equation}
where $f(\xi)$ is polynomial. $k_i=0,1$ implies $k_i^2=k_i$. Substituting (\ref{Lame wavefunction}) into (\ref{A1 Lame ODE 1}), we get the gauged transformed ODE. 
\begin{equation}\label{A1 Lame ODE}
    \begin{split}
        4\big[\xi ^3-\xi ^2 (a_1+a_2+a_3)+\xi  (a_1 a_2+a_1 a_3+a_2 a_3)-4 a_1 a_2 a_3\big]\varphi''(\xi)+2\big[\xi ^2 (2 (k_1+k_2+k_3)+3)\\-2 \xi  ( a_1 (k_2+k_3+1)+ a_2 (k_1+k_3+1)+ a_3 (k_1+k_2+1))+ (2 a_1 a_2 k_3+a_1 a_2+2 a_1 a_3 k_2+a_1 a_3\\+2 a_2 a_3 k_1+a_2 a_3)\big]\varphi'(\xi)+\big[\xi  \big(2 \big(k_1^2+ k_1k_2+k_1k_3+k_2^2+k_2 k_3+k_3^2\big)-m (m+1)\big)-a_1 (k_2+k_3)^2\\-a_2 (k_1+k_3)^2-a_3 (k_1+k_2)^2+\varepsilon\big]\varphi(\xi)=0
    \end{split}
\end{equation}
The ODE operator is
\begin{equation}\label{Lame ODE operator T1}
    \begin{split}
        T_1=4\big[\xi ^3-\xi ^2 (a_1+a_2+a_3)+\xi  (a_1 a_2+a_1 a_3+a_2 a_3)-4 a_1 a_2 a_3\big]\frac{d^2}{d\xi^2}+2\big[\xi ^2 (2 (k_1+k_2+k_3)+3)\\-2 \xi  ( a_1 (k_2+k_3+1)+ a_2 (k_1+k_3+1)+ a_3 (k_1+k_2+1))+ (2 a_1 a_2 k_3+a_1 a_2+2 a_1 a_3 k_2+a_1 a_3\\+2 a_2 a_3 k_1+a_2 a_3)\big]\frac{d}{d\xi}+\big[\xi  \big(2 \big(k_1^2+ k_1k_2+k_1k_3+k_2^2+k_2 k_3+k_3^2\big)-m (m+1)\big)-a_1 (k_2+k_3)^2\\-a_2 (k_1+k_3)^2-a_3 (k_1+k_2)^2\big]
    \end{split}
\end{equation}
The parameter constrain can be written as 
\begin{equation}\label{Lame parameter constrain}
    m=2n+k_1+k_2+k_3. 
\end{equation}
The energy is then given by the formula
\begin{equation}
    \begin{split}
        \varepsilon=-2\sum_{i=1}^n \xi_i(2 k_1+2 k_2+2 k_3+4 n-1)+4 n \big[a_1 (k_2+k_3)+a_3 (k_1+k_2)\big]+4 n^2 (a_1+a_3)\\+a_1 k_2^2+2 a_1 k_2 k_3+a_1 k_3^2+a_2 (k_1+k_3+2 n)^2+a_3 k_1^2+2 a_3 k_1 k_2+a_3 k_2^2. 
    \end{split}
\end{equation}
We then obtain the analytic solutions and corresponding energies for $n=1$. 
\begin{equation}\label{Lame solution n=1}
    \begin{split}
        \varphi_1(\xi)=\xi-\frac{K+a_1 (k_2+k_3+1)+a_2 (k_1+k_3+1)+a_3 (k_1+k_2+1)}{2( k_1+k_2+k_3)+3},\\ \varphi_2(\xi)=\xi+\frac{K-a_1 (k_2+k_3+1)-a_2 (k_1+k_3+1)-a_3 (k_1+k_2+1)}{2 (k_1+ k_2+ k_3)+3}.
    \end{split}
\end{equation}
\begin{equation}\label{Lame solution energy n=1}
\begin{split}
    \varepsilon_1=a_1 [k_2 (2 k_3+3)+3 k_3+2]+2 (a_2+a_3-K)+(2 k_1+3) (a_2 k_3+a_3 k_2)+3 k_1 (a_2+a_3),\\
    \varepsilon_2=a_1 [k_2 (2 k_3+3)+3 k_3+2]+2 (a_2+a_3+K)+(2 k_1+3) (a_2 k_3+a_3 k_2)+3 k_1 (a_2+a_3).
\end{split}
\end{equation}
and $K$ is 
\begin{equation}
\begin{split}
        K=\bigg[(a_1 (k_2+k_3+1)+a_2 (k_1+k_3+1)+a_3 (k_1+k_2+1))^2-(2 k_1+2 k_2\\+2 k_3+3) (a_1 (2 a_2 k_3+a_2+2 a_3 k_2+a_3)+a_2 a_3 (2 k_1+1))\bigg]^{1/2}
\end{split}
\end{equation}
The partner polynomial solution is then
\begin{equation}\label{Lame partner polynomial}
    \varphi^{(2)}(\xi)=\frac{4 K \sqrt{(\xi-a_1 ) (\xi -a_2) (\xi -a_3)}}{a_1 (k_2+k_3+1)+a_2 (k_1+k_3+1)+a_3 (k_1+k_2+1)+K-2 \xi  (k_1+k_2+k_3)-3 \xi }
\end{equation}
The ODE partner potential is expressed as
\begin{equation}\label{Lame partner ODE potential}
    \begin{split}
        V_2=\frac{8 (\xi-a_1 ) (\xi -a_2) (\xi -a_3) (2 k_1+2 k_2+2 k_3+3)^2}{(a_1 (k_2+k_3+1)+a_2 (k_1+k_3+1)+a_3 k_1+a_3 k_2+a_3+K-2 k_1 \xi -2 k_2 \xi -2 k_3 \xi -3 \xi )^2}\\+4 (\xi-a_1 ) (\xi -a_2) (\xi -a_3) \left(\frac{k_1}{(a_1-\xi )^2}+\frac{k_2}{(a_2-\xi )^2}+\frac{k_3}{(a_3-\xi )^2}\right)+2 \big[a_1 a_2+a_1 a_3+a_2 a_3\\-2 \xi  (a_1+a_2+a_3)+3 \xi ^2\big] \left(\frac{k_1}{a_1-\xi }+\frac{k_2}{a_2-\xi }+\frac{k_3}{a_3-\xi }\right)+2 a_1 k_2 k_3+a_1 (k_2+k_3)+a_2 (2 k_1 k_3\\+k_1+k_3)+a_3 (2 k_1 k_2+k_1+k_2)+2 \xi  (2 k_1+2 k_2+2 k_3+3)-4 \big[a_1 (a_2+a_3-2 \xi )+a_2 (a_3-2 \xi )\\+\xi  (3 \xi -2 a_3)\big]\bigg[\xi -\frac{a_1 (k_2+k_3+1)+a_2 (k_1+k_3+1)+a_3 (k_1+k_2+1)+K}{2 k_1+2 k_2+2 k_3+3}\bigg]^{-1}.
    \end{split}
\end{equation}
The partner ODE operator is 
\begin{equation}\label{Lame partner ODE operator T2}
    \begin{split}
        T_2=4\big[\xi ^3-\xi ^2 (a_1+a_2+a_3)+\xi  (a_1 a_2+a_1 a_3+a_2 a_3)-4 a_1 a_2 a_3\big]\frac{d^2}{d\xi^2}+2\big[\xi ^2 (2 (k_1+k_2+k_3)+3)\\-2 \xi  ( a_1 (k_2+k_3+1)+ a_2 (k_1+k_3+1)+ a_3 (k_1+k_2+1))+ (2 a_1 a_2 k_3+a_1 a_2+2 a_1 a_3 k_2+a_1 a_3\\+2 a_2 a_3 k_1+a_2 a_3)\big]\frac{d}{d\xi}+V_2(\xi).
    \end{split}
\end{equation}
The ODE supercharges are taking the form
\begin{align}
    &\begin{aligned}\label{Lame ODE supercharge A}
        A=4 \sqrt{K (\xi-a_1 ) (\xi -a_2) (\xi -a_3)}\left[\frac{d}{d\xi}-\left(\xi -\frac{a_1 (k_2+k_3+1)+a_2 (k_1+k_3+1)+a_3 (k_1+k_2+1)+K}{2 k_1+2 k_2+2 k_3+3}\right)^{-1}\right],
    \end{aligned}\\
    &\begin{aligned}\label{Lame ODE supercharge Ad}
        B=\sqrt{\frac{(\xi-a_1 ) (\xi -a_2) (\xi -a_3)}{K}}\bigg\{\frac{d}{d\xi}-\frac{1}{2 (\xi -a_1) (\xi -a_2) (\xi -a_3)}\big[a_1 (k_2+k_3+1)+a_2 (k_1+k_3+1)+a_3 (k_1\\+k_2+1)+K-[2(k_1+k_2+k_3)+3] \xi \big] \bigg[\xi-\frac{a_1 (k_2+k_3+1)+a_2 (k_1+k_3+1)+a_3 (k_1+k_2+1)-K}{2 k_1+2 k_2+2 k_3+3}\\+\frac{2 (a_1-\xi ) (\xi -a_2) (\xi -a_3) (-2 (k_1+k_2+k_3)-3)}{[a_1 (k_2+k_3+1)+a_2 (k_1+k_3+1)+a_3 (k_1+k_2+1)+K-2 \xi  (k_1+k_2+k_3)-3 \xi ]^2}\\+ \frac{3\xi^2-2 \xi  (a_1+a_2+a_3)+a_1 a_2+a_1 a_3+a_2 a_3}{a_1 (k_2+k_3+1)+a_2 (k_1+k_3+1)+a_3 (k_1+k_2+1)+K-2 \xi  (k_1+k_2+k_3)-3 \xi }\bigg]\bigg\}
    \end{aligned}
\end{align}
The ODE operator can be rewritten in terms of supercharges and the following relations hold
\begin{equation}
     \bar{T}_1=BA,\quad  \bar{T}_2=AB, \quad  \bar{T}_1=T_1+\varepsilon_1,\quad  \bar{T}_2=T_2+\varepsilon_1. 
\end{equation}
The supercharges action between the polynomial states and SUSY polynomial states is givne by the formulas
\begin{equation}
    \begin{split}
        &A\varphi_2(z)=\sqrt{\varepsilon_2-\varepsilon_1}\varphi^{(2)}(z)=2\sqrt{K}\varphi^{(2)}(z),\\
        &B\varphi^{(2)}(z)=-\sqrt{\varepsilon_2-\varepsilon_1}\varphi_2(z)=-2\sqrt{K}\varphi_2(z). 
    \end{split}
\end{equation}
We transform back to the Schr\"odinger form involving the Weierstrass function $\wp(z)$. The wavefunctions are then expressed as
\begin{equation}
    \begin{split}
        \psi_1(z)=\bigg[-\frac{a_1 (k_2+k_3+1)+a_2 (k_1+k_3+1)+a_3 (k_1+k_2+1)+K}{2 k_1+2 k_2+2 k_3+3}+\frac{1}{3} (a_1+a_2+a_3)+\wp(z)\bigg]\\\cdot\bigg(\frac{2 a_1-a_2-a_3}{3}-\wp(z)\bigg)^{k1/2}\bigg(\frac{2 a_2-a_1-a_3}{3}-\wp(z)\bigg)^{k2/2}\bigg(\frac{2 a_3-a_1-a_2}{3}-\wp(z)\bigg)^{k3/2},\\
        \psi_2(z)=\bigg[-\frac{a_1 (k_2+k_3+1)+a_2 (k_1+k_3+1)+a_3 (k_1+k_2+1)-K}{2 k_1+2 k_2+2 k_3+3}+\frac{1}{3} (a_1+a_2+a_3)+\wp(z)\bigg]\\\cdot\bigg(\frac{2 a_1-a_2-a_3}{3}-\wp(z)\bigg)^{k1/2}\bigg(\frac{2 a_2-a_1-a_3}{3}-\wp(z)\bigg)^{k2/2}\bigg(\frac{2 a_3-a_1-a_2}{3}-\wp(z)\bigg)^{k3/2}.
    \end{split}
\end{equation}
The partner wavefunction is given by
\begin{equation}
    \begin{split}
        \psi^{(2)}(z)=\frac{4K\sqrt{[a_2+a_3-2a_1+3 \wp(z)] [a_1+a_2-2 a_3+3 \wp(z)] [a_1-2 a_2+a_3+3 \wp(z)]}}{a_1 (k_2+k_3-2k_1)+a_2 (k_1-2 k_2+k_3)+a_3 (k_1+k_2-2 k_3)+3 K-3 \wp(z) (2 k_1+2 k_2+2 k_3+3)}\\\cdot\bigg(\frac{2 a_1-a_2-a_3}{3}-\wp(z)\bigg)^{k1/2}\bigg(\frac{2 a_2-a_1-a_3}{3}-\wp(z)\bigg)^{k2/2}\bigg(\frac{2 a_3-a_1-a_2}{3}-\wp(z)\bigg)^{k3/2}.
    \end{split}
\end{equation}
The corresponding partner potential is then writtenn as
\begin{equation}
    \begin{split}
        V^{(2)}=\frac{1}{3}\bigg\{-\frac{8 (2 k_1+2 k_2+2 k_3+3)^2 (2 a_1-a_2-a_3-3 \wp(z)) (a_1+a_2-2 a_3+3 \wp(z)) (a_1-2 a_2+a_3+3 \wp(z))}{D^2}\\-\frac{12 (2 k_1+2 k_2+2 k_3+3) \left(a_1^2-a_1 (a_2+a_3)+a_2^2-a_2 a_3+a_3^2-9 \wp(z)^2\right)}{D}-\big[a_1 a_2+a_1a_3+a_2 a_3-a_1^2\\-a_2^2-a_3^2+9 \wp(z)^2\big] \bigg[\frac{3 k_1}{a_2+a_3-2 a_1+3 \wp(z)}+\frac{3 k_2}{a_1+a_3-2 a_2+3 \wp(z)}+\frac{3 k_3}{a_1+a_2-2 a_3+3 \wp(z)}\bigg]\\-2 (2 a_1-a_2-a_3-3 \wp(z)) (a_1+a_2-2 a_3+3 \wp(z)) (a_1-2 a_2+a_3+3 \wp(z)) \bigg[\frac{k_1}{(a_2+a_3-2 a_1+3 \wp(z))^2}\\+\frac{k_2}{(a_1+a_3-2 a_2+3 \wp(z))^2}+\frac{k_3}{(a_1+a_2-2 a_3+3 \wp(z))^2}\bigg]+6 a_1 k_2 k_3+3 a_1 (k_2+k_3)+3 a_2 (2 k_1 k_3+k_1+k_3)\\+\frac{W}{(a_1+a_2-2 a_3+3 \wp(z)) (a_1+a_3-2 a_2+3 \wp(z)) (a_2+a_3-2 a_1+3 \wp(z))}+3 a_3 (2 k_1 k_2+k_1+k_2)\\+2 (2 k_1+2 k_2+2 k_3+3) (a_1+a_2+a_3+3 \wp(z))\bigg\},
    \end{split}
\end{equation}
where
   \begin{equation}
        \begin{split}
            W=\big[(k_1+k_2+k_3) (a_1+a_2+a_3+3 \wp(z))^2-3 a_3 (k_1+k_2) (a_1+a_2+a_3+3 \wp(z))-3 a_2 (k_1+k_3) (a_1\\+a_2+a_3+3 \wp(z))-3 a_1 (k_2+k_3) (a_1+a_2+a_3+3 \wp(z))+9 a_1 a_2 k_3+9 a_1 a_3 k_2+9 a_2 a_3 k_1\big]^2,\\
            D=-2 a_1 k_1+a_1 k_2+a_1 k_3+a_2 k_1-2 a_2 k_2+a_2 k_3+a_3 k_1+a_3 k_2-2 a_3 k_3+3 K-3 \wp(z) (2 k_1+2 k_2\\+2 k_3+3). 
        \end{split}
    \end{equation}
Thus, we obtain the partner Hamiltonian as
\begin{equation}
    H_2=-\frac{d^2}{dz^2}+V^{(2)}(z),\quad H_2\psi^{(2)}(z)=\varepsilon_2 \psi^{(2)}(z). 
\end{equation}
The supercharges in the Schr\"odinger form are 
\begin{equation}
\begin{split}
        &\tilde{A}=fAf^{-1}, \quad \tilde{B}=fBf^{-1},\\ &f=\bigg(\frac{2 a_1-a_2-a_3}{3}-\wp(z)\bigg)^{k1/2}\bigg(\frac{2 a_2-a_1-a_3}{3}-\wp(z)\bigg)^{k2/2}\bigg(\frac{2 a_3-a_1-a_2}{3}-\wp(z)\bigg)^{k3/2}.
\end{split}
\end{equation}
We can verify that 
\begin{equation}
    \tilde{A} \psi_2(x)=\sqrt{E_2-E_1}\psi^{(2)}(x),\quad \tilde{B}\psi^{(2)}(x)=-\sqrt{E_2-E_1}\psi_2(x). 
\end{equation}
\begin{equation}
     \bar{H}_1=-\tilde{B}\tilde{A}, \quad  \bar{H}_2=-\tilde{A}\tilde{B},\quad  \bar{H}_1=H-E_1,\quad  \bar{H}_2=H_2-\varepsilon_1. 
\end{equation}

\subsection{$BC_1$ Lam\'e equation XII}\label{BC1 Lame equation XII}
The initial potential is
\begin{equation}\label{BC1 Lame equation initial Schrodinger potential z}
    V(z)=\kappa_2\wp(2z)+\kappa_3\wp(z),
\end{equation}
where $\wp(z)\equiv \wp(z|g_2,g_3)$ is a Weierstrass function, which is decided by the elliptic functions. 
\begin{equation}
    (\wp'(z))^2=4\wp(z)^3-g_2\wp(z)-g_3,
\end{equation}
Changing the variable to $\tau=\wp(z)$, the potential will be a simple rational function, the 
\begin{equation}
\begin{split}                       V(\tau)=\left(\kappa_2\wp(2z)+\kappa_3\wp(z)\right)|_{\tau=\wp(z)}=\frac{\kappa_2+4\kappa_3}{4}\tau+\frac{\kappa_2}{16}\frac{12g_2\tau^2+36g_3\tau+g_2^2}{4\tau^3-g_2\tau-g_3}, 
\end{split}
\end{equation}
To study the QES potential we set the coupling constants 
\begin{equation}
    \kappa_2=2\mu(\mu-1),\quad \kappa_3=(n+2\mu)(n+2\mu+1).
\end{equation}
The Hamiltonian of the $BC_1$ Lam\'e equation with the potential is then
\begin{equation}
\begin{split}
    H=\left(-\frac{1}{2}\frac{d^2}{dz^2}+V(z)\right)|_{\tau=\wp(z)},
\end{split}
\end{equation}
The wavefunction can be expressed as
\begin{equation}\label{BC1 Lame equation initial wavefunction z}
    \psi(z)=\varphi(z)\left[\wp'(z)\right]^{\mu}, 
\end{equation}
where $\varphi(z)$ is a polynomial solution of the ODE, which is 
\begin{equation}\label{BC1 Lame equation ODE tau}
\begin{split}
       \left(2 \tau ^3 -\frac{g_2 \tau }{2}-\frac{g_3}{2}\right)\varphi''(\tau)+\left[3 (2 \mu +1) \tau ^2-\frac{g_2 (2 \mu +1)}{4} \right]\varphi'(\tau)+\big[E-  (6 \mu +2 n+1)n \tau\big]\varphi(\tau)=0,\\\quad \tau=\wp(z). 
\end{split}
\end{equation}
The ODE operator is taking the form
\begin{equation}
    T_1=\left(2 \tau ^3 -\frac{g_2 \tau }{2}-\frac{g_3}{2}\right)\frac{d^2}{d\tau^2}+\left[3 (2 \mu +1) \tau ^2-\frac{g_2 (2 \mu +1)}{4} \right]\frac{d}{d\tau}-  (6 \mu +2 n+1)n \tau. 
\end{equation}
Using the BAE, we get the polynomial solutions and energies for $n=1$.
\begin{equation}\label{BC1 Lame equation polynomial solution tau}
    \varphi_1(\tau)=\tau -\frac{\sqrt{g_2}}{2 \sqrt{3}},\quad \varphi_2(\tau)=\tau+\frac{\sqrt{g_2}}{2 \sqrt{3}}. 
\end{equation}
\begin{equation}\label{BC1 Lame equation energy}
    E_1=-\frac{\sqrt{3g_2} }{2} (2 \mu +1),\quad E_2=\frac{\sqrt{3g_2} }{2} (2 \mu +1). 
\end{equation}
The polynomial solution for the SUSY partner is 
\begin{equation}\label{BC1 Lame equation partner polynomial solution tau}
    \varphi^{(2)}(\tau)=\sqrt{6g_2}\frac{\sqrt{4 \tau ^3-g_2 \tau -g_3}}{\sqrt{3g_2} -6 \tau }
\end{equation}
The partner potential for the ODE setting and the first order SUSY is given by the formula
\begin{equation}\label{BC1 Lame equation partner potential tau}
\begin{split}
        V_2(\tau)=-\frac{3}{4 \left(\sqrt{3g_2}-6 \tau \right)^2 \left(g_2 \tau +g_3-4 \tau ^3\right)}\bigg[4 \left(g_2 \tau +g_3-4 \tau ^3\right) \left(\sqrt{3} g_2^{3/2}+3 g_2 \tau +12 \left(g_3-\tau ^3\right)\right)\\+\mu  \left(12 \tau ^2-4 \sqrt{3g_2} \tau +g_2\right) \left(g_2^2+24 \tau  \left(g_3+2 \tau ^3\right)\right)\bigg].
\end{split}
\end{equation}
Thus, the partner ODE is derived as 
\begin{equation}\label{BC1 Lame equation partner ODE tau}
    \left(2 \tau ^3 -\frac{g_2 \tau }{2}-\frac{g_3}{2}\right)\varphi''(\tau)+\left[3 (2 \mu +1) \tau ^2-\frac{g_2 (2 \mu +1)}{4} \right]\varphi'(\tau)+\big[E-V_2(\tau)]\varphi(\tau)=0
\end{equation}
The partner ODE operator is then
\begin{equation}
    T_2=\left(2 \tau ^3 -\frac{g_2 \tau }{2}-\frac{g_3}{2}\right)\frac{d^2}{d\tau^2}+\left[3 (2 \mu +1) \tau ^2-\frac{g_2 (2 \mu +1)}{4} \right]\frac{d}{d\tau}-V_2(\tau). 
\end{equation}
The supercharges in the ODE setting are expressed as
\begin{align}
    &\begin{aligned}\label{BC1 Lame equation ODE supercharge A}
        A=(3g_2)^{1/4} \sqrt{2(2 \mu +1) \left(4 \tau ^3-g_2 \tau -g_3\right)}\left[\frac{d}{d\tau}+\frac{6}{\sqrt{3g_2} -6 \tau }\right],
    \end{aligned}\\
    &\begin{aligned}\label{BC1 Lame equation ODE supercharge Ad}
        B=-\frac{1}{(3g_2)^{1/4}\sqrt{2(2 \mu +1)}}\bigg[\sqrt{4 \tau ^3-g_2 \tau -g_3}\frac{d}{d\tau}+\frac{24 (3 \mu +1) \tau ^3 -12 \sqrt{3g_2} \mu  \tau ^2-6 g_2 (\mu +1) \tau -6 g_3+\sqrt{3} g_2^{3/2} \mu}{\left(\sqrt{3g_2}-6 \tau \right) \sqrt{4 \tau ^3-g_2 \tau -g_3}}\bigg].
    \end{aligned}
\end{align}
The relations between ODE supercharges and ODE operators are given by the formulas
\begin{equation}
   \bar{T}_1=BA, \quad  \bar{T}_2=AB, \quad  \bar{T}_1=T_1+E_1,\quad  \bar{T}_2=T_2+E_1. 
\end{equation}
The relation between polynomials states and SUSY polynomials states are obtained and given by 
\begin{equation}
    A\varphi_2(\tau)=\sqrt{E_2-E_1}\varphi^{(2)}(\tau)=(3g_2)^{1/4}\sqrt{2 \mu +1} \varphi^{(2)}(\tau),
\end{equation}
\begin{equation}
    B\varphi^{(2)}(\tau)=-\sqrt{E_2-E_1}\varphi_2(\tau)=-(3g_2)^{1/4}\sqrt{2 \mu +1} \varphi_2(\tau).
\end{equation}
We transform into the Schr\"odinger setting via a change of variable to $z$, the wavefunctions for $n=1$ are then
\begin{equation}\label{BC1 Lame equation wavefunction z}
    \psi_{1}(z)=\left(\wp(z) -\frac{\sqrt{g_2}}{2 \sqrt{3}}\right)\left[\wp'(z)\right]^{\mu}, \quad \psi_{2}(z)=\left(\wp(z) +\frac{\sqrt{g_2}}{2 \sqrt{3}}\right)\left[\wp'(z)\right]^{\mu},
\end{equation}
The partner wavefunction is 
\begin{equation}\label{BC1 Lame equation SUSY wavefunction z}
    \psi^{(2)}(z)=\frac{\sqrt{6g_2\left(4 \wp(z)^3-g_2 \wp(z)-g_3\right)} \left(g_2 \wp(z)+g_3-4 \wp(z)^3\right)^{\mu /2}}{\sqrt{3g_2} -6 \wp(z)}
\end{equation}
The partner potential in Schr\"odinger setting is 
\begin{equation}\label{BC1 Lame equation partner Schrodinger potential z}
    \begin{split}
        V^{(2)}(z)=\frac{\mu  (\mu +1) \left(g_2-12 \wp(z) ^2\right)^2}{8 \left(4 \wp(z) ^3-g_2 \wp(z) -g_3\right)}-\frac{3 \left[\sqrt{3} g_2^{3/2}+3 g_2 \wp(z) +12 \left(g_3-\wp(z) ^3\right)\right]}{\left(\sqrt{3g_2}-6 \wp(z)\right)^2}. 
    \end{split}
\end{equation}
We define a partner Hamiltonian operator is 
\begin{equation}\label{BC1 Lame equation partner Hamiltonian z}
    H_2=-\frac{1}{2}\frac{d^2}{dz^2}+V^{(2)}(z),\quad H_2\psi^{(2)}(z)=E_2\psi^{(2)}(z). 
\end{equation}
SUSY supercharges in the Schr\"odinger setting are
\begin{equation}
        \tilde{A}=fAf^{-1}, \quad \tilde{B}=fBf^{-1},\quad f=\left[\wp'(z)\right]^{\mu}. 
\end{equation}
We can verify that
\begin{equation}
    \tilde{A} \psi_2(z)=\sqrt{E_2-E_1}\psi^{(2)}(z),\quad \tilde{B}\psi^{(2)}(z)=-\sqrt{E_2-E_1}\psi_2(z). 
\end{equation}
\begin{equation}
     \bar{H}_1=-\tilde{B}\tilde{A}, \quad  \bar{H}_2=-\tilde{A}\tilde{B},\quad  \bar{H}_1=H-E_1,\quad  \bar{H}_2=H_2-E_1. 
\end{equation}

\section{QES with Schr\"odinger-like form}\label{Type 2 QES}
Type 2 QES is the group of cases containing a weight function $\rho(x)$ in their Schr\"odinger form equation i.e. $\rho(x)H\psi(x)=E\psi(x)$. 
\subsection{Morse-type potentials II}\label{Morse-type potentials II}
\begin{figure}[H]
  \centering
  \begin{subfigure}{0.48\linewidth}
    \centering
    \includegraphics[width=\linewidth]{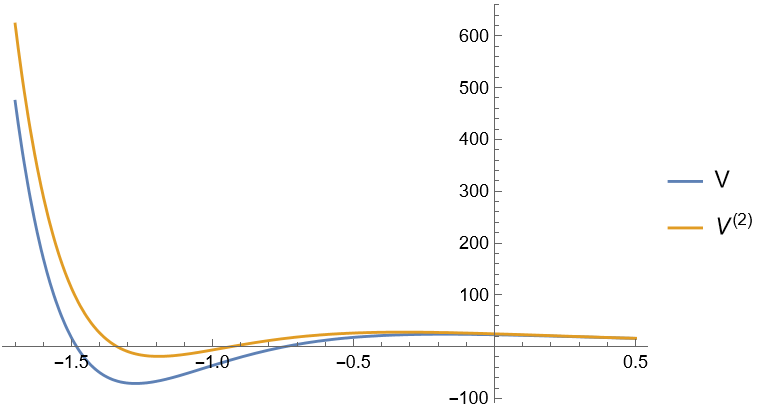}
    \caption{Schr\"odinger-like potential $V$ and partner Schr\"odinger-like potential $V^{(2)}$.}
    \label{f case2 V}
  \end{subfigure}
  \hfill
  \begin{subfigure}{0.48\linewidth}
    \centering
    \includegraphics[width=\linewidth]{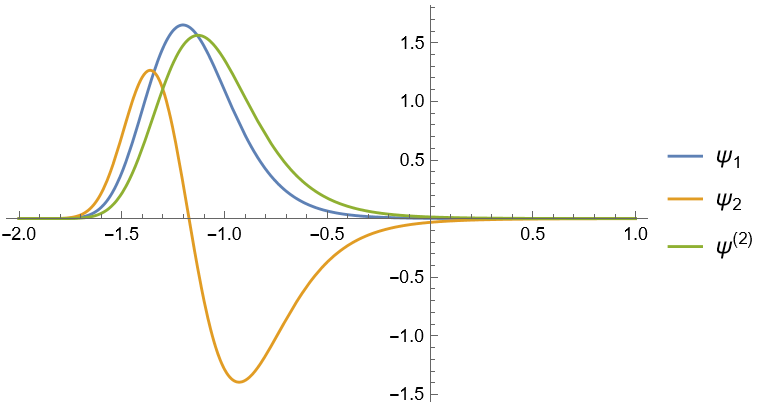}
    \caption{Wavefuntions $\psi_1, \psi_2$ and partner wavefunction $\psi^{(2)}$.}
    \label{f case2 wf}
  \end{subfigure}
  \caption{SUSY transformation in wavefunctions and Schr\"odinger-like potential.}
  \label{fig case2}
\end{figure}

The initial potential is taken as
\begin{equation}\label{Morse 2 initial Schrodinger potential x}
    V=d^2 e^{-4\alpha x}+2ade^{-3\alpha x}+[a^2-2d(b+\alpha+n \alpha)]e^{-2\alpha x}-(2ab+\alpha a)e^{-\alpha x}+b^2. 
\end{equation}
The Schr\"odinger equation with a weight function has the form $\rho(x)H\psi(x)=E\psi(x)$ and the weight function is given by
\begin{equation}\label{Morse 2 weight function x}
    \rho(x)=\frac{e^{\alpha x}}{\alpha}. 
\end{equation}
The wavefunction is then expressed as
\begin{equation}
    \psi(x)=\varphi(x)\exp\left[-\frac{d}{2\alpha}e^{-2\alpha x}-\frac{a}{\alpha}e^{-\alpha x}-bx\right],
\end{equation}
where $\varphi(z)$ is a polynomial and satisfies an ODE which is written as
\begin{equation}\label{Morse 2 initial ODE z}
    \alpha z \varphi''(z)+[2b-2z(a+dz)+\alpha]\varphi'(z)+(E+2ndz)\varphi(z)=0, \quad z=e^{-\alpha x}. 
\end{equation}
The ODE operator $T_1$ is then 
\begin{equation}
    T_1=\alpha z \frac{d^2}{dz^2}+[2b-2z(a+dz)+\alpha]\frac{d}{dz}+2ndz.
\end{equation}
Using BAE, we get polynomial solutions and energies for $n=1$. The explicit formulas are
\begin{equation}\label{Morse 2 ODE polynomial solutions z}
    \varphi_1(z)=z+\frac{a+\sqrt{a^2+4bd+2d\alpha}}{2d},\quad \varphi_2(z)=z+\frac{a-\sqrt{a^2+4bd+2d\alpha}}{2d}. 
\end{equation}
\begin{equation}\label{Morse 2 energy}
    E_1=a-\sqrt{a^2+2d(2b+\alpha)},\quad E_2=a+\sqrt{a^2+2d(2b+\alpha)}. 
\end{equation}
We find the SUSY polynomial partner
\begin{equation}\label{Morse 2 SUSY polynomial z}
    \varphi^{(2)}(z)=\frac{2\sqrt{\alpha z[a^2+2d(2b+\alpha)]}}{2dz+a+\sqrt{a^2+4bd+2\alpha d}}. 
\end{equation}
The SUSY ODE potential is expressed as 
\begin{equation}\label{Morse 2 SUSY ODE potential z}
    V_2(z)=a+\frac{b}{z}+dz+\frac{\alpha}{4z}+\frac{8d^2 \alpha z}{(a+2dz+\sqrt{a^2+4bd+2d\alpha})^2}-\frac{2d\alpha}{a+2dz+\sqrt{a^2+4bd+2d\alpha}}. 
\end{equation}
The SUSY ODE is derived as
\begin{equation}\label{Morse 2 SUSY ODE z}
    \alpha z \varphi''(z)+[2b-2z(a+dz)+\alpha]\varphi'(z)+(E-V_2(z))\varphi(z)=0
\end{equation}
The SUSY partner ODE operator $T_2$ is 
\begin{equation}
    T_2=\alpha z \frac{d^2}{dz^2}+[2b-2z(a+dz)+\alpha]\frac{d}{dz}-V_2(z).
\end{equation}
We find ODE supercharges $A(z)$ and $B(z)$
\begin{equation}\label{Morse 2 SUSY ODE supercharge A z}
    A=\sqrt{2 \alpha  z} \left[a^2+2 d (\alpha +2 b)\right]^{1/4}\left[\frac{d}{dz}+\frac{a + 2 d z-\sqrt{a^2+4 b d+2 \alpha  d}}{\alpha +2 b-2 a z-2 d z^2}\right]. 
\end{equation}
\begin{equation}\label{Morse 2 SUSY ODE supercharge Ad z}
    \begin{split}
     B=\frac{1}{\sqrt{2 \alpha  z} \left[a^2+2 d (\alpha +2 b)\right]^{1/4}}\bigg[2 \alpha  z \frac{d}{dz} +4 b-4 a z+3 \alpha -4 d z^2\\+\frac{2 \alpha z \left(z\sqrt{a^2+4 b d+2 \alpha  d}+az-\alpha -2 b\right)}{-2 z (a+d z)+\alpha +2 b}\bigg].\\
    \end{split}
\end{equation}
The relations between the ODE supercharges and ODE operators are expressed by the formula
\begin{equation}
   \bar{T}_1=BA, \quad  \bar{T}_2=AB, \quad  \bar{T}_1=T_1+E_1,\quad  \bar{T}_2=T_2+E_1. 
\end{equation}
The relation between polynomials and SUSY polynomials are obtained explicitly
\begin{equation}
    A\varphi_2(z)=\sqrt{E_2-E_1}\varphi^{(2)}(z)=\sqrt{2}[a^2+2d(2b+\alpha)]^{1/4} \varphi^{(2)}(z),
\end{equation}
\begin{equation}
    B\varphi^{(2)}(z)=-\sqrt{E_2-E_1}\varphi_2(z)=-\sqrt{2}[a^2+2d(2b+\alpha)]^{1/4}\varphi_2(z).
\end{equation}
We transform to type 2 QES Schr\"odinger form by returning to the $x$ variable. The wavefunctions are then
\begin{equation}\label{Morse 2 wavefunctions x}
\begin{split}
        &\psi_1(x)=\left(e^{-\alpha x}+\frac{a+\sqrt{a^2+4bd+2d\alpha}}{2d}\right)\exp\left[-\frac{d}{2\alpha}e^{-2\alpha x}-\frac{a}{\alpha}e^{-\alpha x}-bx\right],\\
        &\psi_2(x)=\left(e^{-\alpha x}+\frac{a-\sqrt{a^2+4bd+2d\alpha}}{2d}\right)\exp\left[-\frac{d}{2\alpha}e^{-2\alpha x}-\frac{a}{\alpha}e^{-\alpha x}-bx\right]. 
\end{split}
\end{equation}
The SUSY wavefunction is 
\begin{equation}\label{Morse 2 SUSY wavefunction x}
    \psi^{(2)}(x)=\frac{2\sqrt{\alpha e^{-\alpha x}[a^2+2d(\alpha+2b)]}}{a+2de^{-\alpha x}+\sqrt{a^2+2d(\alpha+2b)}}\exp\left[-\frac{d}{2\alpha}e^{-2\alpha x}-\frac{a}{\alpha}e^{-\alpha x}-bx\right]. 
\end{equation}
The SUSY Schr\"odinger-like potential is 
\begin{equation}\label{Morse 2 SUSY potential x}
\begin{split}
        V^{(2)}(x)=&d^2e^{-4\alpha x}+2ade^{-3\alpha x}+(a^2-2bd-d\alpha)e^{-2\alpha x}-2abe^{-\alpha x}+\frac{1}{4}(2b+\alpha)^2\\
        &+\frac{8d^2\alpha^2e^{-2\alpha x}}{\left[2de^{-\alpha x}+a+\sqrt{a^2+4bd+2d\alpha}\right]^2}-\frac{2d\alpha^2e^{-\alpha x}}{2de^{-\alpha x}+a+\sqrt{a^2+4bd+2d\alpha}}. 
\end{split}
\end{equation}
The Fig.\ref{fig case2} provides the potential and the wavefunctions of Schr\"odinger-like equation and their SUSY partners. In this case, we set $a=-6,b=1, d=2, \alpha=1$. The partner potential $V^{(2)}$ is similar to the initial potential $V$ and no singularity in Fig.\ref{f case2 V}. The partner wavefunction $\psi^{(2)}$ and wavefunctions $\psi_1$ are similar and $\psi_2$ has a zero at origin in Fig.\ref{f case2 wf}. 
We define a SUSY Hamiltonian operator is 
\begin{equation}\label{Morse 2 SUSY Hamiltonian operator x}
    H_2=-\frac{d^2}{dx^2}+V^{(2)}(x).
\end{equation}
It is easy to check 
\begin{equation}
    \rho(x)H_2 \psi^{(2)}(x)=E_2 \psi^{(2)}(x).
\end{equation}
Transformed ODE supercharges to Schr\"odinger supercharges is shown
\begin{equation}
    \tilde{A}=fAf^{-1}, \quad \tilde{B}=fBf^{-1},\quad f=\exp\left[-\frac{d}{2\alpha}e^{-2\alpha x}-\frac{a}{\alpha}e^{-\alpha x}-bx\right]. 
\end{equation}
The transformation of differential operator. We have $\frac{d}{dz}=-\frac{e^{\alpha x}}{\alpha} \frac{d}{dx}$. We can verify that 
\begin{equation}
    \tilde{A} \psi_2(x)=\sqrt{E_2-E_1}\psi^{(2)}(x),\quad \tilde{B}\psi^{(2)}(x)=-\sqrt{E_2-E_1}\psi_2(x). 
\end{equation}
\begin{equation}
     \bar{H}_1=-\tilde{B}\tilde{A}, \quad  \bar{H}_2=-\tilde{A}\tilde{B},\quad  \bar{H}_1=\rho(x)H-E_1,\quad  \bar{H}_2=\rho(x)H_2-E_1. 
\end{equation}

\subsection{Morse-type potentials III}\label{Morse-type potentials III}
\begin{figure}[H]
  \centering
  \begin{subfigure}{0.48\linewidth}
    \centering
    \includegraphics[width=\linewidth]{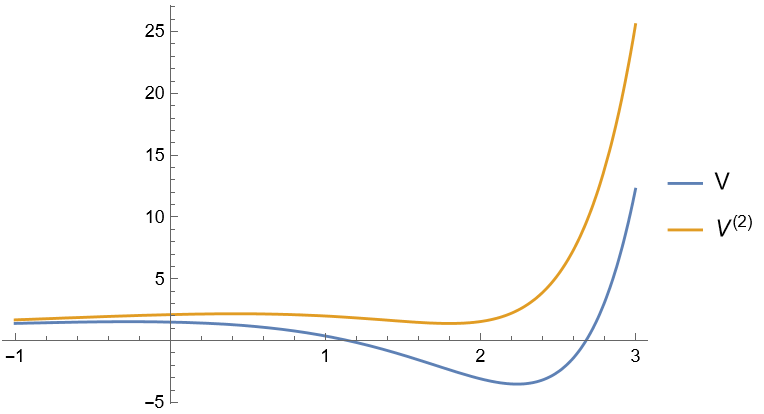}
    \caption{Schr\"odinger-like potential $V$ and partner Schr\"odinger-like potential $V^{(2)}$.}
    \label{f case3 V}
  \end{subfigure}
  \hfill
  \begin{subfigure}{0.48\linewidth}
    \centering
    \includegraphics[width=\linewidth]{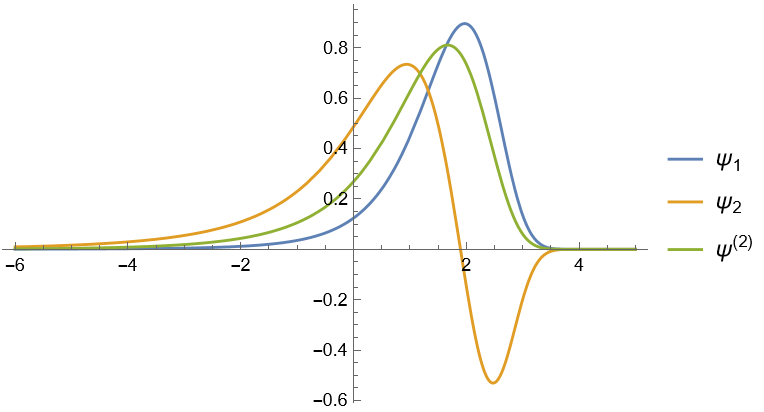}
    \caption{Wavefuntions $\psi_1, \psi_2$ and partner wavefunction $\psi^{(2)}$.}
    \label{f case3 wf}
  \end{subfigure}
  \caption{SUSY transformation in wavefunctions and Schr\"odinger-like potential.}
  \label{fig case3}
\end{figure}

The initial potential for this case if given by the formula
\begin{equation}
    V=d^2e^{4\alpha x}+2ade^{3\alpha x}+[a^2-2d(\alpha+b)]e^{2\alpha x}+b^2+\alpha n(\alpha n-2b). 
\end{equation}
The Schr\"odinger equation with a weight function has the form $\rho(x)H\psi(x)=E\psi(x)$ and the weight function is 
\begin{equation}
    \rho(x)=\frac{e^{-\alpha x}}{\alpha}. 
\end{equation}
The wavefunction is taking the form
\begin{equation}
    \psi(x)=\varphi(x)\exp\left[-\frac{d}{2\alpha}e^{2\alpha x}-\frac{a}{\alpha}e^{-\alpha x}+bx\right],
\end{equation}
where $\varphi(z)$ is a polynomial and satisfies the ODE given by the following expression 
\begin{equation}
    \alpha z^3 \varphi''(z)+[2d+z(2a-2bz+\alpha z)]\varphi'(z)+[E+nz(2b-n\alpha)]\varphi(z)=0, \quad z=e^{-\alpha x}. 
\end{equation}
The ODE operator $T_1$ is 
\begin{equation}
    T_1=\alpha z^3 \frac{d^2}{dz^2}+[2d+z(2a-2bz+\alpha z)]\frac{d}{dz}+nz(2b-n\alpha).
\end{equation}
Applying BAE, we get polynomials and energies for $n=1$
\begin{equation}\label{Morse 3 ODE polynomial solutions z}
    \varphi_1(z)=z-\frac{a-\sqrt{a^2+4bd-2\alpha d}}{2b-\alpha},\quad \varphi_2(z)=z-\frac{a+\sqrt{a^2+4bd-2\alpha d}}{2b-\alpha}.
\end{equation}
\begin{equation}\label{Morse 3 energy}
    E_1=-a-\sqrt{a^2+4bd-2d\alpha},\quad E_2=-a+\sqrt{a^2+4bd-ad\alpha}. 
\end{equation}
We use the ground state as a seed function and find the SUSY polynomial for the partner ODE.
\begin{equation}\label{Morse 3 SUSY polynomial z}
    \varphi^{(2)}(z)= \frac{2z^{3/2}\sqrt{\alpha(a^2+4bd-2d\alpha)}}{(2b-\alpha)z-a+\sqrt{a^2+4bd-2d\alpha}}. 
\end{equation}
Partner potential for partner ODE is 
\begin{equation}\label{Morse 3 SUSY ODE potential z}
    V_2(z)=a+\frac{3d}{z}-bz+\frac{5\alpha z}{4}-\frac{3\alpha(\alpha-2b)z^2}{a-2bz+\alpha z-\sqrt{a^2+4bd-2d\alpha}}+\frac{2\alpha z^3}{\left(z-\frac{a-\sqrt{a^2+4bd-2d\alpha}}{2b-\alpha}\right)^2}. 
\end{equation}
The SUSY partner ODE operator $T_2$ is
\begin{equation}
    T_2=\alpha z^3 \frac{d^2}{dz^2}+[2d+z(2a-2bz+\alpha z)]\frac{d}{dz}-V_2(z).
\end{equation}
The ODE supercharges are
\begin{align}\label{Morse 3 ODE supercharges z}
    &\begin{aligned}
A=\sqrt{2 \alpha } z^{3/2} \left(a^2+4 b d-2 \alpha  d\right)^{1/4}\bigg[\frac{d}{dz}+\frac{(2 b-\alpha)  z-a-\sqrt{a^2+4 b d-2 \alpha  d}}{2 a z-2 b z^2+2 d+\alpha  z^2}\bigg],
    \end{aligned}\\
    &\begin{aligned}
        B=\frac{1}{\sqrt{2 \alpha } z^{3/2} \left(a^2+4 b d-2 \alpha  d\right)^{1/4}}\bigg[2 \alpha  z^3 \frac{d}{dz}+4 az-z^2 (\alpha +4 b)+4d\\+\frac{2 \alpha  z^3 \left(\sqrt{a^2+4 b d-2 \alpha  d}+a-2 b z+\alpha  z\right)}{z (2 a+z (\alpha -2 b))+2 d}\bigg].
    \end{aligned}
\end{align}
The ODE operator can be related the to supercharges  
\begin{equation}
     \bar{T}_1=BA,\quad  \bar{T}_2=AB, \quad  \bar{T}_1=T_1+E_1,\quad  \bar{T}_2=T_2+E_1.
\end{equation}
The ODE supercharges and their action on the SUSY polynomial states can be determined explicitly via the formulas
\begin{align}
    &\begin{aligned}
            A\varphi_2(z)=\sqrt{E_2-E_1}\varphi^{(2)}(z)=\sqrt{2}(a^2+4bd-2d\alpha)^{1/4}\varphi^{(2)}(z)
    \end{aligned}\\
    &\begin{aligned}
        B\varphi^{(2)}(z)=-\sqrt{E_2-E_1}\varphi_2(z)=-\sqrt{2}(a^2+4bd-2d\alpha)^{1/4}\varphi_2(z). 
    \end{aligned}
\end{align}
Using a change of variable form $z$ to $x$, all operators and polynomial states can be written in terms of x. The wavefunctions are now given as
\begin{equation}\label{Morse 3 wavefunctions x}
\begin{split}
        &\psi_1(x)=\left(e^{-\alpha x}-\frac{a-\sqrt{a^2+4bd-2\alpha d}}{2b-\alpha}\right)\exp\left(-\frac{d}{2\alpha}e^{2\alpha x}-\frac{a}{\alpha}e^{\alpha x}+bx\right),\\
        &\psi_2(x)=\left(e^{-\alpha x}-\frac{a+\sqrt{a^2+4bd-2\alpha d}}{2b-\alpha}\right)\exp\left(-\frac{d}{2\alpha}e^{2\alpha x}-\frac{a}{\alpha}e^{\alpha x}+bx\right). 
\end{split}
\end{equation}
The SUSY wavefunction of the partner Hamiltonian is then 
\begin{equation}\label{Morse 3 SUSY wavefunction x}
    \psi^{(2)}(x)=\frac{2e^{-\sqrt{\alpha}x}\sqrt{\alpha(a^2+4bd-2d\alpha)}}{2b-\alpha+e^{\alpha x}(\sqrt{a^2+4bd-2d\alpha}-a)}\exp\left(-\frac{a}{\alpha}e^{\alpha x}-\frac{d}{2\alpha}e^{2\alpha x}+b x\right)
\end{equation}
The partner Schr\"odinger-like potential is 
\begin{equation}\label{Morse 3 SUSY Schrodinger potential x}
\begin{split}
        V^{(2)}(x)&=d^2e^{4\alpha x}+2ade^{3\alpha x}+(a^2-2bd+d\alpha)e^{2\alpha x}-2abe^{\alpha x}+\frac{1}{4}(4b^2-4b\alpha+5\alpha^2)\\
        &-\frac{3\alpha^3e^{\alpha x}\left(a-\sqrt{a^2+4bd-2d\alpha}\right)+a^2(\alpha^2+2b\alpha-8b^2)}{\left[\alpha-2b+\left(a-\sqrt{a^2+4bd-2d\alpha}\right)e^{\alpha x}\right]^2}+\frac{6\alpha^2b}{\alpha-2b+\left(a-\sqrt{a^2+4bd-2d\alpha}\right)e^{\alpha x}}.
\end{split}
\end{equation}
The Fig.\ref{fig case3} provides the potential and the wavefunctions of Schr\"odinger-like equation and their SUSY partners. In this case, we set $a=-1, b=1, d=0.5, \alpha=0.5$. The partner potential $V^{(2)}$ is similar to the initial potential $V$ and no singularity in Fig.\ref{f case3 V}. The partner wavefunction $\psi^{(2)}$ and wavefunctions $\psi_1$ are positive and $\psi_2$ has a zero in Fig.\ref{f case3 wf}. 
Then, we define the partner Hamiltonian $H_2$ is 
\begin{equation}\label{Morse 3 SUSY Hamiltonian x}
    H_2=-\frac{d^2}{dx^2}+V^{(2)}(x). 
\end{equation}
It is easy to verify that the SUSY wavefunction is an eigenstate of $\rho(x)H_2$, where $\rho(x)$ is a weight function i.e.
\begin{equation}
    \rho(x)H_2\psi^{(2)}(x)=E_2 \psi^{(2)}(x).
\end{equation}
The supercharges can be obtained in the following way
\begin{equation}
    \tilde{A}=fAf^{-1}, \quad \tilde{B}=fBf^{-1},\quad f=\exp\left[-\frac{d}{2\alpha}e^{2\alpha x}-\frac{a}{\alpha}e^{-\alpha x}+bx\right],
\end{equation}
where $\frac{d}{dz}=-\frac{e^{\alpha x}}{\alpha} \frac{d}{dx}$. Then, we obtain the action of the supercharges on the SUSY wavefunction and the first excited state. The Hamiltonian and SUSY Hamiltonian can also be directly related to the supercharges. the formulas are given by
\begin{equation}
    \tilde{A} \psi_2(x)=\sqrt{E_2-E_1}\psi^{(2)}(x),\quad \tilde{B}\psi^{(2)}(x)=-\sqrt{E_2-E_1}\psi_2(x). 
\end{equation}
\begin{equation}
     \bar{H}_1=-\tilde{B}\tilde{A}, \quad  \bar{H}_2=-\tilde{A}\tilde{B},\quad  \bar{H}_1=\rho(x)H-E_1,\quad  \bar{H}_2=\rho(x)H_2-E_1. 
\end{equation}


\subsection{P\"oschl-Teller-type potentials V}\label{Poschl-Teller-type potentials V}
\begin{figure}[H]
  \centering
  \begin{subfigure}{0.48\linewidth}
    \centering
    \includegraphics[width=\linewidth]{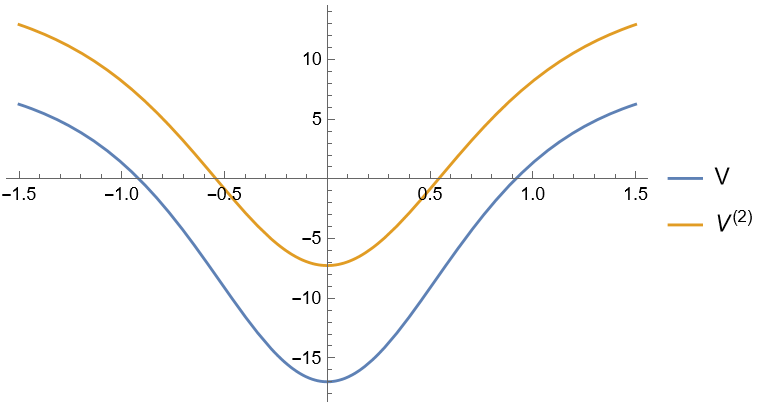}
    \caption{Schr\"odinger-like potential $V$ and partner Schr\"odinger-like potential $V^{(2)}$.}
    \label{f case5 V}
  \end{subfigure}
  \hfill
  \begin{subfigure}{0.48\linewidth}
    \centering
    \includegraphics[width=\linewidth]{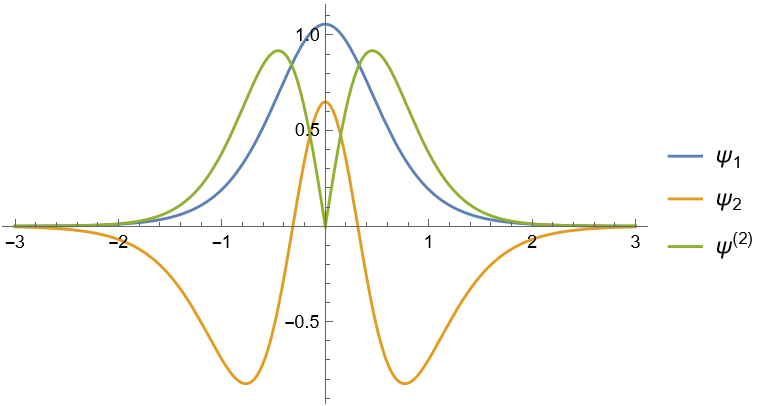}
    \caption{Wavefuntions $\psi_1, \psi_2$ and partner wavefunction $\psi^{(2)}$.}
    \label{f case5 wf}
  \end{subfigure}
  \caption{SUSY transformation in wavefunctions and Schr\"odinger-like potential.}
  \label{fig case5}
\end{figure}

The orginal potential is 
\begin{equation}\label{Poschl-teller V Schrodinger orginal potential x}
    \begin{split}
        V=&-b^2\cosh{\alpha x}^{-6}-b(2a+3b+\alpha+4n\alpha+2p\alpha)\cosh{\alpha x}^{-4}-\big[a^2+2ab+a\alpha(2p-2n-1)\\
        &+\alpha(2b(n+p-1)+\alpha(n+2n^2+2np+(p-1)p)\alpha)\big]\cosh{\alpha x}^{-2}+[a+2b+\alpha(p-1)]^2
    \end{split}
\end{equation}
The Schr\"odinger equation has a weight function $\rho(x)$ given by
\begin{equation}
    \rho(x)H\psi(x)=E\psi(x),\quad \rho(x)=\frac{\cosh{\alpha x}^2}{\alpha}. 
\end{equation}
There are two available values for $p$, which are 0 and 1. In this section, we choose $p=1$ to analyse. The wavefunction is 
\begin{equation}
    \psi(x)=\varphi(x)\exp\left[\frac{a+2b}{2\alpha}\ln\left(\cosh{\alpha x}^2\right)+\frac{b}{2\alpha}\cosh{\alpha x}^2\right], 
\end{equation}
where $P_3(z)$ is a polynomial. The transformed ODE is 
\begin{equation}\label{Poschl-teller V ODE z}
\begin{split}
        (4\alpha z-4\alpha z^2)\varphi''(z)+[-4bz^2-(4a+4b-6\alpha)z+4\alpha+4a+8b]\varphi'(z)+[4bnz+\alpha n(3+2n)\\+2an+2bn+E]\varphi(z)=0. 
\end{split}
\end{equation}
The ODE operator $T_1$ is 
\begin{equation}
\begin{split}
        T_1=(4\alpha z-4\alpha z^2)\frac{d^2}{dz^2}+[-4bz^2-(4a+4b-6\alpha)z+4\alpha+4a+8b]\frac{d}{dz}+[4bnz+\alpha n(3+2n)\\+2an+2bn].
\end{split}
\end{equation}
Applying BAE, we get polynomial solutions and energies for $n=1$  and they are given by the formulas
\begin{equation}\label{Poschl-teller V polynomial solutions z}
    \begin{split} &\varphi_1(z)=z+\frac{2(a+b)+3\alpha+\sqrt{4(a+3b)^2+4\alpha(3a+7b)+9\alpha^2}}{4b}, \\
    &\varphi_2(z)=z+\frac{2(a+b)+3\alpha-\sqrt{4(a+3b)^2+4\alpha(3a+7b)+9\alpha^2}}{4b}.
    \end{split}
\end{equation}
\begin{equation}\label{Poschl-teller V energies}
    E_1=-2\alpha -\sqrt{4(a+3b)^2+4\alpha(3a+7b)+9\alpha^2},\quad E_2=-2\alpha +\sqrt{4(a+3b)^2+4\alpha(3a+7b)+9\alpha^2}.
\end{equation}
The SUSY polynomial solution lead to the following eigenstate 
\begin{equation}\label{Posch-teller V SUSY polynomial z}
    \varphi^{(2)}(z)=\frac{4\sqrt{\alpha(z-z^2)[4(a+3b)^2+4\alpha(3a+7b)+9\alpha^2]}}{4bz+2(a+b)+3\alpha+\sqrt{4(a+3b)^2+4\alpha(3a+7b)+9\alpha^2}}. 
\end{equation}
The SUSY ODE potential is written as
\begin{equation}\label{Poschl-teller V SUSY ODE potential z}
    \begin{split}
        V_2(z)=&-2a-4b-5\alpha+2(2a+4b+\alpha)-\frac{2a+4b+\alpha}{z}-\frac{2(z-1)(2a+4b+\alpha)}{z}\\
        &+\frac{16b\alpha(2z-1)}{2(a+b)+3\alpha+4bz+\sqrt{4a^2+36b^2+28\alpha b+9\alpha^2+12(2b+\alpha)}}\\
        &-\frac{8\alpha z(z-1)}{\left[z+\frac{2(a+b)+3\alpha+\sqrt{4(a+3b)^2+4\alpha(3a+7b)+9\alpha^2}}{4b}\right]^2}.
    \end{split}
\end{equation}
The SUSY partner ODE operator $T_2$ is 
\begin{equation}
    \begin{split}
        T_2=(4\alpha z-4\alpha z^2)\frac{d^2}{dz^2}+[-4bz^2-(4a+4b-6\alpha)z+4\alpha+4a+8b]\frac{d}{dz}-V_2(z).
    \end{split}
\end{equation}
The supercharges can be determined explicitly and they take the following form 
\begin{align}
    &\begin{aligned}
        A=2\lambda\sqrt{1-z}\bigg\{\frac{d}{dz}+\frac{\sqrt{9 \alpha ^2+4 \alpha  (3 a+7 b)+4 (a+3 b)^2}-2 a-3 \alpha -4 b z-2 b}{4 (z-1) [a+b (z+2)]+2 \alpha  (3 z-2)}\bigg\}, 
    \end{aligned}\\
    &\begin{aligned}
       B=\frac{\sqrt{1-z}}{\lambda}\bigg\{2 \alpha  z\frac{d}{dz}+2 a+3 \alpha +2 b z+4 b+\frac{\alpha  [\alpha  (4-3 z)-2 a (z-2)-2 b (z-4)]}{2 (z-1) [a+b (z+2)]+\alpha  (3 z-2)}\\-\frac{\alpha z\sqrt{9 \alpha ^2+4 \alpha  (3 a+7 b)+4 (a+3 b)^2}}{2 (z-1) [a+b (z+2)]+\alpha  (3 z-2)}\bigg\}. 
    \end{aligned}
\end{align}
where $\lambda=\sqrt{2 \alpha  z} \left[9 \alpha ^2+4 \alpha  (3 a+7 b)+4 (a+3 b)^2\right]^{1/4}$.
The ODE operator can be rewritten in terms of supercharges and the following relations hold
\begin{equation}
     \bar{T}_1=BA,\quad  \bar{T}_2=AB, \quad  \bar{T}_1=T_1+E_1,\quad  \bar{T}_2=T_2+E_1. 
\end{equation}
The supercharges action between the polynomial states and SUSY polynomial states is given by the formulas below 
\begin{equation}
    \begin{split}
        &A\varphi_2(z)=\sqrt{E_2-E_1}\varphi^{(2)}(z)=\sqrt{2} \left[9 \alpha ^2+4 \alpha  (3 a+7 b)+4 (a+3 b)^2\right]^{1/4}\varphi^{(2)}(z),\\
        &B\varphi^{(2)}(z)=-\sqrt{E_2-E_1}\varphi_2(z)=-\sqrt{2} \left[9 \alpha ^2+4 \alpha  (3 a+7 b)+4 (a+3 b)^2\right]^{1/4}\varphi_2(z). 
    \end{split}
\end{equation}
We transform the different operators and states back to the variable $x$ related to the Schr\"odinger form. The wavefunctions are given explictly by
\begin{align}\label{Poschl-teller V wavefunctions x}
    &\begin{aligned}
        \psi_1(x)&=\left(\cosh{\alpha x}^{-2}+\frac{2(a+b)+3\alpha+\sqrt{4(a+3b)^2+4\alpha(3a+7b)+9\alpha^2}}{4b}\right)\\
        &\cdot \exp\left[\frac{a+2b}{2\alpha}\ln\left(\cosh{\alpha x}^2\right)+\frac{b}{2\alpha}\cosh{\alpha x}^2\right], 
    \end{aligned}\\
    &\begin{aligned}
        \psi_2(x)&=\left(\cosh{\alpha x}^{-2}+\frac{2(a+b)+3\alpha-\sqrt{4(a+3b)^2+4\alpha(3a+7b)+9\alpha^2}}{4b}\right)\\
        &\cdot \exp\left[\frac{a+2b}{2\alpha}\ln\left(\cosh{\alpha x}^2\right)+\frac{b}{2\alpha}\cosh{\alpha x}^2\right]. 
    \end{aligned}
\end{align}
The SUSY wavefunction is then taking the form
\begin{equation}\label{Poschl-teller V SUSY wavefunction x}
    \psi^{(2)}(x)=\frac{4\cosh{\alpha x}^{-\left(1+\frac{a+2b}{\alpha}\right)}\tanh{\alpha x}\sqrt{\alpha[4(a+3b)^2+4\alpha(3a+7b)+9\alpha^2]}}{2(a+b)+3\alpha+4b\cosh{\alpha x}^{-2}+\sqrt{4(a+3b)^2+4\alpha(3a+7b)+9\alpha^2}}\exp \left[\frac{b}{\alpha(\cosh{2\alpha x}+1)}\right]. 
\end{equation}
The partner Schr\"odinger-like potential is 
\begin{equation}\label{Poschl-teller V SUSY Schrodinger potential x}
\begin{split}
        V^{(2)}(x)&=-b^2\cosh{\alpha x}^{-6}-b(2a+3b+3\alpha)\cosh{\alpha x}^{-4}-(a^2+2ab+3a\alpha+4b\alpha+5\alpha^2)\cosh{\alpha x}^{-2}\\
        &+(a+2b+\alpha)^2+\frac{128b^2\alpha^2 \cosh{\alpha x}^{-4}\tanh{\alpha x}^2}{\left[4b\cosh{\alpha x}^{-2}+2a+2b+3\alpha+\sqrt{4(a+3b)^2+4\alpha(3a+7b)+9\alpha^2}\right]^2}\\
        &-\frac{8b\alpha^2 (\cosh{2\alpha x}-3)\cosh{\alpha x}^{-4}}{4b\cosh{\alpha x}^{-2}+2a+2b+3\alpha+\sqrt{4(a+3b)^2+4\alpha(3a+7b)+9\alpha^2}}.
\end{split}
\end{equation}
The Fig.\ref{fig case5} provides the potential and the wavefunctions of Schr\"odinger-like equation and their SUSY partners. In this case, we set $a=b=\alpha=1$ and $p=1$. The partner potential $V^{(2)}$ is similar to the initial potential $V$ and no singularity in Fig.\ref{f case5 V}. The partner wavefunction $\psi^{(2)}$ and wavefunctions $\psi_1$ and $\psi_2$ are symmetry in Fig.\ref{f case5 wf}. $\psi^{(2)}$ has a zero at origin and $\psi_2$ has two zeros symmetry at $y$-axis. 
We can obtain partner Hamiltonian 
\begin{equation}
    H_2=-\frac{d^2}{dx^2}+V^{(2)},\quad \rho(x)H_2\psi^{(2)}(x)=E_2\psi^{(2)}(x). 
\end{equation}
Supercharges expressed in Schr\"odinger form are
\begin{equation}
    \tilde{A}=fAf^{-1}, \quad \tilde{B}=fBf^{-1},\quad f=\exp\left[\frac{a+2b}{2\alpha}\ln\left(\cosh{\alpha x}^2\right)+\frac{b}{2\alpha}\cosh{\alpha x}^2\right].
\end{equation}
We have $\frac{d}{dz}=-\frac{\cosh{\alpha x}^2}{2\alpha\tanh{\alpha x}} \frac{d}{dx}$. We can verify that 
\begin{equation}
    \tilde{A} \psi_2(x)=\sqrt{E_2-E_1}\psi^{(2)}(x),\quad \tilde{B}\psi^{(2)}(x)=-\sqrt{E_2-E_1}\psi_2(x). 
\end{equation}
\begin{equation}
     \bar{H}_1=-\tilde{B}\tilde{A}, \quad  \bar{H}_2=-\tilde{A}\tilde{B},\quad  \bar{H}_1=\rho(x)H-E_1,\quad  \bar{H}_2=\rho(x)H_2-E_1. 
\end{equation}


\section{QES with radial Shcr\"odinger(-like) form}\label{Radial type QES}
A third and last type of QES systems that was classified in regard of underlying $sl(2)$ hidden symmetry algebra ocncerns radial type QES for which a Hamiltonian in expressed in terms of spherical coordinates i.e.$H_S=-\frac{d^2}{dr^2}-\frac{d-1}{r}\frac{d}{dr}-\frac{l(l+d-2)}{r^2}+V(r)$. 

\subsection{Harmonic oscillator-type potentials VII}\label{Harmonic oscillator-type potentials VII}
\begin{figure}[H]
  \centering
  \begin{subfigure}{0.48\linewidth}
    \centering
    \includegraphics[width=\linewidth]{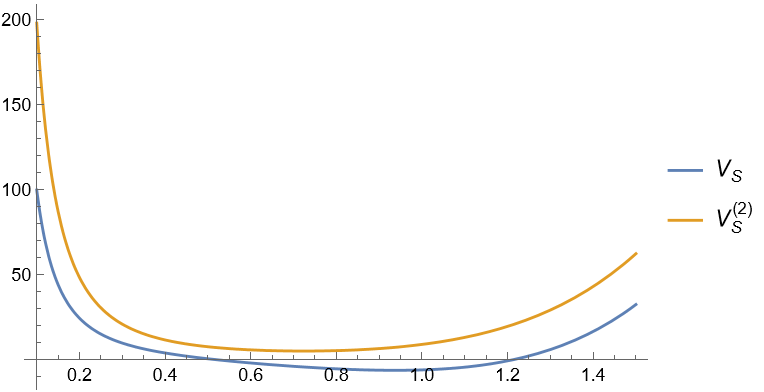}
    \caption{Radial Schr\"odinger potential $V$ and partner radial Schr\"odinger potential $V^{(2)}$ in spherical coordinate.}
    \label{f case7 V}
  \end{subfigure}
  \hfill
  \begin{subfigure}{0.48\linewidth}
    \centering
    \includegraphics[width=\linewidth]{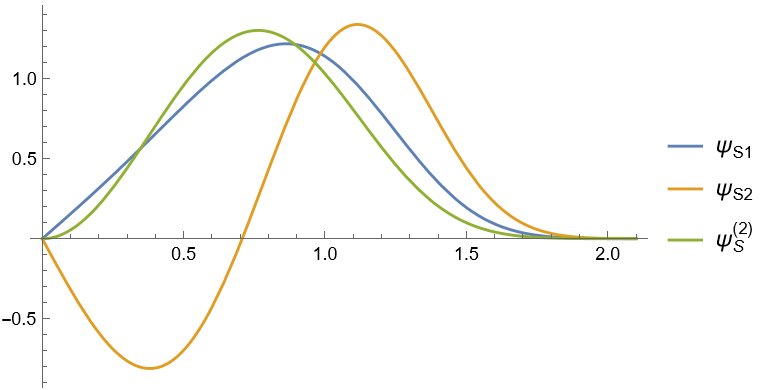}
    \caption{Radial wavefuntions $\psi_1, \psi_2$ and partner radial wavefunction $\psi^{(2)}$ in spherical coordinate.}
    \label{f case7 wf}
  \end{subfigure}
  \caption{SUSY transformation in radial wavefunctions and radial Schr\"odinger potential in spherical coordinate.}
  \label{fig case7}
\end{figure}

We consider a first potential among this type of QES system, which corresponds to the harmonic oscillator-type potential and given by the formula
\begin{equation}
    V=a^2r^6+2abr^4+[b^2-a(4n+2l+d-2c+2)]r^2+\frac{c(c-2l-d+2)+2l(l+d-2)}{r^2}. 
\end{equation}
In the spherical coordinates, the Hamiltonian is 
\begin{equation}
    H_S=-\frac{d^2}{dr^2}-\frac{d-1}{r}\frac{d}{dr}-\frac{l(l+d-2)}{r^2}+V. 
\end{equation}
The wavefunction is 
\begin{equation}
    \psi(r)=\varphi(r)\exp\left[-\frac{a}{4}r^4-\frac{b}{2}r^2-(c-l)\ln{r}\right],
\end{equation}
$\varphi(r)$ is a polynomial and satisfies ODE, which is 
\begin{equation}\label{Coulomb-type VII ODE z}
    4z\varphi''(z)+2[-2c+d+2l-2z(b+az)]\varphi'(z)+[E+4anz+b(2c-d-al)]=0, \quad z=r^2. 
\end{equation}
The ODE operator is
\begin{equation}\label{Coulomb-type VII orginal ODE operator z}
    T_1=4z\frac{d^2}{dz^2}+2[-2c+d+2l-2z(b+az)]\frac{d}{dz}+4anz+b(2c-d-al). 
\end{equation}
Using BAE, we get polynomial solutions and energies for $n=1$
\begin{equation}\label{Coulomb-type VII polynomials z}
    \varphi_1(z)=z+\frac{b+\sqrt{b^2+2a(2l-2c+d)}}{2a},\quad \varphi_2(z)=z+\frac{b-\sqrt{b^2+2a(2l-2c+d)}}{2a}. 
\end{equation}
\begin{equation}\label{Coulomb-type VII energies}
    E_1=b(2-2c+2l+d)-2\sqrt{b^2+2a(2l-2c+d)},\quad E_2=b(2-2c+2l+d)+2\sqrt{b^2+2a(2l-2c+d)}. 
\end{equation}
The SUSY polynomial is 
\begin{equation}\label{Coulomb-type VII SUSY polynomial z}
    \varphi^{(2)}(z)=\frac{4\sqrt{z[b^2+2a(2l-2c+d)]}}{2az+b+\sqrt{b^2+2a(2l-2c+d)}}. 
\end{equation}
The partner potential takes the form
\begin{equation}\label{Coulomb-type VII SUSY potential z}
    \begin{split}
        V_2(z)&=b(2-2c+2l+d)+\frac{2l+d-2c-1}{z}-\frac{8a}{2az+b+\sqrt{b^2+2a(2l-2c+d)}}\\
        &+2az\left[1+\frac{16a}{\left(b++2az+\sqrt{b^2+2a(2l-2c+d)}\right)^2}\right].
    \end{split}
\end{equation}
The partner ODE operator is then given by
\begin{equation}\label{Coulomb-type VII SUSY ODE operator z}
    T_2=4z\frac{d^2}{dz^2}+2[-2c+d+2l-2z(b+az)]\frac{d}{dz}-V_2(z). 
\end{equation}
Supercharges of the ODE are expressed as
\begin{align}\label{Coulomb-type VII ODE supercharges z}
    &\begin{aligned}
        A=4\sqrt{z}[b^2+2a(2l-2c+d)]^{1/4}\left[\frac{d}{dz}-\frac{2az+b-\sqrt{b^2+2a(2l-2c+d)}}{2c-d-2l+2z(b+az)}\right],
    \end{aligned}\\
    &\begin{aligned}
        B=\frac{1}{\sqrt{z}[b^2+2a(2l-2c+d)]^{1/4}}\bigg[z\frac{d}{dz}-\frac{\left(b+\sqrt{2 a d+4 a l+b^2-4 a c}\right)z+2 c - d - 2 l}{2 a z^2+2 b z+2 c-d-2 l}\\-a z^2+\frac{1+d+2 l-2 c}{2}\bigg]. 
    \end{aligned}
\end{align}
The ODE operator can be rewritten in terms of supercharges.
\begin{equation}
     \bar{T}_1=BA,\quad  \bar{T}_2=AB, \quad  \bar{T}_1=T_1+E_1,\quad  \bar{T}_2=T_2+E_1. 
\end{equation}
The supercharges action with the polynomial states and SUSY polynomial states are given by the formulas 
\begin{equation}
    \begin{split}
        &A\varphi_2(z)=\sqrt{E_2-E_1}\varphi^{(2)}(z)=2[b^2+2a(2l-2c+d)]^{1/4}\varphi^{(2)}(z),\\
        &B\varphi^{(2)}(z)=-\sqrt{E_2-E_1}\varphi_2(z)=-2[b^2+2a(2l-2c+d)]^{1/4}\varphi_2(z). 
    \end{split}
\end{equation}
The wavefunction for the Schr\"odinger form of the system is given by
\begin{equation}\label{Coulomb-type VII Schrodinger wavefunction r}
\begin{split}
        &\psi_1(r)=\left(r^2+\frac{b+\sqrt{b^2+2a(2l-2c+d)}}{2a}\right)\exp\left[-\frac{a}{4}r^4-\frac{b}{2}r^2-\frac{2(c-l)-d+1}{2}\ln{r}\right],\\
        &\psi_2(r)=\left(r^2+\frac{b-\sqrt{b^2+2a(2l-2c+d)}}{2a}\right)\exp\left[-\frac{a}{4}r^4-\frac{b}{2}r^2-\frac{2(c-l)-d+1}{2}\ln{r}\right].
\end{split}
\end{equation}
The SUSY wavefunction for the Schr\"odinger form of the system is then
\begin{equation}\label{Coulomb-type VII SUSY Schrodinger wavefunction r}
    \psi^{(2)}(r)=\frac{4\sqrt{b^2+2a(2l-2c+d)}}{2ar^2+b+\sqrt{b^2+2a(2l-2c+d)}}\exp\left[-\frac{1}{4}r^2(ar^2+2b)+\frac{2(l-c)+d+1}{2}\ln{r}\right]. 
\end{equation}
The partner potential of Schr\"odinger form is 
\begin{equation}\label{Coulomb-type VII SUSY Schrodinger potential r}
\begin{split}
        V^{(2)}(r)&=a^2r^6+2abr^4+[b^2+a(2c-d-2l)]r^2+2b+\frac{32a^2r^2}{\left(2ar^2+b+\sqrt{b^2+2a(2l-2c+d)}\right)^2}\\
        &+\frac{(2l-2c+d)^2-1}{4r^2}-\frac{8a}{2ar^2+b+\sqrt{b^2+2a(2l-2c+d)}}.
\end{split}
\end{equation}
We can construct the partner Hamiltonian with $V_2$
\begin{equation}\label{Coulomb-type VII SUSY Schrodinger Hamiltonian r}
    H_2=-\frac{d^2}{dr^2}+V_2,\quad H_2 \psi^{(2)}(r)=E_2\psi^{(2)}(r). 
\end{equation}
The radial Schr\"odinger partner Hamiltonian is 
\begin{equation}\label{Coulomb-type VII SUSY radial Schrodinger Hamiltonian r}
    H_{S2}=hH_2h^{-1},\quad h=r^{-\frac{d-1}{2}}. 
\end{equation}
The wavefunction is radial Schr\"odinger form is 
\begin{equation}\label{Coulomb-type VII radial Schrodinger wavefunction r}
\begin{split}
    &\psi_{S1}(r)=\psi_1(r)h=\left(r^2+\frac{b+\sqrt{b^2+2a(2l-2c+d)}}{2a}\right)\exp\left[-\frac{a}{4}r^4-\frac{b}{2}r^2-(c-l)\ln{r}\right],\\
    &\psi_{S2}(r)=\psi_2(r)h=\left(r^2+\frac{b-\sqrt{b^2+2a(2l-2c+d)}}{2a}\right)\exp\left[-\frac{a}{4}r^4-\frac{b}{2}r^2-(c-l)\ln{r}\right].
\end{split}
\end{equation}
The SUSY wavefunction for radial Schr\"odinger form is 
\begin{equation}\label{Coulomb-type VII SUSY radial Schrodinger wavefunction r}
    \psi_S^{(2)}(r)= \frac{4r\sqrt{b^2+2a(2l-2c+d)}}{2ar^2+b+\sqrt{b^2+2a(2l-2c+d)}}\exp\left[-\frac{1}{4}(2b+ar^2)r^2+(l-c)\ln{r}\right].
\end{equation}
The partner potential of radial Schr\"odinger form is 
\begin{equation}\label{Coulomb-type VII SUSY radial Schrodinger potential r}
\begin{split}
        V_S^{(2)}(r)&=a^2r^6+2abr^4+[b^2+a^2(2c-d-2l)]r^2+2b+\frac{32a^2r^2}{\left(2ar^2+b+\sqrt{b^2+2a(2l-2c+d)}\right)^2}\\
        &-\frac{8a}{2ar^2+b+\sqrt{b^2+2a(2l-2c+d)}}+\frac{(c-d+1) (c-2 l-1)+2 l^2}{r^2}. 
\end{split}
\end{equation}
The Fig.\ref{fig case7} provides the potential and the wavefunctions of Schr\"odinger equation and their SUSY partners in spherical coordinate. In this case, we set $a=2, b=c=l=1$ and $d=2$. Due to $r\geq 0$, we only plot the part of $r>0$. The partner potential $V_S^{(2)}$ is similar to the initial potential $V_S$ and they have the same singularity at origin in Fig.\ref{f case7 V}. Thus, no new singularity appears in $V_S^{(2)}$, which is consistent with the structure of partner potentials in SUSYQM. The partner wavefunction $\psi_S^{(2)}$ and wavefunctions $\psi_{S1}$ and $\psi_{S2}$ are in Fig.\ref{f case7 wf}. $\psi_S^{(2)}$ has a zero at origin and $\psi_{S2}$ has a zeros symmetry at $0.8$. 
We directly obtain 
\begin{equation}
   H_{S2}=-\frac{d^2}{dr^2}-\frac{d-1}{r}\frac{d}{dr}-\frac{l(l+d-2)}{r^2}+V_S^{(2)},\quad  H_{S2}\psi_S^{(2)}(r)=E_2\psi_S^{(2)}(r). 
\end{equation}
The supercharges in the radial Schr\"odinger form is 
\begin{equation}
    \tilde{A}_S(r)=hfA(z)f^{-1}h^{-1}, \quad \tilde{A}_S(r)^{\#}=hfB(z)f^{-1}h^{-1},\quad f=\exp\left[-\frac{a}{4}r^4-\frac{b}{2}r^2-(c-l)\ln{r}\right], \quad h=r^{-\frac{d-1}{2}}.
\end{equation}
We have $\frac{d}{dz}=\frac{1}{2r} \frac{d}{dr}$. We can then verify that  the following relations hold
\begin{equation}
    \tilde{A}_S \psi_{S2}(r)=\sqrt{E_2-E_1}\psi_S^{(2)}(r),\quad \tilde{B}_S\psi_S^{(2)}(r)=-\sqrt{E_2-E_1}\psi_{S2}(r). 
\end{equation}
\begin{equation}
     \bar{H}_{S1}=\tilde{A}_S(r)^{\#}\tilde{A}_S(r), \quad  \bar{H}_{S2}=\tilde{A}_S(r) \tilde{A}_S(r)^{\#},\quad  \bar{H}_1=H_S-E_1,\quad  \bar{H}_{S2}=H_{S2}-E_1. 
\end{equation}

\subsection{Coulomb-type potentials VIII}\label{Coulomb-type potentials VIII}

\begin{figure}[H]
  \centering
  \begin{subfigure}{0.48\linewidth}
    \centering
    \includegraphics[width=\linewidth]{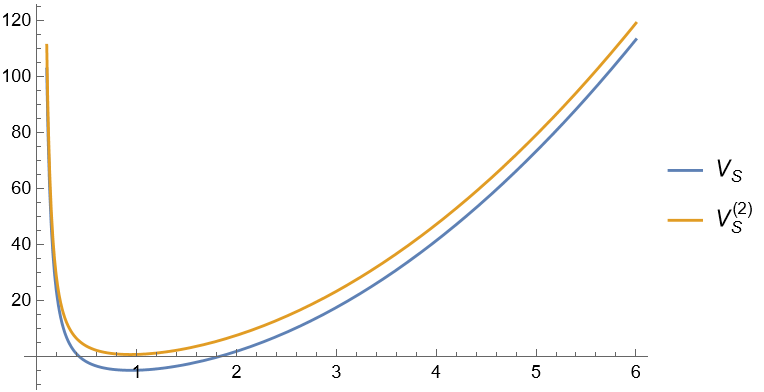}
    \caption{Radial Schr\"odinger-like potential $V$ and partner radial Schr\"odinger-like potential $V^{(2)}$ in spherical coordinate.}
    \label{f case8 V}
  \end{subfigure}
  \hfill
  \begin{subfigure}{0.48\linewidth}
    \centering
    \includegraphics[width=\linewidth]{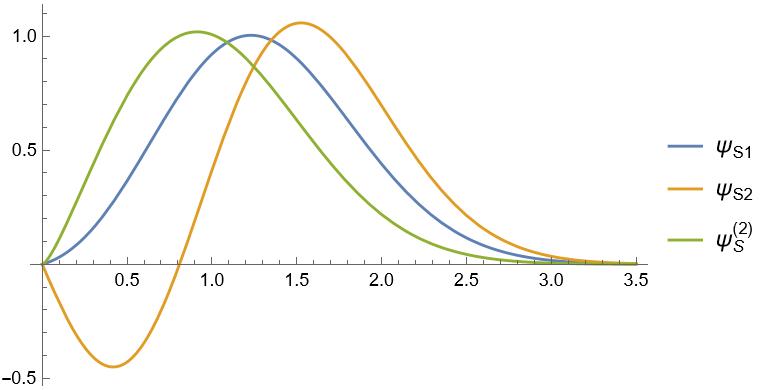}
    \caption{Radial wavefuntions $\psi_1, \psi_2$ and partner radial wavefunction $\psi^{(2)}$ in spherical coordinate.}
    \label{f case8 wf}
  \end{subfigure}
  \caption{SUSY transformation in radial wavefunctions and radial Schr\"odinger-like potential in spherical coordinate.}
  \label{fig case8}
\end{figure}

The initial potential is 
\begin{equation}
    V=a^2r^2+2abr-\frac{b(Dc-1)}{r}+\frac{c(c-2l-d+2)+2l(d+l-2)}{r^2}+b^2-a(Dc+2n), \quad Dc=d+2l-2c. 
\end{equation}
In the spherical coordinates, Hamiltonian is 
\begin{equation}
   H_{S\rho}=\rho(r)\left[-\frac{d^2}{dr^2}-\frac{d-1}{r}\frac{d}{dr}-\frac{l(l+d-2)}{r^2}+V\right],\quad \rho(r)=r.
\end{equation}
The wavefunction is 
\begin{equation}
    \psi(r)=\varphi(r)\exp\left[-\frac{a}{2}r^2-br-(c-l)\ln{r}\right],
\end{equation}
$\varphi(r)$ is a polynomial and satisfies ODE, which is 
\begin{equation}\label{Coulomb-type VIII ODE z}
    z\varphi''(z)-[2z(b+az)+2c-d-2l+1]\varphi'(z)+(E+2anz)\varphi(z)=0, \quad z=r. 
\end{equation}
The ODE operator is 
\begin{equation}\label{Coulomb-type VIII ODE operator z}
    T_1=z\frac{d^2}{dz^2}-[2z(b+az)+2c-d-2l+1]\frac{d}{dz}+2anz.
\end{equation}
Using BAE, polynomial solutions and energies for $n=1$ are
\begin{equation}\label{Coulomb-type VIII polynomial solutions z}
    \varphi_1(z)=z-\frac{b+\sqrt{b^2+2a(2l-2c+d-1)}}{2a},\quad \varphi_2(z)=z-\frac{b-\sqrt{b^2+2a(2l-2c+d-1)}}{2a}
\end{equation}
\begin{equation}\label{Coulomb-type VIII energies}
    E_1=b-\sqrt{b^2+2a(2l-2c+d-1)},\quad E_2=b+\sqrt{b^2+2a(2l-2c+d-1)}
\end{equation}
The corresponding SUSY polynomial of ODE is given by
\begin{equation}\label{Coulomb-type VIII SUSY polynomial z}
    \varphi^{(2)}(z)=\frac{2\sqrt{z[b^2+2a(2l-2c+d-1)]}}{2az+b+\sqrt{b^2+2a(2l-2c+d-1)}},
\end{equation}
while the partner ODE potential is 
\begin{equation}\label{Coulomb-type VIII SUSY ODE potential z}
\begin{split}
        V_2(z)&=az\left[1+\frac{8a}{\left(2az+b+\sqrt{b^2+2a(2l-2c+d-1)}\right)^2}\right]-\frac{2a}{2az+b+\sqrt{b^2+2a(2l-2c+d-1)}}\\
        &+\frac{4(l-c)+2d-3}{4z}+b. 
\end{split}
\end{equation}
The partner ODE operator is then expressed
\begin{equation}\label{Coulomb-type VIII SUSY ODE operator z}
    T_2=z\frac{d^2}{dz^2}-[2z(b+az)+2c-d-2l+1]\frac{d}{dz}-V_2(z).
\end{equation}
with the supercharges of the ODE taking the following form
\begin{align}\label{Coulomb-type VIII ODE supercharges z}
    &\begin{aligned}
        A=\sqrt{2z}[b^2+2a(2l-2c+d-1)]^{1/4}\left(\frac{d}{dz}+\frac{\sqrt{2 a d+4 a l-2 a+b^2-4 a c}-2az-b}{2 a z^2+2 b z+2 c-d-2 l+1}\right),
    \end{aligned}\\
    &\begin{aligned}
        B=-\frac{1}{\sqrt{2z}[b^2+2a(2l-2c+d-1)]^{1/4}}\bigg[-z\frac{d}{dz}+2 a z^2+2bz+\frac{4 c-2 d-4 l+1}{2}\\+\frac{\left(b+\sqrt{2 a d+4 a l-2 a+b^2-4 a c}\right)z+2 c-d-2 l+1}{2 a z^2+2 b z+2 c-d-2 l+1}\bigg]. 
    \end{aligned}
\end{align}
The ODE operator can be rewritten in terms of supercharges.
\begin{equation}
     \bar{T}_1=BA,\quad  \bar{T}_2=AB, \quad  \bar{T}_1=T_1+E_1,\quad  \bar{T}_2=T_2+E_1. 
\end{equation}
The supercharges action between polynomial state and SUSY polynomial state is 
\begin{equation}
    \begin{split}
        &A\varphi_2(z)=\sqrt{E_2-E_1}\varphi^{(2)}(z)=\sqrt{2}[b^2+2a(2l-2c+d-1)]^{1/4}\varphi^{(2)}(z),\\
        &B\varphi^{(2)}(z)=-\sqrt{E_2-E_1}\varphi_2(z)=-\sqrt{2}[b^2+2a(2l-2c+d-1)]^{1/4}\varphi_2(z). 
    \end{split}
\end{equation}
We transform it to Schr\"odinger form firstly. The wavefunctions are
\begin{equation}\label{Coulomb-type VIII Schrodinger wavefunctions r}
    \begin{split}
        &\psi_1(r)=\left(r-\frac{b+\sqrt{b^2+2a(2l-2c+d-1)}}{2a}\right)\exp\left[-\frac{a}{2}r^2-br-\frac{2(c-l)-d+1}{2}\ln{r}\right],\\
        &\psi_2(r)=\left(r-\frac{b-\sqrt{b^2+2a(2l-2c+d-1)}}{2a}\right)\exp\left[-\frac{a}{2}r^2-br-\frac{2(c-l)-d+1}{2}\ln{r}\right],
    \end{split}
\end{equation}
SUSY wavefunction in Schr\"odinger form is 
\begin{equation}\label{Coulomb-type VIII Schrodinger SUSY wavefunctions r}
    \psi^{(2)}(r)=\frac{2\sqrt{r[b^2+2a(2l-2c+d-1)}]}{b+2ar+\sqrt{b^2+2a(2l-2c+d-1)}}\exp\left[-\frac{a}{2}r^2-br-\frac{2(c-l)-d+1}{2}\ln{r}\right]
\end{equation}
The partner Schr\"odinger potential is
\begin{equation}\label{Coulomb-type VIII SUSY Schrodinger potential r}
\begin{split}
       V^{(2)}(r)&=a^2r^2+2abr+a(1+2c-d-2l)+b^2+\frac{8a^2}{\left(2ar+b+\sqrt{b^2+2a(2l-2c+d-1)}\right)^2}\\
       &-\frac{2a}{r\left(2ar+b+\sqrt{b^2+2a(2l-2c+d-1)}\right)}+\frac{(2+2c-d-2l)(2c-d-2l+4br)}{4r^2}.
\end{split}
\end{equation}
We construct partner Hamiltonian in Schr\"odinger form.
\begin{equation}\label{Coulomb-type VIII SUSY Schrodinger Hamiltonian r}
     H_2=-\frac{d^2}{dr^2}+V^{(2)}(r),\quad \rho(r)H_2 \psi^{(2)}(r)=E_2\psi^{(2)}(r).
\end{equation}
Then, we transform to radial Schr\"odinger Hamiltonian in spherical coordinate. The partner Hamiltonian in spherical coordinate is 
\begin{equation}\label{Coulomb-type VIII SUSY radial Schrodinger Hamiltonian r}
    H_{S2}=h\rho(r)H_2h^{-1},\quad h=r^{-\frac{d-1}{2}}. 
\end{equation}
The wavefunctions are
\begin{equation}\label{Coulomb-type VIII radial Schrodinger wavefunctions r}
    \begin{split}
        &\psi_{S1}(r)=h\psi_1(r)=\left(r-\frac{b+\sqrt{b^2+2a(2l-2c+d-1)}}{2a}\right)\exp\left[-\frac{a}{2}r^2-br-(c-l)\ln{r}\right],\\
        &\psi_{S2}(r)=h\psi_(2)=\left(r-\frac{b-\sqrt{b^2+2a(2l-2c+d-1)}}{2a}\right)\exp\left[-\frac{a}{2}r^2-br-(c-l)\ln{r}\right],
    \end{split}
\end{equation}
The SUSY wavefunction in radial Schr\"odinger form is
\begin{equation}\label{Coulomb-type VIII SUSY radial Schrodinger wavefunctions r}
    \psi_S^{(2)}(r)=\frac{2\sqrt{r[b^2+2a(2l-2c+d-1)}]}{b+2ar+\sqrt{b^2+2a(2l-2c+d-1)}}\exp\left[-\frac{a}{2}r^2-br-(c-l)\ln{r}\right].
\end{equation}
The partner potential in radial Schr\"odinger form is
\begin{equation}\label{Coulomb-type VIII SUSY radial Schrodinger potential r}
    \begin{split}
        V_S^{(2)}(r)&=a^2 r^2+a(2 b r+2 c-d-2 l+1)+ b^2+\frac{8a^2 }{\left(\sqrt{2 a (d+2 l-2 c-1)+b^2}+2 a r+b\right)^2}\\
        &-\frac{2a}{r \left(\sqrt{2 a (d+2 l-2 c-1)+b^2}+2 a r+b\right)}+\frac{ c^2-c (d+2 l-2 b r-1)-b r (d+2 l-2)}{r^2}\\&+\frac{4l (2 l-3)+d (8 l+2)-3}{4r^2}. 
    \end{split}
\end{equation}
The Fig.\ref{fig case8} provides the potential and the wavefunctions of Schr\"odinger-like equation and their SUSY partners in spherical coordinate. In this case, we set $a=2, b=-1, c=1,l=1$ and $d=2$. Due to $r\geq 0$, we only plot the part of $r>0$. The partner potential $V_S^{(2)}$ is similar to the initial potential $V_S$ and they have the same singularity at origin in Fig.\ref{f case8 V}. Thus, no new singularity appears in $V^{(2)}$, which is consistent with the structure of partner potentials in SUSYQM. The partner wavefunction $\psi^{(2)}$ and wavefunctions $\psi_{S1}$ and $\psi_{S2}$ are in Fig.\ref{f case8 wf}. $\psi_S^{(2)}$ has a zero at origin and $\psi_{S2}$ has a zeros symmetry at $0.4$. 
It is also easy to verify that 
\begin{equation}
   H_{S \rho 2}=\rho(r)\left[-\frac{d^2}{dr^2}-\frac{d-1}{r}\frac{d}{dr}-\frac{l(l+d-2)}{r^2}+V_S^{(2)}\right],\quad  H_{S\rho2}\psi_S^{(2)}(r)=E_2\psi_S^{(2)}(r). 
\end{equation}
Supercharges in radial Schr\"odinger form is 
\begin{equation}
    \tilde{A}_S(r)=hfA(z)f^{-1}h^{-1}, \quad \tilde{A}_S(r)^{\#}=hfB(z)f^{-1}h^{-1},\quad f=\exp\left[-\frac{a}{2}r^2-br-(c-l)\ln{r}\right], \quad h=r^{-\frac{d-1}{2}}.
\end{equation}
We have $\frac{d}{dz}=\frac{d}{dr}$. We can verify that 
\begin{equation}
    \tilde{A}_S \psi_{S2}(r)=\sqrt{E_2-E_1}\psi_S^{(2)}(r),\quad \tilde{B}_S\psi_S^{(2)}(r)=-\sqrt{E_2-E_1}\psi_{S2}(r). 
\end{equation}
\begin{equation}
     \bar{H}_{S\rho1}=\tilde{B}_S\tilde{A}_S, \quad \bar{H}_{S\rho2}=\tilde{A}_S \tilde{B}_S,\quad  \bar{H}_{S\rho1}=H_{S\rho}-E_1,\quad \bar{H}_{S\rho2}=H_{S\rho2}-E_1. 
\end{equation}

\subsection{Coulomb-type potentials IX}\label{Coulomb-type potentials IX}

\begin{figure}[H]
  \centering
  \begin{subfigure}{0.48\linewidth}
    \centering
    \includegraphics[width=\linewidth]{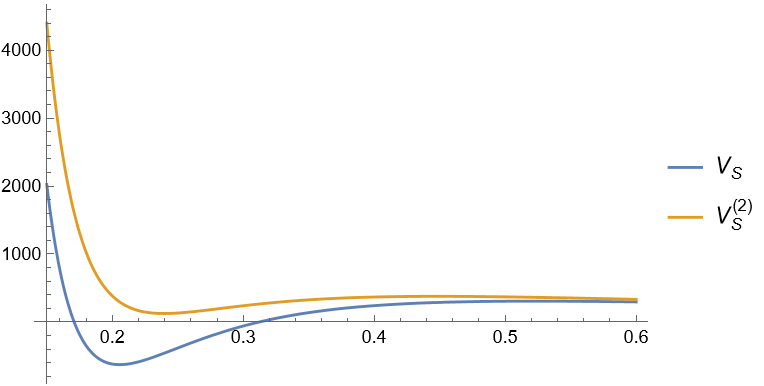}
    \caption{Radial Schr\"odinger-like potential $V_S$ and partner radial Schr\"odinger-like potential $V^{(2)}$ in spherical coordinate.}
    \label{f case9 V}
  \end{subfigure}
  \hfill
  \begin{subfigure}{0.48\linewidth}
    \centering
    \includegraphics[width=\linewidth]{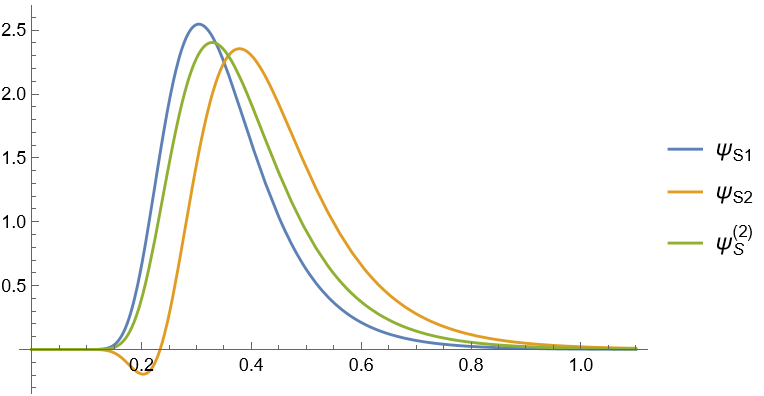}
    \caption{Radial wavefuntions $\psi_1, \psi_2$ and partner radial wavefunction $\psi^{(2)}$ in spherical coordinate.}
    \label{f case9 wf}
  \end{subfigure}
  \caption{SUSY transformation in radial wavefunctions and radial Schr\"odinger-like potential in spherical coordinate.}
  \label{fig case9}
\end{figure}

The initial potential is given by the formula 
\begin{equation}
    V=\frac{b^4}{r^4}+\frac{b(Dc-3)}{r^3}+\frac{c(c-2l-d+2)+2ab+2l(l+d-2)}{r^2}-\frac{a(2n+Dc-1)}{r}+a^2, \quad Dc=d+2l-2c. 
\end{equation}
In the spherical coordinates, the Hamiltonian is expressed as 
\begin{equation}
    H_{S\rho}=\rho(r)\left[-\frac{d^2}{dr^2}-\frac{d-1}{r}\frac{d}{dr}-\frac{l(l+d-2)}{r^2}+V\right]\quad \rho(r)=r^2,
\end{equation}
and the wavefunction cis of the form 
\begin{equation}
    \psi(r)=\varphi(r)\exp\left[-ar-\frac{b}{r}-(c-l)\ln{r}\right],
\end{equation}
where $\varphi(r)$ is a polynomial and satisfies the following ODE
\begin{equation}\label{Coulomb-type IX ODE z}
    z^2\varphi''(z)-[(1+2c-d-2l+2az)z-2b]\varphi'(z)+(E+2anz-4ab)\varphi(z)=0, \quad z=r. 
\end{equation}
The ODE operator is then  
\begin{equation}\label{Coulomb-type IX ODE operator z}
    T_1=z^2\frac{d^2}{dz^2}-[(1+2c-d-2l+2az)z-2b]\frac{d}{dz}+(2anz-4ab).
\end{equation}
Using BAE, polynomial solutions and energies for $n=1$ are
\begin{equation}\label{Coulomb-type IX polynomial solutions z}
    \begin{split}
        &\varphi_1(z)=z-\frac{2l-2c+d-1-\sqrt{16ab+(2l-2c+d-1)^2}}{4a},\\ &\varphi_2(z)=z-\frac{2l-2c+d-1+\sqrt{16ab+(2l-2c+d-1)^2}}{4a}.
    \end{split}
\end{equation}
\begin{equation}\label{Coulomb-type IX energies}
\begin{split}
        &E_1=\frac{1}{2}\left[1+8ab+2c-d-2l-\sqrt{16ab+(2l-2c+d-1)^2}\right],\\
        &E_2=\frac{1}{2}\left[1+8ab+2c-d-2l+\sqrt{16ab+(2l-2c+d-1)^2}\right].
\end{split}
\end{equation}
The SUSY polynomial solution of ODE is 
\begin{equation}\label{Coulomb-type IX SUSY polynomial solutions z}
    \varphi^{(2)}(z)=\frac{2z\sqrt{16ab+(2l-2c+d-1)^2}}{4az-2(l-c)-d+1+\sqrt{16ab+(2l-2c+d-1)^2}}
\end{equation}
The partner ODE potential is 
\begin{equation}\label{Coulomb-type IX SUSY ODE potential z}
    \begin{split}
        V_2(z)&=4ab+\frac{2b}{z}+\frac{32a^2z^2}{\left[4az+1+2c-d-2l+\sqrt{16ab+(2l-2c+d-1)^2}\right]^2}\\
        &-\frac{8az}{4az+1+2c-d-2l+\sqrt{16ab+(2l-2c+d-1)^2}}. 
    \end{split}
\end{equation}
The partner ODE operator is 
\begin{equation}\label{Coulomb-type IX SUSY ODE operator z}
    T_2=z^2\frac{d^2}{dz^2}-[(1+2c-d-2l+2az)z-2b]\frac{d}{dz}-V_2(z). 
\end{equation}
Supercharges of ODE are
\begin{align}\label{Coulomb-type IX SUSY polynomial supercharges z}
    &\begin{aligned}
        A=z\left[16ab+(2l-2c+d-1)^2\right]^{1/4}\left[\frac{d}{dz}-\frac{4 a}{\sqrt{16 a b+(-2 c+d+2 l-1)^2}+4 a z+2 c-d-2 l+1}\right],
    \end{aligned}\\
    &\begin{aligned}
    B=\frac{1}{\left[16ab+(2l-2c+d-1)^2\right]^{1/4}}\bigg[z \frac{d}{dz}-\frac{4 b}{z \left(\sqrt{16 a b+(-2 c+d+2 l-1)^2}-2 c+d+2 l-1\right)+4 b}\\-2 a z+\frac{2 b}{z}-2 c+d+2 l-1\bigg].
    \end{aligned}
\end{align}
The ODE operator cna be rewritten in terms of supercharges.
\begin{equation}
     \bar{T}_1=BA,\quad  \bar{T}_2=AB, \quad  \bar{T}_1=T_1+E_1,\quad  \bar{T}_2=T_2+E_1. 
\end{equation}
The supercharges action between polynomial state and SUSY polynomial state is 
\begin{equation}
    \begin{split}
        &A\varphi_2(z)=\sqrt{E_2-E_1}\varphi^{(2)}(z)=[16ab+(2l-2c+d-1)]^{1/4}\varphi^{(2)}(z),\\
        &B\varphi^{(2)}(z)=-\sqrt{E_2-E_1}\varphi_2(z)=-[16ab+(2l-2c+d-1)]^{1/4}\varphi_2(z). 
    \end{split}
\end{equation}
Transform into Schr\"odinger form, we have wavefunctions
\begin{equation}\label{Coulomb-type IX Schrodinger wavefunctions r}
    \begin{split}
        &\psi_1(r)=\left(r-\frac{2l-2c+d-1-\sqrt{16ab+(2l-2c+d-1)^2}}{4a}\right)\exp\left[-ar-\frac{b}{r}+\frac{2(l-c)+d-1}{2}\ln{r}\right],\\
        &\psi_2(r)=\left(r-\frac{2l-2c+d-1+\sqrt{16ab+(2l-2c+d-1)^2}}{4a}\right)\exp\left[-ar-\frac{b}{r}+\frac{2(l-c)+d-1}{2}\ln{r}\right].
    \end{split}
\end{equation}
The SUSY wavefunction for the Schr\"odinger form is
\begin{equation}\label{Coulomb-type IX Schrodinger SUSY wavefunctions r}
    \psi^{(2)}(r)=\frac{2r\sqrt{16ab+(2l-2c+d-1)^2}}{4ar+2c-2l-d+2+\sqrt{16ab+(2l-2c+d-1)^2}}\exp\left[-ar-\frac{b}{r}+\frac{2(l-c)+d-1}{2}\ln{r}\right]
\end{equation}
The partner potential can be expressed in a Schr\"odinger form as
\begin{equation}\label{Coulomb-type IX Schrodinger SUSY potential r}
    \begin{split}
        V^{(2)}(r)&=a^2+\frac{a(2c-2l-d+1)}{r}+\frac{8ab+(1+2c-d-2l)(3+2c-d-2l)}{4r^2}+\frac{b(2l-2c+d-1)}{r^3}\\
        &+\frac{32a^2}{\left[1+2c-2l-d+4ar+\sqrt{16ab+(2l-2c+d-1)^2}\right]^2}\\
        &-\frac{8a}{r\left[1+2c-2l-d+4ar+\sqrt{16ab+(2l-2c+d-1)^2}\right]},
    \end{split}
\end{equation}
and the related Hamiltonian can be written as
\begin{equation}
    H_2=-\frac{d^2}{dr^2}+V^{(2)}(r),\quad \rho(r)H_2 \psi^{(2)}(r)=E_2\psi^{(2)}(r). 
\end{equation}
The partner Hamiltonian is given by the following formula
\begin{equation}\label{Coulomb-type IX SUSY radial Schrodinger Hamiltonian r}
    H_{S2}=h\rho(r)H_2h^{-1},\quad h=r^{-\frac{d-1}{2}},
\end{equation}
and the radial SUSY wavefunctions are expressed as
\begin{equation}\label{Coulomb-type IX radial Schrodinger SUSY wavefunctions r}
    \begin{split}
        &\psi_{S1}(r)=\left(r-\frac{2l-2c+d-1-\sqrt{16ab+(2l-2c+d-1)^2}}{4a}\right)\exp\left[-ar-\frac{b}{r}+(l-c)\ln{r}\right],\\
        &\psi_{S2}(r)=\left(r-\frac{2l-2c+d-1+\sqrt{16ab+(2l-2c+d-1)^2}}{4a}\right)\exp\left[-ar-\frac{b}{r}+(l-c)\ln{r}\right].
    \end{split}
\end{equation}
The SUSY wavefunction for radial Schr\"odinger form is then
\begin{equation}\label{Coulomb-type IX SUSY radial Schrodinger wavefunction r}
    \psi_S^{(2)}(r)= \frac{2r\sqrt{16ab+(2l-2c+d-1)^2}}{4ar+2c-2l-d+2+\sqrt{16ab+(2l-2c+d-1)^2}}\exp\left[-ar-\frac{b}{r}+(l-c)\ln{r}\right].
\end{equation}
while the partner potential for radial Schr\"odinger is 
\begin{equation}\label{Coulomb-type IX radial Schrodinger SUSY potential r}
    \begin{split}
        V_S^{(2)}(r)&=a^2+\frac{a(2c-2l-d+1)}{r}+\frac{2 a b+c (c-d-2 l+2)+2 l (d+l-2)}{r^2}+\frac{b(2l-2c+d-1)}{r^3}\\
        &+\frac{32a^2}{\left[1+2c-2l-d+4ar+\sqrt{16ab+(2l-2c+d-1)^2}\right]^2}\\
        &-\frac{8a}{r\left[1+2c-2l-d+4ar+\sqrt{16ab+(2l-2c+d-1)^2}\right]}.
    \end{split}
\end{equation}
The Fig.\ref{fig case9} provides the potential and the wavefunctions of Schr\"odinger-like equation and their SUSY partners in spherical coordinate. In this case, we set $a=\frac{1}{2}, b=4, c=15, l=1$ and $d=-5$. Due to $r\geq 0$, we only plot the part of $r>0$. The partner potential $V_S^{(2)}$ is similar to the initial potential $V_S$ and they have the same singularity at origin in Fig.\ref{f case9 V}. Thus, no new singularity appears in $V_S^{(2)}$, which is consistent with the structure of partner potentials in SUSYQM. The partner wavefunction $\psi_S^{(2)}$ and wavefunctions $\psi_{S1}$ and $\psi_{S2}$ have zero at origin in Fig.\ref{f case9 wf}.  
It is also easy to verify that 
\begin{equation}
   H_{S \rho 2}=\rho(r)\left[-\frac{d^2}{dr^2}-\frac{d-1}{r}\frac{d}{dr}-\frac{l(l+d-2)}{r^2}+V_S^{(2)}\right],\quad  H_{S\rho2}\psi_S^{(2)}(r)=E_2\psi_S^{(2)}(r). 
\end{equation}
Supercharges in radial Schr\"odinger form is 
\begin{equation}
    \tilde{A}_S(r)=hfA(z)f^{-1}h^{-1}, \quad \tilde{A}_S(r)^{\#}=hfB(z)f^{-1}h^{-1},\quad f=\exp\left[-ar-\frac{b}{r}-(c-l)\ln{r}\right], \quad h=r^{-\frac{d-1}{2}}.
\end{equation}
We have $\frac{d}{dz}=\frac{d}{dr}$. We can verify that 
\begin{equation}
    \tilde{A}_S \psi_{S2}(r)=\sqrt{E_2-E_1}\psi_S^{(2)}(r),\quad \tilde{B}_S\psi_S^{(2)}(r)=-\sqrt{E_2-E_1}\psi_{S2}(r). 
\end{equation}
\begin{equation}
     \bar{H}_{S\rho1}=\tilde{B}_S\tilde{A}_S, \quad  \bar{H}_{S\rho2}=\tilde{A}_S \tilde{B}_S,\quad  \bar{H}_{S\rho1}=H_{S\rho}-E_1,\quad  \bar{H}_{S\rho2}=\rho(r)H_{S\rho2}-E_1. 
\end{equation}

\section{SUSYQM for higher level n}\label{SUSYQM for higher level n}
\begin{figure}[H]
  \centering
  \begin{subfigure}{0.48\linewidth}
    \centering
    \includegraphics[width=\linewidth]{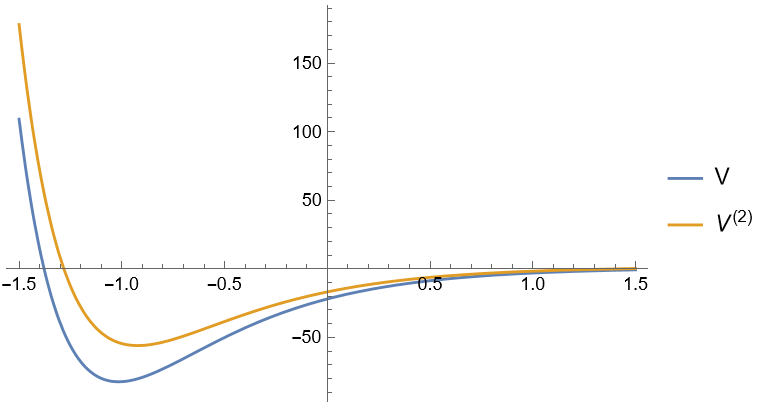}
    \caption{Schr\"odinger-like potential $V$ and partner Schr\"odinger potential $V^{(2)}$.}
    \label{f case2 n=10 V}
  \end{subfigure}
  \hfill
  \begin{subfigure}{0.48\linewidth}
    \centering
    \includegraphics[width=\linewidth]{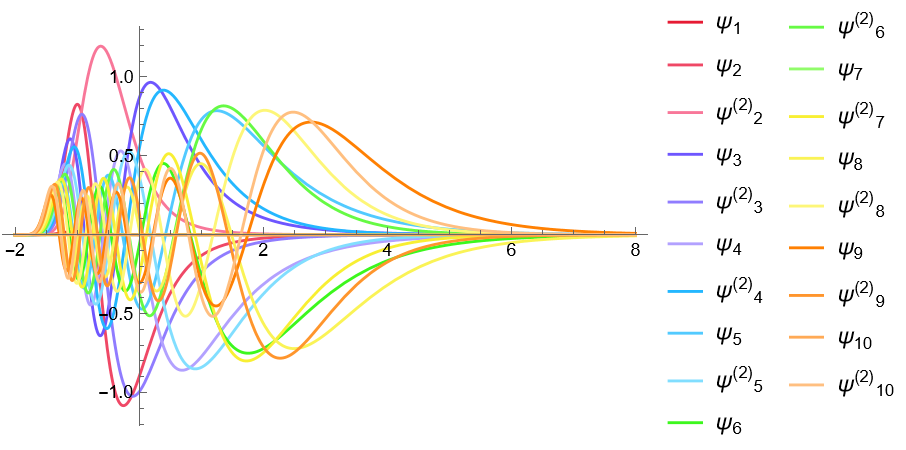}
    \caption{Wavefuntions $\psi_1, \psi_2$ and partner wavefunction $\psi^{(2)}$.}
    \label{f case2 n=10 wf}
  \end{subfigure}
  \caption{SUSY transformation in wavefunctions and Schr\"odinger potential.}
  \label{fig case2 n=10}
\end{figure}

\begin{figure}[H]
  \centering
  \begin{subfigure}{0.48\linewidth}
    \centering
    \includegraphics[width=\linewidth]{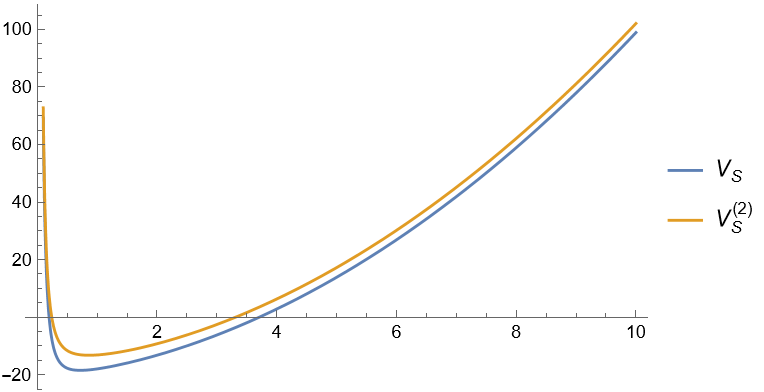}
    \caption{Radial Schr\"odinger-like potential $V_S$ and partner Schr\"odinger-like potential $V_S^{(2)}$.}
    \label{f case8 n=10 V}
  \end{subfigure}
  \hfill
  \begin{subfigure}{0.48\linewidth}
    \centering
    \includegraphics[width=\linewidth]{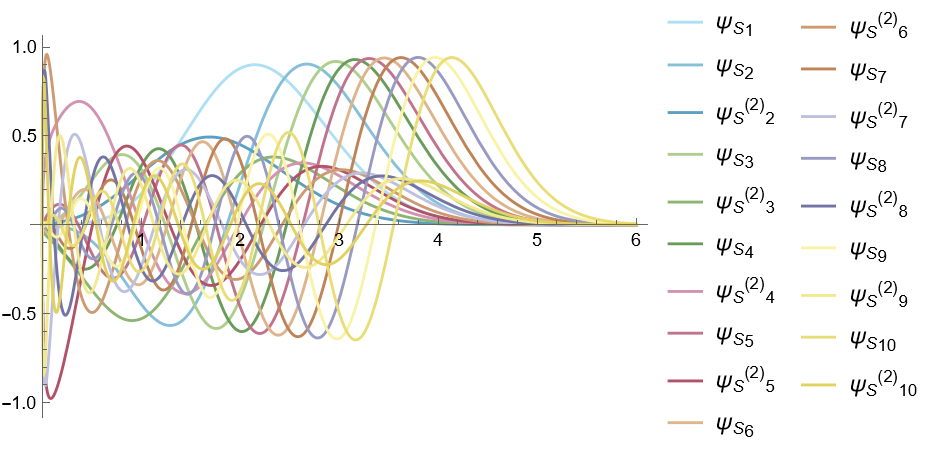}
    \caption{Radial wavefuntions $\psi_{S1}, \psi_{S2}$ and partner wavefunction $\psi_S^{(2)}$.}
    \label{f case8 n=10 wf}
  \end{subfigure}
  \caption{SUSY transformation in radial wavefunctions and radial Schr\"odinger-like potential in spherical coordinate.}
  \label{fig case8 n=10}
\end{figure}

Here we want to show how the SUSY transformation can be applied for a higher level $n$. Let us emphasis that the use of higher excited states for exactly solvable systems can be done in different ways depending on choice of chains and then issues of regular and singular Hamiltonians arise and need to be carefully studied. Here we want to point out how there is regular Hamiltonian and physical states that can be obtain via higher values of $n$. This depends upon choices of parameters which satisfy the parameter constraints. The case of $n$ means that $n+1$ states will be obtained through the Bethe Ansatz approach and related set of algebraic equations for the roots. This provides further possibilities for the seed solutions and then for the supercharges. Due to the difficulty of solving Bethe ansatz then numerical methods need to be used to obtain Bethe roots. We use the cases in \ref{Morse-type potentials II} and \ref{Coulomb-type potentials VIII} as examples to investigate the SUSY transformation for $n=10$ in Fig.\ref{fig case2 n=10} and Fig.\ref{fig case8 n=10}, respectively. There are 11 independent wavefunctions for $n=10$. The wavefunction with the lowest energy is chosen as a seed function $\psi_{\Lambda}$ and one of the wavefuntions from the rest is $\psi_E$.  

We set parameters $a=b=d=\alpha=1$ for \ref{Morse-type potentials II} and present the comparison of potential and wavefunctions of Schr\"odinger-like equation to their SUSY partners in Fig.\ref{fig case2 n=10}. The shapes of $V$ and $V^{(2)}$ are similar and no singularity appears in the $V^{(2)}$ in Fig.\ref{f case2 n=10 V}. We compare the seed wavefunction $\psi_1$, $\psi_2$ and the partner wavefunction $\psi^{(2)}$ in Fig.\ref{f case2 n=10 wf}. The shape of partner wavefunction $\psi^{(2)}$ is similar to the seed function $\psi_1$.

We also use Fig.\ref{fig case8 n=10} to illustrate SUSY transformation of radial Schr\"odinger-like equation for $n=10$ in spherical coordinate. We set $a=b=c=l=1$ and $d=2$ and plot the figure in $r>0$. Fig.\ref{f case8 n=10 V} is the radial Schr\"odinger-like potential $V_S$ and its partner potential $V^{(2)}_S$. $V_s$ and $V^{(2)}_S$ have the same singularity at origin. Therefore, the absence of new singularities in $V^{(2)_S}$ is consistent with the properties of SUSYQM partner potentials. Fig.\ref{f case8 n=10 wf} is the wavefunction of $\psi_{S1}$, $\psi_{S2}$ and $\psi^{(2)}_S$. $\psi_{S1}$ and $\psi_S^{(2)}$ are similar. 

\section{Conclusion}
We have developed an approach to construct new quasi-exactly solvable systems based on supersymmetric quantum mechanics and the Bethe Ansatz method. The Bethe Ansatz method consists in finding a set of algebraic equations and parameter constraints which provide existence of polynomial solutions. We have applied the framework to QES systems with $sl(2)$ hidden algebra which were classified in the literature and studied from different perspectives. Those QES systems can take different forms as they may be in the Schrodigner form but also in radial form or even involve weight function. In this context, we have used their corresponding standard ODE form in order to provide a more unified treatment. This also offers a formalism of interest beyond the scope of the current paper which may be applicable to other quasi-exactly solvable systems which do not possess an underlying $sl(2)$ hidden symmetry algebra. The ODE setting beyond its wider applicability is a convenient formalism for the Bethe Ansatz method, as its rely on such a standard form for calculating polynomial solutions in terms of the Bethe root. We have given explicit analysis of the SUSY transformation in QES models for $n=1$ and plotted the relationships between the initial potentials and wavefunction and their SUSY partners. For higher level states, we have analyzed the QES models with the roots obtained numerically and presented the results for $n=10$. We have also look for singularity in Hamiltonian and for all cases ensured that no new singularity is created for range of parameters. This is important as the QES systems do not have solutions in terms of a well-known classical orthogonal polynomials for which the patterns of zeroes have been studied extensively. This shows how supersymmetry in quantum mechanics can be a useful tool and used in a complementary way with the Bethe Ansatz method to create a variety of new quantum systems which are quasi-exactly solvable. 

\section*{Acknowledgement}

IM was supported by the Australian Research Council Future Fellowship FT180100099. YZZ was supported by the Australian Research Council Discovery Project DP190101529.

\bibliographystyle{unsrt}
\bibliography{Reference.bib}

\end{document}